\documentclass[twoside,12pt]{article}

\newif\iflongversion
\usepackage[utf8]{inputenc}
\usepackage{graphicx}
\usepackage[all]{xy}
\usepackage{amssymb}
\usepackage{amsmath}
\usepackage{amsthm}
\usepackage{mathtools}
\usepackage{verbatim}
\usepackage{latexsym}
\usepackage{array}
\usepackage{apacite}
\renewcommand{\Cup}{\bigcup}
\renewcommand{\Cap}{\bigcap}

\newcommand{\e}{\varepsilon}

\newcommand{\f}{\varphi}

\newcommand{\s}{\sigma}

\newcommand{\es}{\varnothing}

\newcommand{\A}{\mathcal{A}}

\newcommand{\CC}{\mathcal{C}}

\newcommand{\E}{\mathcal{E}}

\newcommand{\FF}{\mathcal{F}}
\renewcommand{\L}{\mathcal{L}}
\newcommand{\M}{\mathcal{M}}
\newcommand{\I}{\mathcal{I}}

\newcommand{\X}{\mathcal{X}}
\newcommand{\U}{\mathcal{U}}

\newcommand{\N}{\mathbb{N}} %natural numbers
\newcommand{\Z}{\mathbb{Z}} %integers
\newcommand{\R}{\mathbb{R}} %reals
\newcommand{\cat}{{}^\frown}  %catenate

\newcommand{\PP}{\mathcal{P}}             %power set
\newcommand{\Po}{\mathcal{P}}             %power set

\newcommand{\la}{\langle}
\newcommand{\ra}{\rangle}

\newcommand{\LTS}{{\operatorname{LTS}}}
\newcommand{\suf}{{\operatorname{suf}}} % sufficient
\newcommand{\deginsuf}{{\operatorname{degins}}}

\newcommand{\Pred}{\operatorname{Pred}} %regular

\newcommand{\rest}{\!\restriction\!}
\newcommand{\restl}{\restriction}  % restriction for lower case
\newcommand{\SM}{\operatorname{SM}}
\newcommand{\aSM}{\operatorname{aSM}}
\newcommand{\dom}{\operatorname{dom}}
\newcommand{\ran}{\operatorname{ran}}

\renewcommand{\le}{\leqslant} 

\renewcommand{\subset}{\subseteq}
\newcommand{\LL}{\mathcal{L}} 

\newcommand{\Land}{\bigwedge}
\newcommand{\Lor}{\bigvee}

\newtheorem*{Thm*}{Theorem}
\newtheorem{Thm}{Theorem}[section]
\newtheorem{Lemma}[Thm]{Lemma}
\newtheorem{Cor}[Thm]{Corollary}
\newtheorem{Prop}[Thm]{Proposition}

\usepackage{algorithm} 
\usepackage[noend]{algpseudocode} 
\usepackage{xcolor}
\newcommand{\transf}{\tau}
\newcommand{\labelf}{h}
\newcommand{\obs}{s}
\newcommand{\stateSp}{X}
\newcommand{\action}{m}

\graphicspath{{./figures/}}
\usepackage[scriptsize]{subfigure}

\theoremstyle{definition}
\newtheorem{Def}[Thm]{Definition}

\newtheorem{Ex}[Thm]{Example}
\newtheorem{Exs}[Thm]{Examples}
\newtheorem*{claim}{Claim}

\theoremstyle{remark}
\newtheorem*{Remark}{Remark}
\newtheorem{RemarkN}[Thm]{Remark}

\title{An Enactivist-Inspired Mathematical \\ Model of Cognition}

\author{Vadim Weinstein, Basak Sakcak, Steven M. LaValle%
  \thanks{This work was supported by a European Research Council
    Advanced Grant (ERC AdG, ILLUSIVE: Foundations of Perception
    Engineering, 101020977), Academy of Finland (projects PERCEPT
    322637, CHiMP 342556), and Business Finland (project HUMOR
    3656/31/2019).  All authors are with the Center of Ubiquitous
    Computing, Faculty of Information Technology and Electrical
    Engineering, University of Oulu, Finland {\tt\small (e-mail:
      firstname.lastname@oulu.fi).}}%
}

\date{\today}

\usepackage[text={17cm, 23.5cm}, centering]{geometry}

\usepackage{tikz}

\usetikzlibrary{scopes}

\begin{document}

\maketitle
\thispagestyle{empty}

\vspace*{-1.0cm}

\begin{abstract}
  In this paper we start from the philosophical position in cognitive
  science known as enactivism. We formulate five basic enactivist
  tenets that we have carefully identified in the relevant literature
  as the main underlying principles of that philosophy.  We then
  develop a mathematical framework to talk about cognitive systems
  (both artificial and natural) which complies with these enactivist
  tenets. In particular we pay attention that our mathematical
  modeling does not attribute contentful symbolic representations to
  the agents, and that the agent's brain, body and environment are
  modeled in a way that makes them an inseparable part of a greater
  totality.  The purpose is to create a mathematical foundation for
  cognition which is in line with enactivism. We see two main benefits
  of doing so: (1) It enables enactivist ideas to be more accessible
  for computer scientists, AI researchers, roboticists, cognitive
  scientists, and psychologists, and (2) it gives the philosophers a
  mathematical tool which can be used to clarify their notions and
  help with their debates.
  
  Our main notion is that of a sensorimotor system which is a special
  case of a well studied notion of a transition system. We also
  consider related notions such as labeled transition systems and
  deterministic automata. We analyze a notion called
  \emph{sufficiency} and show that it is a very good candidate for a
  foundational notion in the ``mathematics of cognition from an
  enactivist perspective''. We demonstrate its importance by proving a
  uniqueness theorem about the minimal sufficient refinements (which
  correspond in some sense to an optimal attunement of an organism to
  its environment) and by showing that sufficiency corresponds to
  known notions such as sufficient history information spaces.
  \iflongversion We then develop other related notions such as
  \emph{degree of insufficiency}, \emph{universal covers},
  \emph{hierarchies}, \emph{strategic sufficiency}.  \fi In the end,
  we tie it all back to the enactivist tenets.
\end{abstract}

\clearpage
\tableofcontents
\clearpage

% FIXME: Add English
% Concatenation
% Meaning emerge
% Coupling
% Meaningless
% Extremes: quasipolicy, quasifilter.
% Foundational
% Number of definitions / number of theorems
% 5.3
% Key tenets of enactivism. What are the mathematical ways to
% model them?

\section{Introduction: Mathematizing Enactivism}

The premise of this paper is to lay down a logical framework for
analyzing agency in a novel way, inspired by enactivism. Classically,
mathematical and logical models of cognition are in line with the
cognitivist paradigm in that they rely on the notion of symbolic
representation and do not emphasize embodiment or enactment
\cite{newellsimon72, fodor2008lot, computational_brain2009,
  bsmpresc16}.  Cognitivism presumes that the world possesses
objective structure and the contentful information of this structure
is acquired and represented by the cognitive agent. This aligns well
with the classical model-theoretic paradigm. Given a first-order
relational vocabulary $L$, an $L$-model is a tuple
$\M=(M,R_0,\dots,R_k)$ where $R_i\subset M^{\# R_i}$ is a relation
which corresponds to the relational symbol~$R_i$. Here $\# R$ is the
arity $n$ of the relation $R$ and $M^{\# R}=M^n$ is the set of $n$-tuples
of elements of~$M$ which itself is just a set, as is common in model
theory. The first-order language over the vocabulary $L$ then
describes relationships that can hold or not in the model. For example
if $R_0$ and $R_1$ are unary relations, the formula
$\exists x(R_0(x)\land R_1(x))$ says that there exists an element $x$
which belongs to the interpretations of both relations $R_0$ and
$R_1$. In the cognitivist analogy, the agent possesses (``in its
head'') formulas of the language and $\M$ is the world or the
environment of the agent. If the formulas possessed by the agent hold
in $\M$, then the agent's representation of the world is correct;
otherwise, it is incorrect. Such view of cognitive agency is rejected
by the enactivists either weakly or strongly depending on the branch
of enactivism. For example, radical enactivism
\cite{huttomyin1,huttomyin2} rejects this view strongly. Our question
for this paper is: What would the mathematical logic of cognition look
like, if even the radical enactivists were to accept~it?

We do not take part in the cognitivist-enactivist, or the
represen\-tationalist-anti\-repre\-senta\-tionalist debate
\cite{embodiment_debate11, ORegan2012-OREDOJ,
  gallagher2018decentering, Fuchs20}.  Rather, we take a somewhat
extreme enactivist and antirepresentational view as our axiomatic
starting point and as a theoretical explanatory target. Then we
develop a mathematical theory that attempts to account for cognition
in a way congruent with enactivism.  Even though most forms of
enactivism (even radical ones) have room for representation, it is not
our main goal at the moment to bridge the gap between ``basic minds''
and ``scaffolded minds'', to use terminology of
\cite{huttomyin2}. Thus, in this terminology, we are going to explore
a mathematical/logical model (only) of \emph{basic minds}.

\begin{figure}
    \centering
    \includegraphics[width=0.6\textwidth]{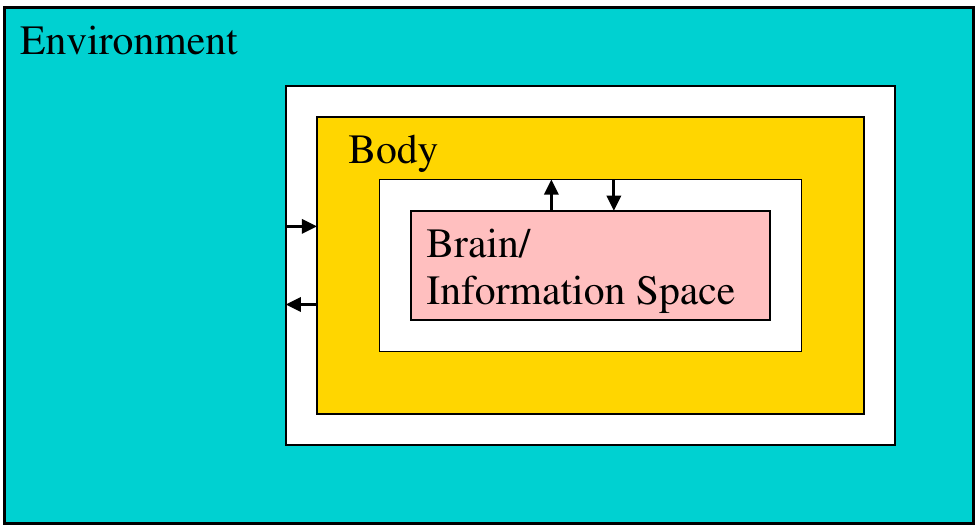}
    \caption{The environment, body, and brain will be modeled as inseparable parts of a coupled transition system.}
    \label{fig:ebb}
\end{figure}

The following ``axioms'' we take as fundamentals for our work:

\begin{enumerate}
\item[(EA1)] Embodiment. ``From a third-person perspective the organism-environment 
  is taken as the explanatory unit'' \cite{gallagher2017enactivist}. 
  The environment, the body, and the brain are inseparable
  parts of the system they form by coupling; see Figure \ref{fig:ebb}. They cannot be
  meaningfully understood in isolation from each other. ``Mentality is
  in all cases concretely constituted by, and thus literally consists
  of, the extensive ways in which organisms interact with their
  environments, where the relevant ways of interacting involve, but
  are not exclusively restricted to, what goes on in
  brains.'' (Embodiment Thesis~\cite{huttomyin1})
\item[(EA2)] Groundedness. The brain does not ``acquire'' or ``possess''
  contentful states, representations, or manipulate semantic information in any
  other way. ``Mentality-constituting interactions are grounded in,
  shaped by, and explained by nothing more, or other, than the history
  of an organism's previous interactions. Nothing other than its
  history of active engaging structures or explains an organism's
  current interactive tendencies.'' (Developmental-Explanatory
  Thesis~\cite{huttomyin1}).
\item[(EA3)] Emergence. The crucial properties of the brain-body-environment
  system from the point of view of cognition emerge from the
  embodiment, the brain-body-environment coupling, the situatedness,
  and the skills of the agent. The agent's and the environment's prior
  structure come together to facilitate new structure which emerges
  through the sensorimotor engagement. ``[T]he mind and world arise
  together in enaction, [but] their manner of arising is not arbitrary''
  (i.e., structured)~\cite{varela1992embodied} 
\item[(EA4)] Attunement. Agents differ in their ways of attunement and adaptation
  to their environments, and in the skills they have.  A \emph{skill}
  is a potential possibility to engage \emph{reliably} in complex sensorimotor
  interactions with the environment.~\cite{gallagher2017enactivist}
\item[(EA5)] Perception. Sensing and perceiving are not the same thing.
 Perception arises from skillful sensorimotor activity. To perceive is to become
 better attuned to the environment.~\cite{oregannoe2001sensorimotor, noe2004action}
 ``Perception and action, sensorium and motorium, are linked together as successively
 emergent and mutually selecting patterns''.~\cite{varela1992embodied}.
\end{enumerate}
\noindent The mathematics we use to capture those ideas is a mixture of 
known and new concepts from logic, theoretical
robotics, (non-)deterministic automata and transition
systems theory, Kripke model theory, team semantics, and dynamical
systems.  It will also build upon the \emph{information spaces} framework, introduced in \cite{lavalle06} as a unified way to model sensing, actuation, and planning in robotics; the framework itself builds upon earlier ideas such as dynamic games with imperfect information \cite{BasOls95,VonMor44}, control with imperfect state information \cite{Ber01,KumVar86}, knowledge states \cite{Erd93,LozMasTay84}, perceptual equivalence classes \cite{DonJen91,Don95}, maze and graph-exploring automata \cite{BluKoz78,FraIlcPeePelPel05,Sha52}, and belief spaces \cite{KaeLitCas98,RoyGor03}. 

Although information spaces refer to ``information'', they are not directly related to Shannon's \emph{information theory} \cite{Sha48}, which came later than von Neumann's use of information in the context of sequential game theory. Neither does ``information'' here refer to content-bearing information. One important intuition behind the information in information spaces is 
that more information corresponds to narrowing down the space of possibilities (for example
of future sensorimotor interactions).

The main mathematical concept of this paper is a \emph{sensorimotor
  system} (SM-system), which is a special case of a transition
system. Sensorimotor systems can describe the body-brain system, the
body-environment system as well as other parts of the
brain-body-environment system. Given two SM-systems they can be
\emph{coupled} to produce another (third) SM-system. Mathematically,
the coupling operation is akin to a direct product. Now we need a way
to compare different ways in which the agent can be coupled to the
environment. Classically, agents are compared by the number of
``correct'' statements that they make about the environment in which
they are. Since our aim is to avoid representational talk, we do
something else. Instead, we introduce several notions that describe
the coupling of the agent and the environment from an outside
perspective. The main notion is that of \emph{sufficiency}. This
notion does not compare ``internal'' models of the agent to
``external'' states of affairs. Rather it asks whether the way in
which the agent engages in sensorimotor patterns is well
structured. In some sense \emph{sufficiency} compares the
sensorimiotor capacity of the agent \emph{to itself} by asking whether
the sensorimotor patterns (in a given environment) are a reliable
predictor (from an outside perspective) of the future of the very same
sensorimotor behaviour. We then introduce several related notions.
The \emph{degree of insufficiency} is a measure by which various
agents can be compared in their coupling versatility
(Def~\ref{def:DegreeOfInsufficiency}). \emph{Minimal sufficient
  refinement} is a concept that can be used in the most vivid ways to
illustrate how the sensorimotor interaction ``enacts'' properties of
the brain-body-environment system. The notion of minimal sufficient
refinement ties together mathematics of sensorimotor systems and the
philosophical ideas of emergence, structural coupling and enactment of
the ``world we inhabit'' \cite[cf.]{varela1992embodied}; see
Example~\ref{ex:Mini}. We prove the uniqueness of minimal sufficient
refinements (Theorem~\ref{thm:MinSufRef1}) and point out their
connection to the notions of bisimulation and sufficient information
mappings. \emph{Strategic sufficiency} is a mathematically more
challenging concept, but has appealing properties in the philosophical
and practical sense. A sensor mapping is strategically sufficient
\emph{for} some subset of the state space $G$, if that sensor can (in
principle) be used by the agent to reach $G$. Again, any sensor
mapping has minimal strategic refinements, but this time they are not
unique. Different minimal refinements in this case can be thought of
as different adaptations to the same environmental demands.

Mathematically, sufficiency is a relative concept to some known
notions in theoretical computer science and robotics: that of
\emph{bisimulation} in automata and Kripke model theory
\cite{GORANKO2007249}, and \emph{sufficient information mappings} in
information spaces theory \cite{lavalle06}.

Minimal sufficient refinements lead to unique classifications of
agent-environment states that ``emerge'' from the way in which the
agent is coupled to the environment, not merely from the way the
environment is structured on its own. Thus, the world is
simultaneously objectively existing (from the global ``god''
perspective), but also ``brought about'' by the agent.

This should be enough to answer the two questions that, according to
\cite{dipaoloenactivecon2018}, any embodied theory of cognition should be able to
provide precise answers to: What is its conception of bodies? What
central role do bodies play in this theory different from the roles
they play in traditional computationalism?

Section~\ref{sec:TransSyst} introduces the basics
of transition and SM-systems, their coupling, and other
mathematical constructs such as quotients. Section~\ref{sec:IllustrSM}
illustrates the introduced notions with detailed
examples. Section~\ref{sec:SRE} introduces the notion of
sufficiency, sufficient refinements, and minimal sufficient refinements. We will prove the uniqueness theorem for the latter
and illustrate the notions in a computational setting. We
will explore the importance of sufficiency and related notions
 for the enactivist way of looking at cognitive
organization.
\iflongversion
Section~\ref{sec:Coverings} formalizes the idea
that the agent can never distinguish between returning to the 
same place again and visiting a very similar place. Section~\ref{sec:OtherImportantNotions}
collects other important ideas that stem from this research,
but which are still in a preliminary phase. It contains
weakenings of the notion of sufficiency, a logical approach
to the present framework, and considerations about
a fractal-like hierarchy in transition systems (what if the
states of a transition system are themselves transition systems?)
\fi
Finally, Section~\ref{sec:Discussion} ties  
the mathematics back to the philosophical premises.

\section{Transition Systems and Sensorimotor Systems}
\label{sec:TransSyst}

At the most abstract level, the central concept for our 
mathematical theory is that of a \emph{transition system}:

\begin{Def}
  A \emph{transition system} is a triple $(X,U,T)$ where $X$ is the
  \emph{state space} (mathematically it is just a set), $U$ is the set
  of names for outgoing transitions (another set), and
  $T\subset X\times U\times X$ is a ternary relation.
\end{Def}

\noindent The intuitive interpretation of $(X,U,T)$ is that it is
possible to transition from the state $x_1\in X$ to the state $x_2\in X$
via $u\in U$ iff $(x_1,u,x_2)\in T$. We use the notation
$x_1\stackrel{u}{\rightarrow} x_2$ to mean that $(x_1,u,x_2)\in T$.  Our
notion of transition system is often called a \emph{labeled}
transition system because each potential transition has a name or
label, $u \in U$.  However, we drop the term ``labeled'' because in
Section \ref{sec:lts} we will introduce a version of transition
systems in which the states are relabeled, thereby introducing a new
kind of labeling.  Note that when working with such transition
systems, we are safely within the realm of the
Developmental-Explanatory Thesis~(EA2).

\begin{Def}\label{def:IsoBisim}
  Let $\X=(X,U,T)$ and $\X'=(X',U',T')$ be transition systems.  An
  \emph{isomorphism} is a bijective function $f\colon X\to X'$ such
  that for all $x_1,x_2\in X$ and $u\in U$ we have
  $(x_1,u,x_2)\in T\iff (f(x_1),u,f(x_2))\in T'$.  A \emph{bisimulation} is a
  relation $R\subset X\times X'$ such that for all $(x_1,x_1')\in R$, all
  $u\in U$ and all $x_2\in X$, we have that if
  $(x_1,u,x_2)\in T$, then there exists $x_2'\in X'$ with
  $(x_1',u, x_2')\in T'$ and $(x_2,x_2')\in R$.

  The notation $\X\cong \X'$ means that $\X$, $\X'$ are isomorphic,
  and $\X\sim \X'$ means that there is a bisimulation $R$ such that
  $X= \dom(R)$ and $X'= \ran(R)$. We speak of \emph{automorphism}
  and \emph{autobisimulation}, if $\X=\X'$.
\end{Def}

We are ready to make the first observation:

\begin{Prop}
  If $\X\cong\X'$, then $\X\sim\X'$. 
\end{Prop}
\begin{proof}
  Let $f$ be an isomorphism $f\colon X\to X'$.  Then
  $R=\{(x_1,x_2)\in X\times X'\mid x_2=f(x_1)\}$ is a bisimulation.
\end{proof}

\subsection{Transition systems as a unifying concept}

\label{ssec:TS}

There are several ways in which transition systems and their relatives
appear in the literature relevant to us. 
\begin{Exs}\label{exs:TS_LitRelExs} 
  Here are some examples how transition systems are related to 
  concepts that appear in relevant literature.
  Let $(X,U,T)$ be a transition system.
  \begin{enumerate}
  \item Let $x_0\in X$ and $F\subset X$. Let
    $\hat T\colon X\times U\to \Po(X)$ be defined by
    $\hat T(x,u)=\{x_2\in X\mid x_1\stackrel{u}{\rightarrow} x_2\}$. Then
    $(X,U,\hat T,x_0,F)$ is a \emph{nondeterministic automaton}. If
    in addition $X$ and $U$ are finite, then it is a
    \emph{nondeterministic finite automaton}~(NFA).
  \item \label{rem:TS2} Let $\tilde T\colon X\times X\to \Po(U)$ be
    the function
    $\tilde T(x_1,x_2)=\{u\in U\mid x_1\stackrel{u}{\rightarrow} x_2\}$.  Then
    $\tilde T(x_1,x_2)$ is the set of all $u$ that take $x_1$
    to~$x_2$. Then, $(X,\tilde T)$ is a labeled directed graph in which the
    labels are subsets of~$U$. Another way to think of it is as a
    labeled directed multigraph: the multiplicity of the edge from $x_1$
    to $x_2$ is $n=|\tilde T(x_1,x_2)|$ and these $n$ edges are labeled by
    the labels from the set~$\tilde T(x_1,x_2)$.
  \item \label{rem:TS3} If for all $x_1\in X$ and $u\in U$ there is a 
    unique $x_2\in X$
    with $x_1\stackrel{u}{\rightarrow} x_2$, let
    $\tau\colon X\times U\to X$ be the function defined such that
    $\tau(x_1,u)=x_2$ iff $x_1\stackrel{u}{\rightarrow} x_2$. 
    Let $x_0\in X$ and $F\subset X$.  Then $(X,U,\tau,x_0,F)$ is a
    \emph{deterministic automaton}, and if $X$ and $U$ are finite,
    then it is a \emph{deterministic finite automaton} (DFA). Without $F$, 
    $(X,U,\tau,x_0)$ also satisfies
    the definition of the \emph{temporal filter} of
    \cite[§4.2.3]{lavalle12}. In this case $X$ is the
    \emph{information space} or the \emph{I-space} (usually denoted by
    $\I$ instead of~$X$), and $U$ is the observation space (usually
    denoted by~$Y$ instead of~$U$).
  \end{enumerate}
\end{Exs}

\subsection{Information spaces and Filters}
\label{sec:InfoSpaces} 

We can reformulate the notion of a \emph{history information space} 
introduced by \cite{lavalle06} as follows.  In this 
context, $X$ is an external state space that characterizes the robot's 
configuration, velocity, and environment, $U$ is an action space, $f$ is a 
state transition mapping that produces a next state from a current state and
action, $h$ is a sensor mapping that maps states to observations, and $Y$ 
is a sensor observation space. As
in \cite{lavalle06}, for each $x\in X$, let $\Psi(x)$ be a finite set
of ``nature sensing actions'' and for each $x\in X$ and $u\in U$ let
$\Theta(x,u)$ be a finite set of ``nature actions''.  Let
$X_\Psi=\{(x,\psi)\mid \psi\in \Psi(x)\}$ and let
$h\colon X_\Psi\to Y$ be a ``sensor mapping'' where $Y$ is a set
called the ``observation space''.  Let
$X_{\Theta}=\{(x,u,\theta)\mid \theta\in \Theta(x,u)\}$ and let
$f\colon X_{\Theta}\to X$ be the ``transition function''. 

\begin{Def}
    A \emph{valid history I-state for $X,\Psi,\Theta,f$} is a sequence
    $(u_0,y_0,\dots,u_{k-1},y_{k-1})$ of length $2k$ for which there
    exist $\bar x=(x_0,\dots,x_{k-1})$,
    $\bar\psi=(\psi_0,\dots,\psi_{k-1})$ and
    $\bar\theta=(\theta_0,\dots,\theta_{k-2})$ such that for all $i<k$
    we have
    \begin{enumerate}
    \item $\theta_i\in \Theta(x_i,u_i)$,
    \item if $i<k-1$, then $x_{i+1}=f(x_i,u_i,\theta_i)$,
    \item $\psi_i\in \Psi(x_i)$,
    \item $y_i=h(x_i,\psi_i)$.
    \end{enumerate}
    In this case we say that $(u_0,y_0,\dots,u_{k-1},y_{k-1})$ is
    \emph{witnessed} by $\bar x$, $\bar\psi$ and $\bar \theta$. 
\end{Def}

Now let
$\I$ be the set of all valid paths for $X,\Psi,\Theta,f$.  For all
$k\in\N$, all $\bar x\in X^{k-1}$, all
$\bar\psi=(\psi_0,\dots,\psi_{k-1})$ and all
$\bar\theta=(\theta_0,\dots,\theta_{k-2})$, let
$\I^k(\bar x,\bar\psi,\bar\theta)$ be the set of all valid paths
$(u_0,y_0,\dots,u_{k-1},y_{k-1})$ witnessed by $\bar x$, $\bar\psi$,
and $\bar\theta$.  Now let $T\subset \I\times (U\times Y)\times\I$
be defined by
\begin{align*}
  T=\Big\{(\eta,(u,y),\eta')\mid &\text{ there exist }k\in\N,
                                   \bar x=(x_0,\dots,x_{k-1}),
                                   \bar\psi=(\psi_0,\dots,\psi_{k-1}),\\
                                 &\bar\theta=(\theta_0,\dots,\theta_{k-2}),
                                   \theta\in \Theta(x_{k-1},u)\text{ and }
                                   \psi\in \Psi(f(x_{k-1},u,\theta))\\
                                 &\text{such that }\\
                                 &\eta\in \I^k(\bar x,\bar\psi,\bar\theta)
                                   \land
                                   \eta'\in \I^{k+1}
                                   (\bar x',\bar\psi',\bar\theta'),\\
                                 &\text{where }
                                   \bar x'=\bar x\cat(f(x_{k-1},u,\theta)),
                                   \ \bar\psi'=\psi\cat (\psi),\text{ and }
                                   \bar\theta'=\theta\cat (\theta)\Big\}.
\end{align*}
Here, $x\cat y$ is the concatenation of sequences $x$ and~$y$.  Then
$(\I, U\times Y, T)$ is the \emph{history I-space transition system}.
In the terminology of \cite[Ch~11]{lavalle06}, 
the valid path $(u_0,y_0,\dots,u_{k-1},y_{k-1})$

Suppose for each $x,y\in X$ there is at most
one $u\in U$ with $x\stackrel{u}{\rightarrow} y$. Let
$$E_T=\{(x,y)\in X^2\mid \exists u\in U(x\stackrel{u}{\rightarrow}y)\},$$
and let $l\colon E_T\to U$ be defined so that $l((x,y))$ is the unique
$u$ such that $x\stackrel{u}{\rightarrow} y$.  Then $(X,E_T,l,x_0)$
with $x_0\in X$ is a passive I-state graph as in
\cite[Def~1]{okaneshell}.

\begin{Def}\label{def:AutomatonStar}
  Let $\X=(X,U,T)$ be a transition system. If for all
  $(x,u)\in X\times U$ there is a unique $y\in X$ with $(x,u,y)\in T$, then
  we denote the function $(x,u)\mapsto y$ by $\tau$, and write
  $(X,U,\tau)$ instead of $(X,U,T)$. In this case we call $\X$ an
  \emph{automaton}. Note that usually in computer science literature an automaton
  is finite and also has an initial state and a set of accepting states, but we
  do not have those in our definition.

  For automata we also use the notation $x*u=\tau(x,u)$ and if
  $\bar u=(u_0,\dots,u_{k-1})$, then $x*\bar u$ is defined by
  induction for $k>1$ as follows:
  $x*(u_0,\dots,u_{k-1})=(x*(u_0,\dots,u_{k-2}))*u_{k-1}$
\end{Def}

\begin{Exs} Here are some more examples of how automata and transition
  systems can model agent-environment and related dynamics.
  \begin{enumerate}
    \item If $(X,\cdot)$ is a group, $U\subset X$ is a set of
    generators, and $\tau(x,u)=x\cdot u$, then $(X,U,\tau)$ is an
    automaton. For example, consider the situation in which $X=\Z\times \Z$
    and $U=\{a,b,a^{-1},b^{-1}\}$ in which $a=(1,0)$ and $b=(0,1)$. Thus, $X$ is presented with generators $a$, $b$, and relation $a \cdot b = b \cdot a$.  This models an agent moving without rotation in an infinite 2D-grid and the agent can move left, right, up and
    down.  There are no obstacles.  The standard Cayley graph is equivalent to the graph based representation of the automaton.
  \item \label{item1:MonoidExample} Let $U^*$ be the set of
    all finite sequences (``strings'') of elements of $U$. If $\bar u=(u_0,\dots,u_{k-1})\in U^*$ 
    and $u_k\in U$, we denote by $\bar u\cat u_k$ the \emph{concatenation}
    $(u_0,\dots,u_{k-1},u_k)$. If $\bar u_0,\bar u_1\in U^*$, then 
    $\bar u_0\cat \bar u_1$ is similarly the concatenation of two strings.
    The operation of concatenation turns $U^*$ into a monoid. Suppose
    $\tau\colon X\times U^*\to X$ is an action of the monoid $U^*$ on $X$ meaning
    that it satisfies $\tau(\tau(x,\bar u),\bar u')=\tau(x,\bar u\cat \bar u')$ and $\tau(x,\es)=x$.
    Then the automaton $(X,U,\tau)$ is a discrete-time
    control system. A sequence of \emph{controls} $\bar u= (u_0,\dots,u_{k-1})$
    produces a unique \emph{trajectory} $(x_0,\dots,x_{k})$, given the initial
    state $x_0$ by induction: $x_{i+1}=\tau(x_i,u_i)$ for all $i < k$.
  \item Consider an automaton $(X,U,\tau)$ in which $U$ is a group, and $\tau$
    is a group action of $U$ on $X$. In some situations it can be
    natural to consider the set of motor-outputs of an agent to be a
    group: the neutral element is no motor-output at all, every
    motor-output has an ``inverse'' for which the effect is the opposite, or
    negating (say, moving right as opposed to moving left), the
    composition of movements is many movements applied
    consecutively. The action $\tau$ of $U$ on $X$ is then the
    realization of those motor-outputs in the environment. In
    realistic scenarios, however, this is not a good way to model the
    sensorimotor interaction because of the following reason. Suppose
    the agent has actions ``left'' and ``right'', but it is standing
    next to an obstacle on its left.  Then moving ``left'' will result
    in staying still (because of the obstacle), but moving ``right''
    will result in actually moving right, if there is no obstacle at
    the right of the agent. In this situation the sequence
    ``left-right'' results in a different position of the agent than
    the sequence ``right-left'', so if ``left'' and ``right'' are each
    other's inverses in $G$, then the axioms of group action are
    violated.
  \item Note that if $T=\es$, then
    $(X,U,T)$ is a transition system.
  \item Let $X=\{0,1\}^*$ as in \eqref{item1:MonoidExample},
    $U=\{0\}$, and $(x,0,y)\in T$ if and only if $|y|=|x|+1$, then
    $(X,U,T)$ is a transition system, where $|x|$ is the length of the
    string~$x$.
  \item If $(X,U,T)$ is a transition system and $E\subset X$ an
    equivalence relation, then $(X/E,U,T/E)$ is a transition system,
    where $X/E=\{[x]_E\mid x\in X\}$ and
    $T/E=\{([x]_E,u,[y]_E)\mid (x,u,y)\in T\}$, and $/$ denotes a
    quotient space; see Definition~\ref{def:Quotient}.
  \end{enumerate}
\end{Exs}

\subsection{Sensorimotor systems}

Next, we will define a \emph{sensorimotor system}, which is a special
case of a transition system. Following the tenet (EA1) that
``environment is inseparable from the body which is inseparable from
the brain'', our sensorimotor systems can model any part of the
environment-body-brain coupling. The model that describes the
environment differs from the one that describes the agent merely in
the type of information it contains, but not in an essential
mathematical way.

SM-systems can be thought of as a partial specification of (some part of) the
brain-body-environment coupling. Physicalist determinism demands that
under full specification%
\footnote{This means a full specification of the environment, the
  agent's body, its brain, their coupling, as well as the initial
  states.}
we are left with a deterministic system. A specification is partial
when it leaves room for unknowns in some, or all, parts of the
system.

\begin{Def}
  A \emph{sensorimotor system} (or \emph{SM-system}) is a transition system
  $(X,U,T)$ where $U=S\times M$ for some sets $S$ and~$M$, which we call
  in this context the \emph{sensory set} and the \emph{motor set},
  respectively.
\end{Def}

The interpretation is that if $x\stackrel{(s,m)}{\longrightarrow}{y}$,
then $s$ is the sensation that either occurs at $x$, or along the
transition to the next state $y$, and $m$ the motor action which leads to the transition. We will show later how SM-systems can be
connected together (Definition~\ref{def:Coupling}) to form coupled
systems. Sometimes an SM-system is modeling a brain-body totality,
and other times it is modeling body-environment totality. A coupling between these two will model the brain-body-environment totality. 
This is a
flexible framework which enables enactivist-style analysis.

In fact the sensory and motor components can be decoupled which
might be more natural in some cases. The following alternative
definition shows how, and the following theorem shows that
they can be used interchangeably.

\begin{Def}
  An \emph{asynchronous SM-system} is a transition system 
  $(X,U,T)$ such that there exist partitions $U=S\cup M$
  and $X=X_s\cup X_m$ such that for all $(x,u,y)\in T$
  we have
  \begin{enumerate}
  \item if $x\in X_s$, then $u\in S$,
  \item if $x\in X_m$, then $u\in M$, and
  \item $x\in X_m\iff y\in X_s$.
  \end{enumerate}
  Thus, the state space of a sequential SM-system
  contains separate \emph{sensory states} and \emph{motor states}.
\end{Def}

\begin{Def}\label{def:E-preserving}
  Suppose $E$ is an equivalence relation on a set $X$. We say that
a map $f\colon X\to X$ is \emph{$E$-preserving} if for all $x,y\in X$,
we have $xEy\iff f(x)Ef(y)$. 
\end{Def}

There is a natural correspondence between SM-systems and their
asynchronous counterpart:
\begin{Thm}\label{thm:SM_asynchSM}
  Let $\SM$ and $\aSM$ be the classes of SM-systems
  and asynchronous SM-systems respectively. There are
  functions $F\colon\SM\to \aSM$
  and $G\colon \aSM\to \SM$ such that 
  \begin{enumerate}
  \item $F$ and $G$ are isomorphism and bisimulation preserving,
  \item restricted to finite systems, $F$ and $G$ are polynomial-time computable, and restricted to the infinite ones they are
  Borel-functions in the sense of classical descriptive set theory~\cite{kechris}.
  \end{enumerate}
\end{Thm}
\begin{proof}
  We will show the construction for $F$ and $G$ and leave the rest of
  the proof to the reader because it would go beyond the scope of the
  present paper. A reader familiar with the corresponding areas
  (computability, descriptive set theory) can readily verify (3) for
  the construction of $F$ below.
  
  Suppose $\X=(X,S\times M,T)$ is an SM-system.  Let
  $X_s=X$ and $X_m=T$. Thus, for each transition in $\X$, 
  we will have a motor state in $F(\X)$. Define
  $$T_0=\{(x,s,t)\mid (\exists y,m)((x,(s,m),y)=t)\}$$
  and
  $$T_1=\{(t,m,y)\mid (\exists x,s)((x,(s,m),y)=t)\}$$
  be disjoint copies of $X$ and define
  $F(\X)=(X_s\cup X_m,S\cup M,T_0\cup T_1)$.
  
  On the other hand, given $\X\in \aSM$, 
  $\X=(X_s\cup X_m,S\cup M,T)$, let
  $G(\X)=(X',S\times M,T')$ where
  $X'=X_s$ and $(x,(s,m),y)\in T'$ if and only if
  there exists $z\in X_m$ such that $(x,s,z)\in T$
  and $(z,m,y)\in T$.
\end{proof}

Another type of a system, which is in a similar way
equivalent to a special case of an SM-system,
is a state-labeled transition system
which we will introduce next, and prove a similar theorem,
Theorem~\ref{thm:QF_to_LTS}.

\subsection{Quasifilters and quasipolicies}
\label{ssec:Quasi}

The amount of information specified in a given SM-system depends on
which part of the brain-body-environment system we are modeling.  At
one extreme, we specify the environment's dynamics down to the small
detail and leave the brain's dynamics completely unspecified.  In this
case the SM-system will have only one sensation corresponding to each
state and the transition to the next state will be completely
determined by knowing the motor action. This is, in a sense, the
environment's perspective. At the other extreme, we specify the brain
completely, but leave the environment unspecified. We ``don't know''
which sensation comes next, but we ``know'' which motor actions are we
going to apply. This is in a sense the perspective of the agent.  The
first extreme case is the perspective often taken in
robotics and other engineering fields when either specifying a planning problem \cite{choset2005,ghallab2004,OkaLav07b}, or designing a
filter \cite{Hag90,lavalle12,sarkka2013,ThrBurFox05} (also known as sensor fusion). This is why we call SM-systems of that sort
\emph{quasifilters} (Definition~\ref{def:QFilter}). The other extreme
is the perspective of a policy. The policy depends on sensory input,
but the motor actions are determined (by the policy). This is why we
call the SM-systems of the latter sort \emph{quasipolicy}.  The
``quasi-'' prefix is used because both are weaker and more general
notions than those that appear in the literature; see
Remarks~\ref{rem:Filter} and~\ref{rem:Policy}.

Another way to look at this is the dichotomy between virtual
reality and robotics.
In virtual reality, scientists are designing the (virtual) environment
for an agent whereas in robotics they are typically designing an agent for an 
environment. In the former case the agent is partially specified:
the type of embodiment is known ($S$ and $M$ are known) and some
types of patterns of embodiment are known (eye-hand coordination).
However, the specific actions to be taken by the agents are left unspecified.
The job of the designer is to specify the environment down to 
the smallest detail, so that every sequence of motor actions of 
the agent yields targeted sensory feedback. The VR-designer
is designing a quasifilter constrained by the partial knowledge of the
 agent's embodiment and internal dynamics. The case for the robot
 designer is the opposite. She has a partial specification of the
 robot's intended environment and usually works with a 
complete specification of the robot's mechanics. She is designing a quasipolicy.
For VR-designers the agent is a black box; for roboticists the agent is a 
white box~\cite{SuoNilLav20} (unless the task is to reverse engineer an 
unknown robot design). For the environment, the roles are reversed. 
%\cite{other_vr_refs?}
A similar dichotomy can be seen between biology (in which the agent
is a black box) and robotics (in which it is usually a white box). 

\begin{Def}\label{def:QFilter}
  Suppose that $(X,S\times M,T)$ is an SM-system 
  with the property that for all $x_1\in X$ there
  exists $s_{x_1}\in S$ such that
  for all $x_2\in X$ and all $(s,m)\in S\times M$ we have that
  $x_1\stackrel{(s,m)}{\longrightarrow} x_2$ implies $s=s_{x_1}$.
  Then, $(X,S\times M,T)$ is a \emph{quasifilter}.
\end{Def}

In a quasifilter the sensory part of the outgoing edge
is unique. The dual
notion (quasipolicy) is when the motor part is unique:

\begin{Def}\label{def:QPolicy}
  Suppose that $(X,S\times M,T)$ is an SM-system 
  with the property that for all $x\in X$ there
  exists $m_x\in M$ such that
  for all $y\in X$ and all $(s,m)\in S\times M$ we have that
  $x\stackrel{(s,m)}{\longrightarrow} y$ implies $m=m_x$.
  Then, $(X,S\times M,T)$ is a \emph{quasipolicy}.
\end{Def}

Before explaining the connections between quasifilter and a filter and
quasipolicy and a policy, let us define projections of the
sensorimotor transition relation to ``motor'' and to ``sensory'':

\begin{Def}\label{def:SMProjections}
  Given an SM-system $(X,S\times M,T)$, let
  \begin{align*}
    T_M&=\{(x,m,y)\in X\times M\times X\mid \exists s\in S(x,(s,m),y)\in T\}\\
    T_S&=\{(x,s,y)\in X\times S\times X\mid \exists m\in M(x,(s,m),y)\in T\}.
  \end{align*}
  These are called the \emph{motor} and the \emph{sensory projections}
  respectively of the sensorimotor transition relation. They are also
  called the \emph{motor transition relation} and the \emph{sensory
    transition relation}, respectively. The corresponding transition
  systems $(X,M,T_M)$ and $(X,S,T_S)$ are called the
  \emph{motor} and the \emph{sensory projection systems}.
\end{Def}

\begin{Def}\label{not:Outgoing}
  Given a transition system $(X,U,T)$, and $x\in X$, let
  $O_T(x)\subset U$ be defined as the set
  $O_T(x)=\{u\in U\mid(\exists y\in
  X)(x\stackrel{u}{\rightarrow}y)\}$.  Combining this notation with
  the one introduced in
  Example~\ref{exs:TS_LitRelExs}\eqref{rem:TS2}, given $x,y\in X$, we
  have
  $$O_T(x)=\Cup_{y\in X}\tilde T(x,y).$$
\end{Def}

For a transition relation 
$T\subset X\times (S\times M)\times X$,
define its \emph{transpose} by
$T^t\subset X\times (S\times M)\times X$ such that
$T^t=\{(x,(m,s),y)\mid (x,(s,m),y)\in T\}$. Note that $(T^t)^t=T$. For
a subset of a Cartesian product $A\subset S\times M$, let $A_1$ be the
projection to the first coordinate
$A_1=\{s\in S\mid (\exists m\in M)((s,m)\in A)\}$
and $A_2$ the projection to the second one:
$A_2=\{m\in M\mid (\exists s\in S)((s,m)\in A)\}$

Mathematically coupling of two transition
systems is symmetric (see Theorem~\ref{thm:CouplingPreservesBisim}(3)),
but from the cognitive perspective there is (usually) an asymmetry
between the agent and the environment (which can be evident from 
some specific properties of the agent and of the environment).
Because of the partial symmetry, many properties of an agent can
dually be held by the environment and vice versa. The following
proposition highlights the duality between quasipolicy
and quasifilters: reversing the roles of the environment and
the agent reverses the roles of a quasipolicy and a quasifilter.

\begin{Prop}\label{prop:QFQP_proj_singl}
  For an SM-system $\X=(X,S\times M,T)$ the following are
  equivalent:
  \begin{enumerate}
  \item $\X$ is a quasifilter,
  \item $\X^t=(X,S\times M,T^t)$ is a quasipolicy,
  \item $O_{T_S}=(O_{T}(x))_2=(O_{T^t}(x))_1$ is a
    singleton for each $x\in X$.
  \end{enumerate}
  Similarly, $\X$ is a quasipolicy if and only if
  $O_{T_M}(x)=(O_T(X))_1$ is a singleton for each $x\in X$.
\end{Prop}
\begin{proof}
  A straightforward consequence of all the definitions.
\end{proof}

\subsection{State-relabeled transition systems}\label{sec:lts}

It will become convenient in the coming framework to assign labels to the states.  The elements $x$ of the state space $X$ are already named; thus, our labeling can be more properly considered as a \emph{relabeling} via a function $h : X \rightarrow L$, in which $L$ is an arbitrary set of \emph{labels}.  This allows partitions to be naturally induced over $X$ by the preimages of $h$.  Intuitively, this will allow the state space $X$ to be characterized at different levels of ``resolution'' or ``granularity''.  Thus, we have the following definition:

\begin{Def}\label{def:LTS}
  A \emph{state-relabeled transition system} (or simply \emph{labeled
    transition system}) is a quintuple $(X,U,T,h,L)$ in which
  $h\colon X\to L$ is a labeling function and $(X,U,T)$ is a
  transition system.
\end{Def}
\noindent We think of \emph{state-relabeled} to be a more
descriptive term, but we shorten it in the remainder of this paper to being simply {\em labeled}.

\begin{RemarkN}\label{rem:LTS_LA}
  In an analogy to Definition~\ref{def:AutomatonStar}, a labeled
  transition system is a \emph{labeled automaton}, if $T$ happens to
  be a function; in other words, for all $(x,u)\in X\times U$ there is
  a unique $y\in X$ with $(x,u,y)\in T$. In this case we may denote
  this function $\tau\colon (x,u)\mapsto y$ and work with the labeled
  automaton $(X,U,\tau,h,L)$. For example, the temporal filter in
  Section~\ref{ssec:TS} is a labeled automaton.
  
  The isomorphism and bisimulation relations are defined similary
  as for transition systems, but in a label-preserving way.
\end{RemarkN}

One intended application of a labeled transition system $(X,U,T,h,L)$ is that
$h$ is a sensor mapping, $L$ is a set of sensor observations, and $U$ is a set of actions.  Thus,
actions $u\in U$ allow the agent to transition between states in $X$
while $h$ tells us what the agent senses in each state.  We intend to show that this can be seen as a special
case of an SM-system by proving a theorem similar to Theorem~\ref{thm:SM_asynchSM},
but stronger, namely these corresponces preserve isomorphism:

\begin{Lemma}\label{thm:QF_to_LTS}
  Let $\FF$ be the class of quasifilters, $\PP$ the class of 
  quasipolicies, and $\LL$ the class of labeled systems. Then
  there are one-to-one maps 
  $$\LTS_P\colon \PP\to\L\text{ and } \LTS_F\colon \FF\to \L$$ 
  such that 
  \begin{enumerate}
  \item $\LTS_P$ and $\LTS_F$ are isomorphism and bisimulation preserving,
  \item restricted to finite systems, $\LTS_P$ and $\LTS_F$ are polynomial-time 
  computable, and restricted to the infinite ones they are
  Borel-functions in the sense of classical descriptive set theory.
  \end{enumerate}
\end{Lemma}
\begin{proof}
    Again, we will only show the constructions
    of the functions and leave the rest of the proof
    to the reader.
    Let $\FF$ be the class of quasifilters, $\PP$ the class of quasipolicies,
    and $\L$ the class of labeled transition systems.
    We will define bijections $\LTS_F\colon \FF\to\L$ and $\LTS_P\colon\PP\to \L$
    The constructions are dual to each other. Suppose
    $\X=(X,S\times M, T)$ is a quasifilter. Let $h\colon X\to S$ be the
    function defined by $h\colon x\mapsto s_x$ where $s_x$ is as in
    Definition~\ref{def:QFilter}. Now let $\LTS_F(\X)=(X,M,T_M,h,S)$.  We use $S$ here as the set of labels
    and $h$ as the labeling function
    to emphasize that the natural interpretation here is that $h$ is a sensor mapping.
    
    Now suppose that $\X=(X,M,A,h,S)$,
    $A\subset X\times M\times X$, is a labeled transition system. Let $T$
    be defined by
    \[T=\{(x,(h(x),m),y)\mid (x,m,y)\in A\}\subset X\times(S\times M)\times X.\]
    Then $\X'=(X,S\times M,T)$ is an SM-system which is a
    quasifilter and in fact $\X'=\LTS_F^{-1}(\X)$. We have now
    $\LTS_F^{-1}(\LTS_F(\X))=\X$ for any quasifilter $\X$ and
    $\LTS_F(\LTS_F^{-1}(\X'))$ for any labeled transition
    system~$\X'$. Similarly if $\X=(X,S\times M,T)$ is a quasipolicy,
    define $\LTS_P(\X)=(X,S,T_S,h,M)$ where $h(x)=m_x$ (see
    Definition~\ref{def:QPolicy}). Analogously to the above we also have
    the inverse function~$\LTS_P^{-1}$. We have found the needed one-to-one correspondences.
\end{proof}

In the end of the proof above we reverse the roles of 
$S$ and $M$, now $M$ being the set of labels, because they
are thought of as the ``outputs'' of the policy. 

\begin{RemarkN}\label{rem:Filter}
  Let $\X=(X,S\times M,T)$ be a quasifilter and
  $\X'=\LTS_F(\X)=(X,M,T_M,h,S)$ as in Lemma~\ref{thm:QF_to_LTS}.
  Suppose further that for each $x,y\in X$ there is at most one
  $u\in U$ with $x\stackrel{u}{\rightarrow} y$. Let
  $$E_T=\{(x,y)\in X^2\mid \exists u\in U(x\stackrel{u}{\rightarrow}y)\},$$
  Then $(X,M,E_T,x_0)$ coincides with the definition of a filter
  \cite[Def~3]{okaneshell}.  If it is also an automaton, meaning that
  above we replace ``at most one'' by ``exactly one'', then every
  sequence of motor actions $(m_0,\dots,m_{k-1})$ determines a unique
  resulting state $x_{k-1}\in X$. This is analogous, and can be proved
  in the same way, as the fact that each sequence of sensory data
  determines a unique resulting state in Remark~\ref{rem:Policy}
  below.
\end{RemarkN}

\begin{RemarkN}\label{rem:Policy}
  Usually, a \emph{policy} is a function which describes how an agent
  chooses actions based on past experience.  Thus,
  if $M$ is the set of motor commands and $S$ is the set of
  sensations, a policy is a function $\pi\colon S^*\to M$ where $S^*$
  is the set of finite sequences of sensory ``histories''; see for
  example~\cite{lavalle06}. Now, suppose that an SM-system
  $\X=(X,S\times M,T)$ is a quasipolicy in the sense of
  Definition~\ref{def:QPolicy} and let $x\mapsto m_x$ be as in that
  Definition. Assume further that $\X$ is an automaton (Section \ref{ssec:TS}) and let
  $\tau\colon X\times (S\times M)\to X$ be the corresponding
  transition function so that for all $x\in X$ and
  $(s,m)\in S\times M$ we have $(x,(s,m),\tau(x,(s,m)))\in T$.  Let
  $x_0\in X$ be an initial state.  We will show how the pair
  $(\X,x_0)$ defines a function $\pi\colon S^*\to M$ in a natural
  way. Let $\bar s=(s_0,\dots,s_{k-1})\in S^k$ be a sequence of
  sensory data. If $k=0$, and so $\bar s=()=\es$, let
  $\pi(\bar s)=m_{x_0}$. If $k>0$, assume that $\pi(s_0,\dots,s_{k-2})$
  and $x_{k-1}$ are both defined (induction hypothesis). Then let
  $x_{k} = \tau(x_{k-1},(m_{x_{k-1}},s_{k-1}))$ and
  $\pi(s_0,\dots,s_{k-2},s_{k-1})=m_{x_k}$.  The idea is that because
  of the uniqueness of $m_x$, a sequence of sensory data determines
  (given an initial state) a unique path through the automaton~$\X$.
\end{RemarkN}

% \subsection{Products and Projections}

% In a quasifilter and a quasipolicy, the sets $O_T(x)$
% (Notation~\ref{not:Outgoing}) have singleton projections
% (Proposition~\ref{prop:QFQP_proj_singl}).  We can talk about
% projections and products more generally:

% \begin{Def}
%   Given $T_1\subset X\times M\times X$ and
%   $T_2\subset X\times S\times X$, their \emph{product}
%   $T_1\cdot T_2$ is the relation on $X\times (S\times M)\times X$
%   defined by
%   $$T_1\cdot T_2=\{(x,(m,s),y)\mid (x,m,y)\in T_1\land (x,s,y)\in T_2\}.$$
% \end{Def}

% \begin{Prop}
%   Let $T_1\subset X\times M\times X$ and
%   $T_2\subset X\times S\times X$. Now we have
%   \begin{enumerate}
%   \item $(T_1\cdot T_2)_M=T_1$ and $(T_1\cdot T_2)_S=T_2$,
%   \item $T_1\cdot T_2=(T_1\cdot (X\times S\times X))\cap ((X\times M\times
%     X)\cdot T_2)$
%   \end{enumerate}
% \end{Prop}
% \begin{proof}
%   Straightforward from all the definitions.
% \end{proof}

% \begin{Def}
%   We say that $T$ is \emph{SM-independent}, if $T=T_M\cdot T_S$.
% \end{Def}

\subsection{Couplings of transition systems}
\label{ssec:CouplingTS}

\begin{figure}[t!]
    \centering
    \subfigure[]{\includegraphics[width=0.4\linewidth, height=3cm, keepaspectratio]{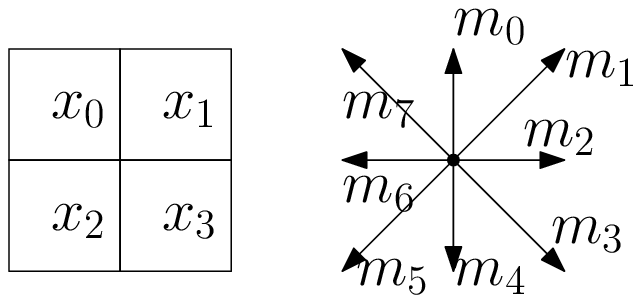}} \hspace{5em}
    \subfigure[]{\includegraphics[width=0.4\linewidth, height=2cm, keepaspectratio]{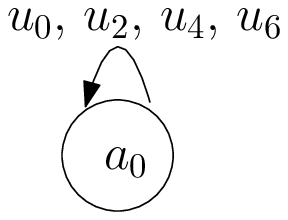}}
    \subfigure[]{\includegraphics[width=0.4\linewidth, height=6cm, keepaspectratio]{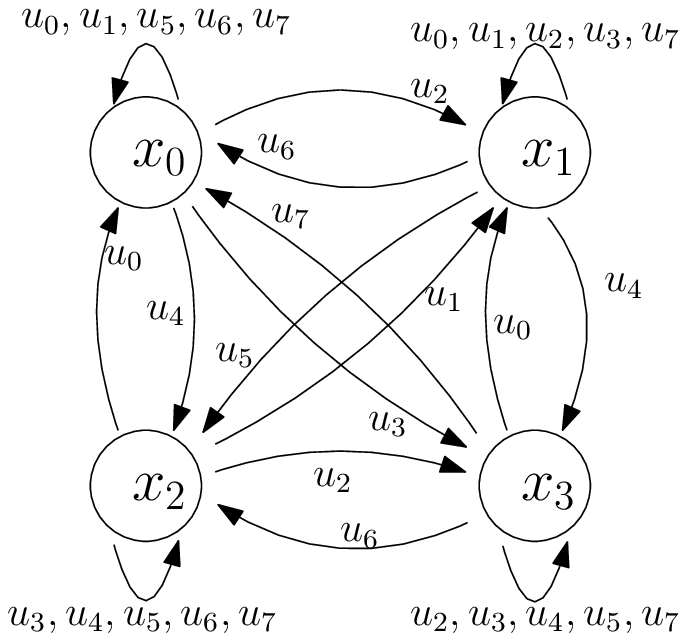}} \hspace{2em}
    \subfigure[]{\includegraphics[width=0.4\linewidth, height=6cm, keepaspectratio]{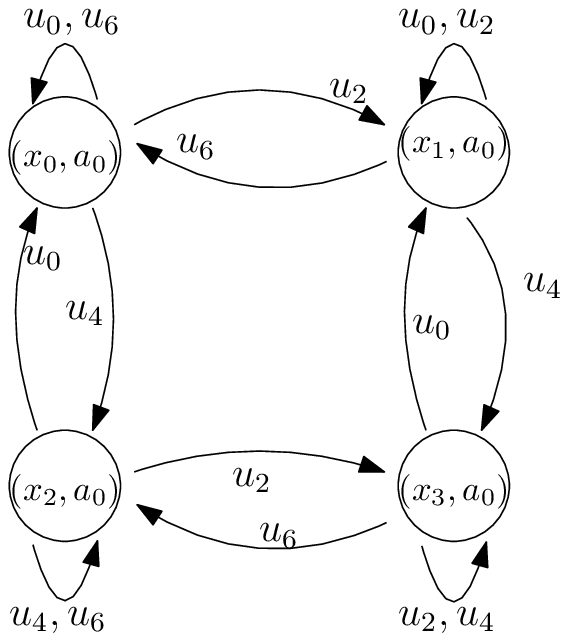}}
    \caption{(a) States and actions for the transition system $\X_0$ that describes a 2-by-2 grid. 8 actions populating the set $M=\{m_0,\dots,m_7\}$ correspond to a move (to a neighbor cell if possible) either sideways or diagonally. Suppose $S$ is a singleton such that $S=\{s\}$. Then, in the following, $u_i$ corresponds to the transition $u_i=(m_i,s)$ for $i=1,\dots,7$. (b) Transition system $\X_1$. (c) Transition system $\X_0$. (d) The coupled system $\X_0*\X_1$.}
    \label{fig:coupled_trans_sys}
\end{figure}

The central concept of this work pertaining to all principles
(EA1)--(EA5) is the coupling of SM-systems. We
define coupling, however, for general transition systems with the 
understanding that our most interesting applications will be for SM-systems where $U_0=U_1=S\times M$.

\begin{Def}\label{def:Coupling}
  Let $\X_0=(X_0,U_0,T_0)$ and $\X_1=(X_1,U_1,T_1)$ be two
  transition systems. The \emph{coupled} system $\X_0*\X_1$ is the
  transition system $(X,U,T)$ defined as follows: $X=X_0\times X_1$,
  $U=U_0\cap U_1$, and
  $$T=T_0*T_1=\{((x_0,x_1),u,(y_0,y_1))\mid (x_0,u,y_0)\in T_0\land (x_1,u,y_1)\in T_1\}.$$
  Equivalently, for all
  $((x_0,x_1),(y_0,y_1))\in (X_0\times X_1)^2$ we have
  $$\tilde T((x_0,y_0),(x_1,y_1))=\tilde T_0(x_0,x_1)\cap \tilde T_1(y_0,y_1)$$
  (recall the $\tilde T$ notation from Example~\ref{exs:TS_LitRelExs}\eqref{rem:TS2}). 
  %the start of Section~\ref{ssec:TS}).
\end{Def}

\begin{Ex}
  A simple example of coupling is illustrated in Figure \ref{fig:coupled_trans_sys}.
\end{Ex}

The coupling is a product of sorts. If we think of one transition
system as ``the environment'' and the other as ``the agent'', then the
coupling tells us about all possible ways in which the agent can
engage with the environment. The fact that the state space of the
coupled system is the product of the state spaces of the two initial
systems reflects the fact that the coupled system includes information
of ``what would happen'' if the environment was in any given state
while the agent is in any given (``internal'') state.

We immediately prove the first theorem concerning coupling:

\begin{Thm}\label{thm:CouplingPreservesBisim}
  Suppose that $\X_i=(X_i,U_i,T_i)$ and $\X'_i=(X'_i,U'_i,T'_i)$ for
  $i\in \{0,1\}$ are four SM-systems. Then the following hold:
  \begin{enumerate}
  \item If $\X_i\cong \X'_i$ for $i\in \{0,1\}$, then
    $\X_0*\X_1\cong \X'_0*\X'_1$.
  \item If $\X_i\sim \X'_i$  for $i\in \{0,1\}$, then
    $\X_0*\X_1\sim \X'_0*\X'_1$.
  \item $\X_0*\X_1\cong \X_1*\X_0$.
  \end{enumerate}
\end{Thm}
\begin{proof}
  In this proof $i$ always ranges over $\{0,1\}$. Every time $i$
  appears below, we drop the phrase ``for $i\in \{0,1\}$''.  
  For part 3 note that $(x,y)\mapsto (y,x)$ is an isomorphism.
  For part 1, if $f_i\colon \X_i\to \X'_i$ are isomorphism,
  let $g\colon (x,y)\mapsto (f_0(x),f_1(y))$. Then $g$ is an isomorphism
  from $\X_0*\X_1$ to $\X'_0*\X'_1$. We leave the verification of 
  these statements to the reader and proceed to prove part~2.
  Let $R_i$ be the
  bisimulation witnessing $\X_i\sim \X'_i$.  Let
  \begin{equation}
    R_0*R_1=\big\{((x_0,x_1),(x'_0,x'_1))\in (X_0\times X_1)\times (X'_0\times X'_1)\mid
    (x_i,x'_i)\in R_i\big\}.\label{eq:CoupledBisim} 
  \end{equation}
  Let us show that $R=R_0*R_1$ witnesses that
  $\X_0*\X_1\sim \X'_0*\X'_1$.  Suppose that
  $((x_0,x_1),(x'_0,x'_1))\in R$, $u\in U_0\cap U_1$ and
  $(y_0,y_1)\in X_0\times X_1$. Suppose further that
  $(x_0,x_1)\stackrel{u}{\rightarrow} (y_0,y_1)$, meaning that
  $$((x_0,x_1),u,(y_0,y_1))\in T_0*T_1.$$ Then by the definition
  of coupling we have that $(x_i,u,y_i)\in T_i$.
  By \eqref{eq:CoupledBisim} we now have $(x_i,x'_i)\in R_i$.
  Since $R_i$ is a bisimulation, there is $y'_i\in X'_i$
  with $(y_i,y'_i)\in R_i$ and $(x'_i,u,y'_i)\in T_i$. By the definition
  of $T_0*T_1$ we have now $((x'_0,x'_1),u,(y'_0,y'_1))\in T_0*T_1$,
  which completes the proof that $R$ is a bisimulation.
\end{proof}

\noindent Coupling provides an interesting way to compare SM-systems 
from the
``point of view'' of other SM-systems. For example, given an SM-system
$\E$ one can define an equivalence relation on SM-systems by saying
that $\I\sim_E\I'$, if $\E*\I=\E*\I'$. If $\E$ is the ``environment''
and $\I$, $\I'$ are ``agents'', this is saying that the agents perform
identically in this particular environment. Or vice versa, for a fixed
$\I$, the relation $\E*\I=\E'*\I$ means that the environments are
indistinguishable by the agent~$\I$.

\begin{RemarkN}\label{rem:DynSyst1}
  In the definition of coupling we see that the two SM-systems
  constrain each other. This is seen from the fact that in the
  definition we take intersections. For example, when an agent is
  coupled to an environment, it chooses certain actions from a large
  range of possibilities. In this way the agent structures its own
  world through the coupling~(EA3). To make this notion further
  connect to enactivist paradigm, we invoke the dynamical systems
  approach to cognition~\cite{schoner2008dynamical}.  An \emph{attractor} in a
  transition system $\X=(X,U,T)$ is a set $A\subset X$ with the
  property that for all infinite sequences
  \[x_0\stackrel{u_0}{\rightarrow} x_1
    \stackrel{u_1}{\rightarrow}\cdots x_{k-1}
    \stackrel{u_{k-1}}{\rightarrow} x_{k}
    \stackrel{u_k}{\rightarrow}\cdots \] there are infinitely
  many indices $n$ such that $x_n\in A$. There could be other possible
  definitions, such as ``for all large enough $n$, $x_n\in A$''. For the
  present illustration purposes it is, however, irrelevant.  It
  could be the case that $A\subset X$ is not an attractor of $\X$, but
  after coupling with $\X'=(X',U',T')$, $A\times X'$ may be an
  attractor of $\X*\X'$. Thus, if $\X$ is the environment and $\X'$ is
  the agent and $A$ is a set of desirable environmental states, then we may
  say that the agent is well attuned to $\X$, if $A$ was not initially
  an attractor, but in $\X*\X'$, then $A\times X'$ becomes one. It could also be
  that the agent needs to arrive to $A$ while being in a certain type
  of an internal state $B\subset X'$, for example, if $A$ is ``food''
  and $B$ is ``hungry''. Then it is not enough that $A\times X'$ is an
  attractor, but it is imperative that $A\times B$ is one.
\end{RemarkN}

\subsection{Unconstrained and fully constrained SM-systems}

As we mentioned before, the information specified in an SM-system
depends on which part of the brain-body-environment system we are
modeling.  In the extreme case we do not specify \emph{anything},
except for the very minimal information.  Consider a body of a robot
for which the set of possible actions (or motor commands) is $M$ and the set of possible
sensor observations is~$S$. Suppose that that is all we know about the robot. We
do not know what kind of environment it is in and we do not know what
kind of ``brain'' (a processor or an algorithm) it is equipped
with. Thus we do not know of any constraints the robot may have in
sensing or moving. We then model this robot as an \emph{unconstrained} SM-system:

\begin{Def}
  An SM-system $(X,S\times M,T)$ is called \emph{unconstrained} iff
  for all $x\in X$, we have $O_T(x)=S\times M$; recall Definition~\ref{not:Outgoing}.
\end{Def}

Unconstrained systems have the role of a neutral element with respect
to coupling (Proposition~\ref{prop:UncNeut}). We now show that
given all unconstrained SM-systems with shared $M$ and~$S$
are mutually bisimulation equivalent:

\begin{Prop}\label{prop:AllUnconstrainedBisim}
  Suppose that $\X=(X,S\times M,T)$ and $\X'=(X',S\times M,T')$ are unconstrained systems.
  Then $\X\sim \X'$.
\end{Prop}
\begin{proof}
  Let $R=X\times X'$. We need to show that $R$ is a bisimulation.  In
  the definition of bisimulation, Definition~\ref{def:IsoBisim}, the
  conclusion of the implication that needs to be satisfied is ``there
  exists $y'\in X'$ with $(x',u, y')\in T'$ and $(y,y')\in R$''.  However,
  because both $R$ and $T'$ contain ``everything'', this is trivially
  true as long as $X\times X'$ is non-empty. However, if it is empty, then the
  premise of that implication is false (no $(x,x')\in R$ exist); thus,
  the definition of bisimulation is again trivially satisfied.
\end{proof}

\begin{Cor}\label{cor:Unconstrained}
  The SM-system
  $\e=(\{0\},\{0\}\times (S\times M)\times \{0\}\})$ is the unique, up to
  bisimulation, unconstrained system.
\end{Cor}

\begin{Prop}\label{prop:UncNeut}
  Let $\e$ be as in Corollary~\ref{cor:Unconstrained} and let
  $\X=(X,S\times M,T)$ be any SM-system. Then $\X*\e\cong \X$.
\end{Prop}

\begin{Cor}
  If $\X$ and $\X'$ are SM-systems and $\X'$ is
  unconstrained, then $\X*\X'\sim \X$. 
\end{Cor}
\begin{proof}
  By Corollary~\ref{cor:Unconstrained} $\X'\sim \e$, So by
  Theorem~\ref{thm:CouplingPreservesBisim} we have $\X*\X'\sim
  \X*\e$. However, by Proposition~\ref{prop:UncNeut}, $\X*\e\sim \X$; thus,
  $\X*\X'\sim \X$.
\end{proof}

The opposite of an unconstrained system is a fully constrained one:

\begin{Def}\label{def:FullyConstrained}
  An SM-system $(X,S\times M,T)$ is \emph{fully constrained}
  iff~$T=\es$. 
\end{Def}

\begin{Prop}
  Dually to the propositions above, we have that (1) all fully
  constrained systems are bisimulation equivalent to each other, (2)
  the simplest example being $\lambda=(\{0\},S\times M,\es)$, and (3)
  if $\X=(X,S\times M,T)$ is another SM-system, then
  $\X*\lambda\sim\lambda$.
\end{Prop}

All transition systems are in some sense between the fully constrained
and the unconstrained, these being the two theoretical extremes.

\subsection{Quotients of transition systems}

When considering labelings and their induced equivalence relations, it
will be convenient to develop a notion of quotient systems, analogous
to quotient spaces in topology.  Suppose $\X = (X,U,T)$ is a
transition system and $E$ is an equivalence relation on~$X$. We can
then form a new transition system, called the \emph{quotient} of $\X$
by~$E$ in which the new states are $E$-equivalence classes and the
transition relation is modified accordingly.

\begin{Def}\label{def:Quotient}
  Suppose $\X=(X,U,T)$ and $E$ are as above. Let
  $X/E=\{[x]_E\mid x\in X\}$, in which each $[x]_E$ is an equivalence
  class of states $x$ under relation $E$, and
  $T/E=\{([x]_E,u,[y]_E)\mid (x,u,y)\in T\}$. Then $\X/E=(X/E,U,T/E)$
  is the \emph{quotient} of $(X,U,T)$ by~$E$.
\end{Def}

\begin{Def}\label{def:Equiv_h}
  Given any function $h\colon X\to L$, denote by $E^h$ the
  inverse-image equivalence: $E^h=\{(x,y)\in X^2\mid h(x)=h(y)\}$.  We
  will denote the equivalence classes of $E^h$ by $[x]_h$ instead of
  $[x]_{E^h}$ if no confusion is possible.
\end{Def}
The equivalence relation $E^h$ partitions $X$ according to the
preimages of $h$, as considered in the sensor lattice theory of
\cite{lavalle2019}.  The partition of $X$ induced by $h$ directly
yields an quotient transition system by applying the previous two
definitions:
\begin{Def}\label{def:LabelQuotient}
  Let $\X=(X,U,T)$ be a transition system and $h\colon X\to L$ be any
  mapping.  Then define $\X/h$ to be $\X/E^h$ where we combine
  Definitions~\ref{def:Equiv_h} and~\ref{def:Quotient}.
\end{Def}

\begin{Prop}\label{prop:OneToOneQuotient}
  If $h$ is one-to-one, then $\X/h\cong \X$. 
\end{Prop}
\begin{proof}
  $h$ is one-to-one if and only if $E^h$ is equality,
  in which case it is straightforward to verify that 
  the function $x\mapsto [x]_{E^h}$ is an isomorphism.
\end{proof}

For $h\colon X\to L$, the transition system $(X/h,U,T/h)$ is
essentially a new state space over the preimages of $h$.  In this case
$\X/h$ is called the \emph{derived information space} (as used
in~\cite{lavalle06}).  More precisely:

\begin{Prop}\label{prop:QTSisonS}
  Let $L'=\ran(h)\subset L$. Define
  \begin{align*}
    T'&=\{(l,u,l')\in L'\times U\times L'\mid (h^{-1}(l),u,h^{-1}(l'))\in T/h\}\\
      &=\{(h(x),u,h(y))\mid (x,u,y)\in T\}.
  \end{align*}
  Then $(X/h,U,T/h)$ is isomorphic to $(L',U,T')$
  via the isomorphism $f\colon [x]_{E^h}\mapsto h(x)$.
\end{Prop}
\begin{proof}
  We now show that $T'$ and $f$ are well defined.  For that note that
  by Definition~\ref{def:Equiv_h} we have $[x]_h=[x]_{E^h}=h^{-1}(x)$, and
  for all $y\in [x]_{h}$ we have $h(y)=h(x)$.  On the other hand,
  if $[y]_{h}\ne [x]_{h}$, then $h(y)\ne h(x)$ and so $f$ is
  injective. It is surjective by the definition of~$S'$.  Finally,
  $([x]_{h},u,[y]_{h})\in T/h$ if and only if
  $(h^{-1}(h(x)),u,h^{-1}(h(y)))\in T/h$ if and only if
  $(h(x),u,h(y))\in T'$ (by the definition of $T'$) if and only if
  $(f([x]_{h}),u,f([y]_{h}))\in T'$ (by the definition of~$f$).
\end{proof}

\section{Illustrative Examples of SM-Systems}
\label{sec:IllustrSM}

We next illustrate how sensorimotor systems model body-environment,
brain-body, and brain-body-environment couplings.  Consider a body in a 
fully understood and specified deterministic
environment.  In this case the body-environment system will be
modeled by a quasifilter, Definition~\ref{def:QFilter}. Instead of using the
quasifilter definition, we work with a labeled transition system which,
according to Proposition~\ref{thm:QF_to_LTS}, is equivalent.
According to the assumption of full specification, we will in fact
work with labeled automata.

The body has a set $M$ of possible motor actions each of which has a
deterministic influence on the body-environment dynamics. Denote the
set of body-environment states by~$E_0$. Whenever a motor action
$m\in M$ is applied at a body-environment state $e\in E$, a new
body-environment state $A(e,m)\in E$ is achieved. At each state
$e\in E$ the body senses data~$\s(e)$. Denote the set of sensations
by~$S$. In this way, the labeled automaton $\E_0=(E,M,A,\s,S)$ models this
body-environment system.  This model is ambivalent towards the agent's
internal dynamics, its strategies, policies and so on, but not
ambivalent towards its embodiment and its environment's structure. In
fact, it characterizes them completely.

Alternatively, consider a brain in a body, and suppose that the
brain is fully understood and deterministic (for example, perhaps it
is designed by us), but we do not know which environment it is in. We
model this by an SM-system which is a quasipolicy. Again, by
the analogous considerations as above, we work directly an equivalent labeled
automaton specification.  Denote the set of internal states of the brain
by~$I$. The agent's internal state is a function of the sensations; therefore,
let $B\colon I\times S\to I$ be a function ($B$ stands for
\emph{brain}) that takes one internal state to another based on new
sensory data.  At each internal state, the agent produces a motor output
which is an element of the set $M$; therefore, let $\mu\colon I\to M$ be a
function assigning a motor output to each internal state. Now,
$\I=(I,S,B,\mu,M)$ is a labeled transition system modeling this agent. It is
ambivalent towards the type of the environment the agent is in, but it
is not ambivalent towards the agent's internal dynamics, policies,
strategies and so on; in fact, it determines them completely.

Now, the coupling of the environment $\E$ and the agent $\A$
is the SM-system obtained as
$$\LTS_F^{-1}(\E)*\LTS_P^{-1}(\A).$$

The sensory and motor sets $S$ and $M$ capture the interface between
the brain and the environment because they characterize the body (but
not the \emph{embodiment}).  

\begin{Ex}\label{ex:SmallAutomata}
  Consider an agent that has four motor outputs, called ``up'' ($U$), 
  ``down'' ($D$),  ``left'' ($L$), and ``right''
  ($R$), and there is no sensor feedback
  (this defines the body). In Corollary~\ref{cor:Unconstrained} we
  gave a minimal example of an unconstrained SM-system. 
  On the other extreme one can give large examples.%
  \iflongversion
  \footnote{An even bigger one is given by the universal covering 
  of, in this case, (any) unconstrained system. 
  See Section~\ref{sec:Coverings}.}\fi
  For instance  the free
  monoid
  generated by the set $M=\{U, D, L, R\}$.

  Let $X$ be the set of
  all possible finite strings in the four ``letter'' alphabet $M$, let
  $T=\{(x,m,y)\mid x\cat m=y\}$. ``No sensor data'' is equivalent to
  always having the same sensor data; thus, we can assume that
  $S=\{s_0\}$ is a singleton and the sensor mapping $h\colon X\to S$
  is constant.  The resulting unconstrained transition system
  $\U=(X,T,M,\s,S)$ can be represented by an infinite quaternary tree, shown in 
  Figure \ref{fig:quat}(a).
  %$$\includegraphics[width=0.65\textwidth]{quaternary_tree1}$$

\begin{figure}[t]
\begin{center}
\begin{tabular}{ccc}
\includegraphics[width=8cm]{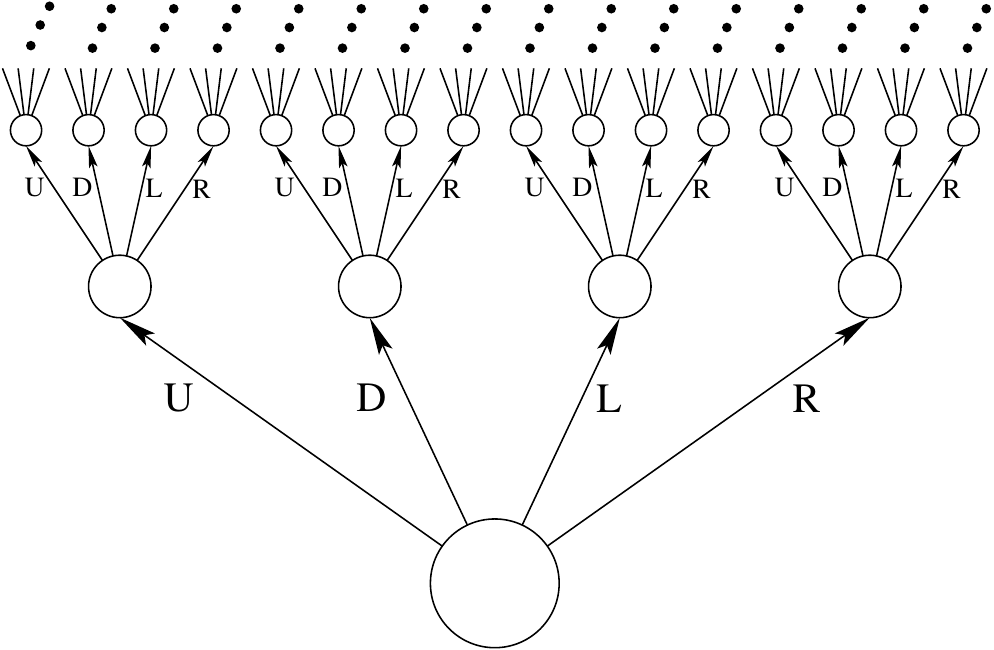} & \hspace*{1.5cm} &
\includegraphics[width=6cm]{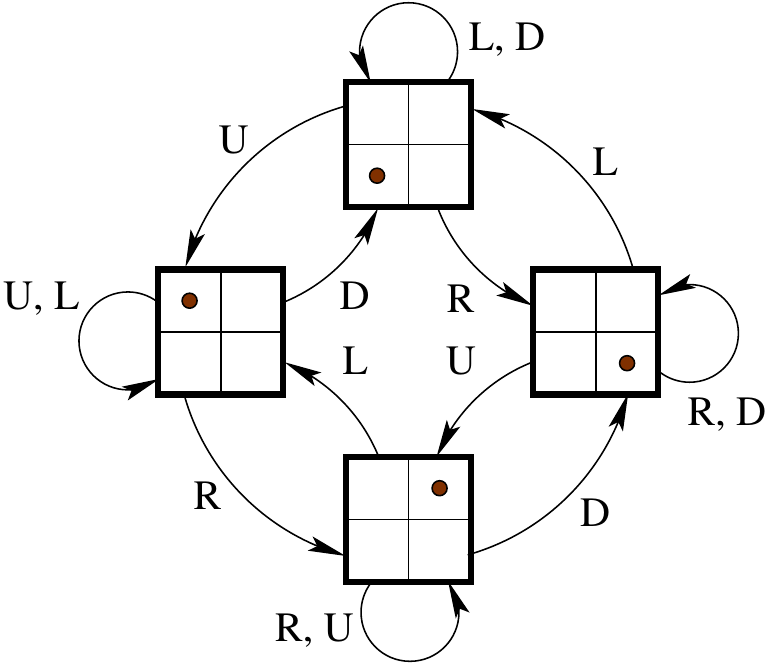} \\
(a) & & (b) 
\end{tabular}
\end{center}
\caption{\label{fig:quat} (a) Having motor commands and no sensory feedback leads to an infinite tree automaton.  (b)  Once the body is coupled with a $2\times 2$ grid environment, a four-state automaton results.}
\end{figure}

  Suppose that this body is situated in a $2\times 2$ grid. The body
  can occupy one of the four grid's squares at a time, and when it
  applies one of the movements, it either moves correspondingly, or,
  if there is a wall blocking the movement, it doesn't. This defines
  the body-environment system. The set of states is now $E$ and has
  four elements corresponding to all the possible positions of the
  body.  The transition function $A\colon E\times M\to E$ tells where
  to move, and the rest is as above.  The system $\E=(E,A,M,\s,S)$
  is shown in Figure \ref{fig:quat}(b).
  %$$\includegraphics[width=0.4\textwidth]{Automaton1}$$
  Let us now look at the agent.  Suppose that it applies the following
  policy: (1) In the beginning move left; (2) if the previous move was
  to the left, then move right, otherwise move left. This can be
  modeled with a two-state automaton $\I=(I,S,B,\mu,M)$ where
  $I=\{L,R\}$, $S=\{s_0\}$, $B(L,s_0)=R$, $B(R,s_0)=L$, $\mu(L)=l$ and
  $\mu(R)=r$.  Now, the coupling $\LTS_F^{-1}(\E)*\LTS_P^{-1}(\I)$ is an automaton that realizes the policy in the environment, as shown in Figure \ref{fig:sixteen}(a).
  
\begin{figure}[t]
\begin{center}
\begin{tabular}{ccc}
\includegraphics[width=3cm]{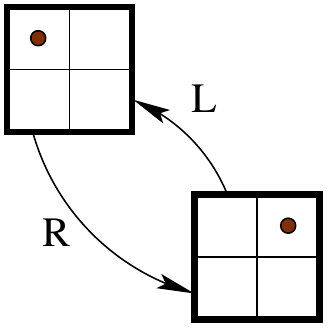} & \hspace*{4cm} &
\includegraphics[width=8cm]{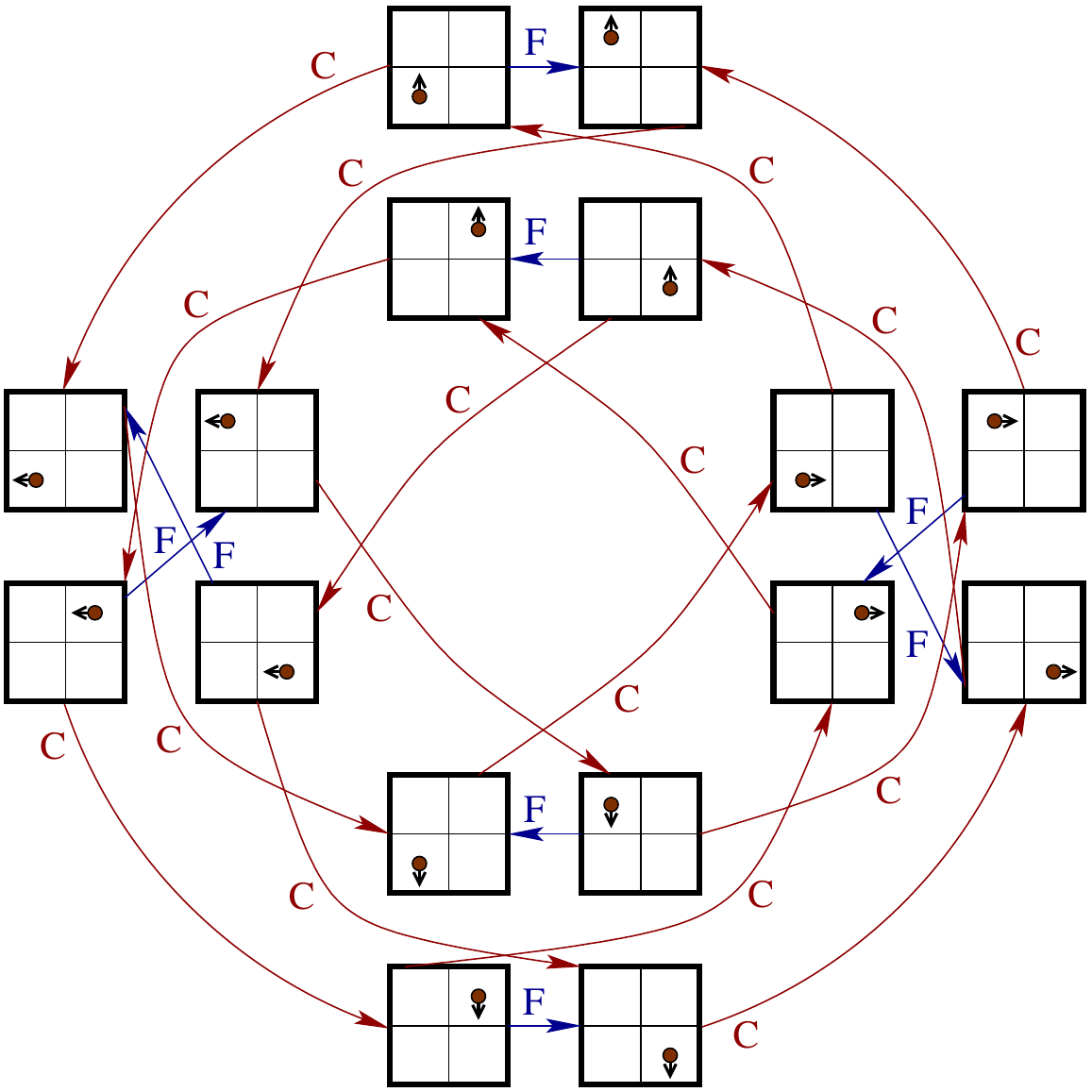} \\
(a) & & (b) 
\end{tabular}
\end{center}
\caption{\label{fig:sixteen} (a) A two-state automaton results from the realized policy.  (b)  If there are only two actions (rotate 90 degrees counterclockwise and going straight) then the second automaton has 16 states instead of four as in Figure \ref{fig:quat}(b).}
\end{figure}

  If the agent has a different embodiment in the same environment,
  then all of the automata will look different. Suppose that instead
  of the previous four actions, the agent has two: ``rotate 90-degrees counterclockwise'' ($C$),``forward one step'' ($F$). Note that these are expressed in the local frame of the robot: It can either rotate relative to its current orientation, or it can move in the direction it is facing; the previous four actions were expressed as if in a global frame or the robot is incapable of rotation.  Under the new embodiment, the  unconstrained automaton with no sensor feedback
 is an infinite \emph{binary tree}, with every node having two outgoing edges, labeled $C$ and $F$, respectively, instead of the quaternary infinite tree depicted on Figure~\ref{fig:quat}(a).
  %$$\includegraphics[width=0.3\textwidth]{binary_tree1}.$$
  Instead of the four-state automaton of Figure~\ref{fig:quat},
  the automaton describing the environment transitions is a 16 state-automaton, because the orientation
  of the agent can now have four different values.  See Figure \ref{fig:sixteen}(b).
  %$$\includegraphics[width=0.6\textwidth]{Automaton2}$$
  Finally the automaton describing the internal mechanics of the agent $\I$ is a quasipolicy in these two actions,
  and finally, the coupling corresponds essentially to taking a
  path in the 16-state automaton above.

  Note that there is a bisimulation between $\U$ and $\E$ which
  reflects the fact that from the point of view of an agent they are
  indistinguishable. This is natural because there is no sensory
  data, so from the agent's viewpoint it is unknowable whether or not it is
  embedded in an environment. A bisimulation is given as
  follows: Let $y_0\in Y$ be the top-right corner and $x_0\in X$ the
  root of the tree. Define $R\subset X\times Y$ be the minimal set
  satisfying the following conditions:
  \begin{enumerate}
  \item $(x_0,y_0)\in R$.
  \item If $(x,y)\in R$ and $m\in M$, then $(T(x,m),U(y,m))\in R$.
  \end{enumerate}
\end{Ex}

\begin{Ex}
  The 16-state automaton of Example~\ref{ex:SmallAutomata} has four
  automorphisms corresponding to the rotation of the environment by 90
  degrees counterclockwise. Each of those automorphisms corresponds
  to an auto-bisimulation. Mirroring is not an 
  automorphism because the agent's rotating action
  fixes the orientation of the automaton. 
\end{Ex}

\begin{figure}[t]
\begin{center}
\includegraphics[width=6cm]{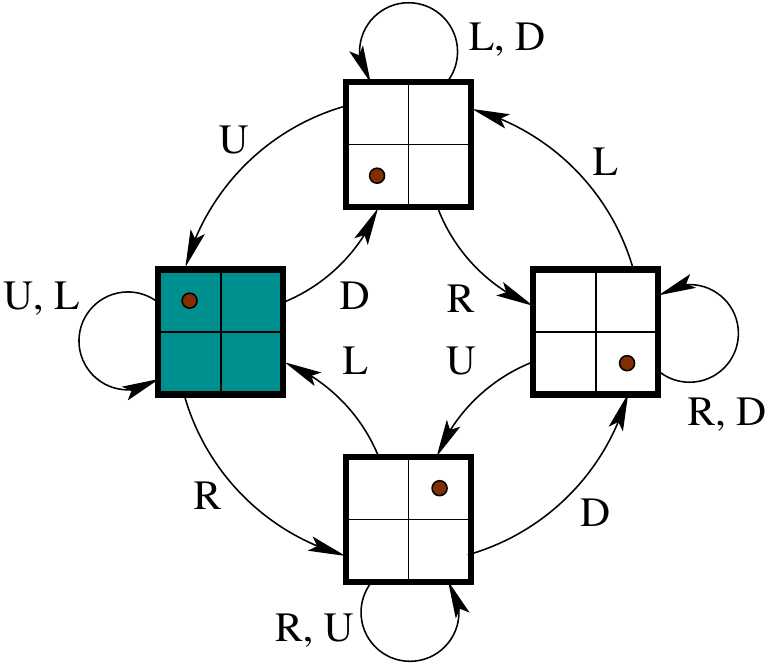}
\end{center}
\caption{\label{fig:scented} Consider the automaton $\E$ of 
  Figure~\ref{fig:quat}(b) from Example~\ref{ex:SmallAutomata},
  but assume that the agent can ``smell'' a different scent in the
  top-left corner. This can be modeled by having a two-element set
  $S=\{0,1\}$ instead of a singleton, and $h\colon X\to \{0,1\}$ such
  that $h(x)=0$ iff $x$ is not the top-left corner. The state with a scent is shaded.}
\end{figure}

\begin{Ex}\label{ex:SmallLabeledAutomaton}
  Figure \ref{fig:scented} shows an example of how
  an automaton with non-trivial sensing could look.
  Jumping a little bit ahead, it will be seen that the labeling
  provided by $h$ in this figure is not sufficient
  (a notion introduced in Definition~\ref{def:SuffRefL}).
\end{Ex}

\section{Sufficient Refinements and Degree of Insufficiency}
\label{sec:SRE}

%%%%%%%%%%%%%%%%%%%%%%%%%%%%%
This section presents the concept of \textit{sufficiency}, which we
present as the main glue between enactivist philosophy and
mathematical understanding of cognition. In
Section~\ref{ssec:Sufficiency} we introduce the main concepts and
explain its profound relevance to enactivist modeling and how it can
be a precursor to the emergence of meaning from meaningless
sensorimotor interactions. We also point out that there is an
intricate relationship between this concept and the celebrated free
energy principle from neuroscience, although we leave that discussion
open. In Section~\ref{ssec:min_suff_ref} we introduce the notion of
minimal sufficient refinements, prove a uniqueness result about them,
and show how they are connected to the classical notions of
bisimulation as well as derived information state spaces.

\subsection{Sufficiency}
\label{ssec:Sufficiency}

\begin{Def}
  Let $(X,U,T)$ be a transition system and $E\subset X\times X$ an
  equivalence relation.  We say that $E$ is \emph{sufficient}
  or \emph{completely sufficient}, if for
  all $(x,y)\in E$ and all $u \in U$, if $(x,u,x')\in T$ and
  $(y,u,y')\in T$, then $(x',y')\in E$.
\end{Def}

This means that if an agent cannot distinguish between states $x$ and $y$, 
then there are no actions it could apply to later distinguish between them. 
To put it differently, if the states are indistinguishable by an instant sensory
reading, then they are in fact indistinguishable even through sensorimotor
interaction. This is related to the equivalence relation known as
Myhill-Nerode congruence in automata theory. 

The equivalence relation
of indistinguishability in the context of sensorimotor interactions is
at its simplest the consequence of indistinguishability by sensors.  
Thus, we define sufficiency for labelings or sensor mappings:

\begin{Def}\label{def:SuffRefL}
  A labeling $h\colon X\to L$ is called \emph{sufficient} (or
  \emph{completely sufficient}) iff for all $x,y,x',y'\in X$ and all
  $u \in U$, the following implication holds:
  $$(h(x)=h(y)\land (x,u,x')\in T\land (y,u,y')\in T) \Rightarrow h(x')=h(y').$$
\end{Def}

\begin{Prop}
  If $(X,U,\tau)$ is an automaton, then $h\colon X\to L$ is sufficient
  if and only if for all $x,y\in X$ and all $u\in U$, we have that if
  $h(x)=h(y)$, then $h(\tau(x,u))=h(\tau(y,u))$. 
\end{Prop}
\begin{proof}
  Checking the definitions.
\end{proof}

There is a connection with the classical notion of bisimulation in
classical transition systems theory (recall
Definition~\ref{def:IsoBisim}):

\begin{Prop}
  An equivalence relation on a state space of an automaton
  $(X,U,\tau)$ is sufficient if and only if it is an autobisimulation.
\end{Prop}
\begin{proof}
  Suppose $E\subset X$ is a bisimulation and let $x_1,x_2\in X$ be
  such that $(x_1,x_2)\in E$ and let $u\in U$. Let
  $x'_1=\tau(x_1,u)$. Since $E$ is a bisimulation, there exists $x'_2$
  such that $\tau(x_2,u)=x'_2$ and $(x'_1,x'_2)\in E$.  But
  $\tau(x_2,u)$ is uniquely determined and so it follows that
  $(\tau(x'_1,u),\tau(x'_2,u))\in E$. The other direction is left to
  the reader.
\end{proof}

Proposition~\ref{prop:SufTrAut} below is an important proposition on
which the idea of derived I-spaces and combinatorial filters builds
upon \cite{lavalle06,lavalle12,okaneshell}, although as far as the
authors are aware, in the literature, only the ``if''-direction is
mentioned. We say that a transition system $(X,U,T)$ is \emph{full},
if for all $x_1\in X$ and all $u\in U$ there exists at least one
$x_2\in X$ with $(x_1,u,x_2)$.

\begin{Prop}\label{prop:SufTrAut} 
  Suppose $\X=(X,U,T)$ is a transition system. Let $h\colon X\to L$
  be a labeling. Then $\X/h$ is an automaton if and only if $\X$ is
  full and $h$ is sufficient.
\end{Prop}
\begin{proof}
  Suppose $\X/h$ is an automaton. That means that $T/h$ is a function
  with domain $X/h\times U$, so $\X/h$ must be full. For sufficiency
  assume that $x_1,x_2\in X$ and $u\in U$ are such that
  $h(x_1)=h(x_2)$, so in particular $[x_1]_h=[x_2]_h$. Suppose further
  that $x'_1,x'_2$ are such that $(x_1,u,x'_1),(x_2,u,x'_2)\in T$. We
  need to show that then $h(x'_1)=h(x'_2)$. We have
  $([x_1]_h,u,[x'_1]_h)\in T/h$ and $([x_2]_h,u,[x'_2]_h)\in T/h$.
  Since $T/h$ is a function, and
  $[x_1]_h=[x_2]_h$, we must have $[x'_1]_h=[x'_2]_h$ which by
  definition means that $h(x'_1)=h(x'_2)$, so $h$ is sufficient. For
  the other direction, suppose that $h$ is sufficient and that
  $\X$ is is full. We want to show that for all $[x_1]_h\in X/h$
  and $u\in U$ there is a unique $[x_2]_h\in X/h$ with
  $([x_1]_h,u,[x_2]_h)\in T/h$. Existence follows from fullness.
  For uniqueness and assume that
  \begin{equation}
    ([x_1]_h,u,[x_2]_h),([x_1]_h,u,[x'_2]_h)\in T/h\label{eq:elements_of_Th}
  \end{equation}
  for some $x_1,x_2,x'_2\in X$. Since they are chosen to arbitrarily,
  it is enough to show that $h(x_2)=h(x'_2)$, because this implies
  that $[x_2]_h=[x'_2]_h$ and so $[x]_h$ together with $u$ uniquely
  determine the element $\xi=[x_2]_h=[x'_2]_h$ such that
  $([x_1]_h,u,\xi)\in T/h$.  By \eqref{eq:elements_of_Th} there must be
  $z_{1},z'_{1}\in [x_1]_h$ and $z_{2}\in [x_2]_h$, $z'_2\in [x'_2]_h$ such
  that $(z_1,u,z_2),(z'_1,u,z'_2)\in T$. Since $h(z_1)=h(z'_1)$ and
  $h$ is sufficient, we have $h(z_2)=h(z'_2)$ and since
  $h(z_2)=h(x_2)$ and $h(z'_2)=h(x'_2)$ we have that
  $h(x_2)=h(x'_2)$ as needed.
\end{proof}

The sufficiency of an information mapping was introduced in
\cite[Ch~11]{lavalle06}, and is encompassed by a sufficient labeling
in this paper. In the prior context, it has meant that the current
sensory perception together with the next action determine the next
sensory perception. The elegance with respect to our principle (EA2)
is that sufficiency is \emph{not} saying that the agent's internal
state corresponds to the environment's state (as is in
representational models). Nor is it saying that the agent
\emph{predicts} the next action. It is saying, rather, that the
agent's current sensation together with a choice of a motor command
\emph{determine} the agent's next sensation; and this statement is
true only as a statement made about the system from outside, not as a
statement which would reside ``in the agent''. The sensation may carry
no meaning at all ``about'' what is actually ``out there''. However,
if the agent has found a way to be coupled to the environment in a
sufficient way, then sensations \emph{begin} to be \emph{about} future
sensation. In this way meaning emerges from sensorimotor
patterns. This relates to (EA3) and somewhat touches on the topic of
perception~(EA5). Furthermore, the property of determining future
outcomes is related to (EA4) because that is what \emph{skill}
is. There is no potential to \emph{reliably} engage with the
environment in complex sensorimotor interactions, if the sensations do
not \emph{reliably} follow certain historical patterns.

Thus, the notion of sufficiency is considered by us to be of
fundamental importance for enactivist-inspired mathematical modeling
of cognition.  The violation of sufficiency means that the current
sensation-action pair does not correlate with the future sensation,
making it harder to be attuned to the environment.  Having a different
sensation following the same pattern can be seen as a primitive notion
of a ``surprise''.  This remarkably aligns with the predictive coding
and the free energy principle from
neuroscience~\cite{RaoBal99,friston2009predictive,Fri10}.  Does the
notion of sufficient labelings capture the same ideas in a more
general way?  This is an open question for further research.

A generalization of sufficiency is $n$-sufficiency, in which the data
of $n$ previous steps is needed to determine the next label. Here, we
define an $n$-chain.

\begin{Def}
  An $n$-\emph{chain} in $\X=(X,U,T)$ is a sequence
  $$c=(x_0,u_0,\cdots,x_{n-1},u_{n-1},x_n)\in (X\times U)^{n}\times X$$ such that
  $x_i\stackrel{u}{\rightarrow}x_{i+1}$ for all $i<n$. If $n=0$,
  then by convention $c=(x_n)$. Let $E\subset X\times X$ be an
  equivalence relation. Let $k<n$. We say that two $n$-chains
  $c=(x_0,u_0,\dots,x_{n-1},u_{n-1},x_n)$,
  $c'=(x'_0,u'_0,\dots,x'_{n-1},u'_{n-1},x'_n)$ are
  $(T,E,k)$-equivalent if for all $i< k$, we have $u_i=u'_i$ and
  $(x_i,x'_i)\in E$. An $\infty$-chain is defined in the same way as
  $n$-chain, except the sequences are infinite, without the ``last''~$x_n$.
\end{Def}

\begin{Def}\label{def:n-suff}
  For a transition system $\X=(X,U,T)$, an equivalence relation $E$ on
  $X$ is called $n$-\emph{sufficient} if there are no two
  $(T,E,n)$-equivalent $n$-chains
  \[c=(x_0,u_0,\dots,x_{n-1},u_{n-1},x_n) \text{ and }
    c'=(x'_0,u'_0,\dots,x'_{n-1},u'_{n-1},x'_n)\]
  such that $(x_n,x'_n)\notin E$.  A labeling $h\colon X\to L$ is
  called $n$-\emph{sufficient} if $E^h$~is $n$-sufficient (Recall
  Definition~\ref{def:Equiv_h}).
\end{Def}

\begin{Prop}
  An equivalence relation $E$ is $0$-sufficient if and only if there
  is only one $E$-equivalence class, and a labeling function $h$ is
  $0$-sufficient if and only if it is constant.
\end{Prop}
\begin{proof}
  The intuition is that if we can predict the future from ``nothing'',
  it means that the future is always the same.  Let us prove the first
  statement. The second statement then follows because $E^h$ has
  exactly one equivalence class if and only if $h$ is constant.
  Plugging in $n=0$ into Definition~\ref{def:n-suff}, it says ``...if
  there are no two $(T,E,0)$-equivalent $0$-chains $(x_0), (x'_0)$
  with $(x_0,x'_0)\notin E$''. However, the $(T,E,n)$-equivalence
  only concerns elements of the sequence before the $n$:th element,
  before $0$:th in this case, which do not exist.  Thus, all
  $0$-chains are equivalent by definition. Hence, the definition of
  $0$-sufficiency becomes ``...if there are no two sequences
  $(x_0), (x'_0)$ with $(x_0,x'_0)\notin E$''. This amounts to saying
  that $E$ has one equivalence class.
\end{proof}

\begin{Prop}
  An equivalence relation $E$ (resp. a labeling~$h$) is sufficient if
  and only if it is $1$-sufficient.
\end{Prop}
\begin{proof}
  $1$-sufficiency says that for all $1$-chains $(x_0,u_0,x_1)$ and
  $(x'_0,u_1,x'_1)$, if $x_0$ and $x'_0$ are equivalent and $u_0=u_1$,
  then $x'_0$ and $x'_1$ are equivalent which is exactly the
  definition of sufficiency.
\end{proof}

\begin{Prop}
  Suppose $n<m$ are natural numbers. If a labeling $h$ is
  $n$-sufficient, then it is $m$-sufficient. The same holds for
  equivalence relations.
\end{Prop}
\begin{proof}
  The intuition here is that ``knowing more doesn't hurt''.
  We prove the statement for equivalence relations.
  Suppose $E$ is $n$-sufficient and $m>n$. Suppose
  to the contrary that $E$ is not $m$-sufficient. This is witnessed by
  some $m$-chains
  $$c=(x_0,u_0,\dots,x_{m-1},u_{m-1},x_m) \text{ and }
  c'=(x'_0,u'_0,\dots,x'_{m-1},u'_{m-1},x'_m)$$
  which are $(T,E,m)$-equivalent and $(x_m,x'_m)\notin E$. Now
  define restrictions $c_0$ and $c'_0$ which are obtained from
  $c$ and $c'$ by ignoring all elements with indices less than $(m-n)$.
  Then $c,c'$ are $n$-chains and are in fact $(T,E,n)$-equivalent.
  The last elements are still $x_m$ and $x'_m$ and $(x_m,x'_m)\notin E$,
  which means that $c_0,c'_0$ witness that $E$ is not $n$-sufficient and
  contradicts our assumption.
\end{proof}

This enables us to define the degree of insufficiency:

\begin{Def}\label{def:DegreeOfInsufficiency}
  The \emph{degree of insufficiency} of the labeled automaton
  $\X=(X,U,\tau,h,L)$ is defined to be the smallest $n$ such that $h$ is
  $n$-sufficient, if such $n$ exists, and $\infty$ otherwise.
  Denote the degree of insufficiency of $\X$ by~$\deginsuf(\X)$,
  or $\deginsuf(h)$ if only the labeling needs to be specified and
  $\X$ is clear from the context.
\end{Def}

The intuition is that the larger the degree of insufficiency of 
an environment $\X$, the harder it is for an agent to 
be attuned to it. We talk more about the connection between
attunement and sufficiency in the following sections.

\subsection{Minimal sufficient refinements}\label{ssec:min_suff_ref}

In this section we prove that the minimal sufficient refinements are
always unique (Theorem~\ref{thm:MinSufRef1}). This will follow from
a deeper result that the sufficient equivalence relations form a
complete sublattice of the lattice of all equivalence relations.  This
does not hold for $n$-sufficient equivalence relations for $n>1$
(Example~\ref{ex:Failure1}).  We will then explore how the
minimal sufficient refinements can be thought of as an enactive
perceptual construct that emerges from the body-environment,
brain-body, and brain-body-environment dynamics.

\begin{Def}\label{def:Refinement}
  An equivalence relation $E$ is a \emph{refinement} of equivalence
  relation $E'$, if $E\subset E'$, also denoted $E'\le_r E$. A
  labeling function $h$ is a refinement of a labeling function $h'$,
  if $E^h$ is a refinement of~$E^{h'}$.
\end{Def}

An important interpretation of the concept of a refinement is that a
better sensor provides the agent with more information about the
environment.%
\footnote{Here we are not talking about contentful or semantic
  information, but merely about correlational information in the
  philosophical sense.}
Each sensor mapping $h$ induces a partition of $X$ via its preimages,
and refinement applies in the usual set-theoretic sense to the
partitions when comparing sensors mappings.  If a sensor mapping $h$
is a refinement of $h'$, then it enables the agent to react in a more
refined way to nuances in the environment. Using the partial ordering
given by refinements, we obtain the \emph{sensor lattice}~\cite{lavalle2019},
so let us diverge for a moment into the theory of lattices.

\subsection{Lattices of equivalence relations}

Let $(L,\le)$ be an ordered set. An \emph{upper bound} of a subset
$L'$ is an element $u$ such that $l\le u$ for all $l\in L$.  It is a
\emph{least upper bound}, it is an upper bound and $u\le x$ for any
other upper bound $x$ of~$L'$. If the least upper bound exists, then
it is unique, for if $u$ and $u'$ are least upper bounds, then by
definition we have $u\le u'$ and $u'\le u$. Symmetrically one defines
the lower bound and the greatest lower bound which also turns out to
be unique. In lattice theory, the least upper bound is often called
\emph{join} and the greatest lower bound is called \emph{meet}.

\begin{Def}
  A \emph{lattice} is an ordered set $(L,\le)$ such that for any two
  elements $a,b\in L$ there is a join $j=a\lor b$ and a meet
  $m=a\land b$. We may denote a lattice as a quadruple
  $(L,\le,\land,\lor)$ where $\land$ and $\lor$ are the join and meet
  operations.
\end{Def}

By induction one can show that in a lattice every finite subset has a
join as well as a meet. A lattice is called \emph{complete}, if this
extends to all, not necessarily finite, subsets:

\begin{Def}
  A lattice $(L,\le)$ is \emph{complete}, if for all $L'\subset L$
  there are a join $j=\Lor L'$ and a meet $m=\Land L'$.
\end{Def}

A lattice $(L,\le,\land,\lor)$ is a \emph{sublattice} of
$(L',\le',\land',\lor')$, if $L\subset L'$,
for all $l_0,l_1\in L$ we have
$$l_0\le l_1\iff l_0\le' l_1,$$
and for all $L_0\subset L$ we have that if $\Land L_0$ exists, then
$\Land L_0=\Land'L_0$ and if $\Lor L_0$ exists, then
$\Lor L_0=\Lor'L_0$.  It is a \emph{complete sublattice}, if $(L,\le)$
is a complete lattice.

Given a set non-empty $A\subset X^2$, the \emph{equivalence relation
  generated by~$A$}, denoted $\la A\ra$, is the smallest equivalence
relation on $X$ which contains all pairs that are in~$A$, so
$$\la A\ra=\Cap \{E\supset A\mid E\text{ is an equivalence relation on }X\},$$
or equivalently, $(x,x')\in \la A\ra$, iff there exist $x_1,\dots,x_n$ such
that the pairs
$$(x,x_1),(x_1,x_2),\dots,(x_{n-1},x_n),(x_n,x')$$
are all in~$A\cup A^T$ where $A^T=\{(x',x)\mid (x,x')\in A\}$.

\begin{Lemma}\label{lemma:EquivSup}
  If $E$ is any equivalence relation such that $A\subset E$, then
  $\la A\ra\subset E$.
\end{Lemma}
\begin{proof}
  Suppose $(x,x')\in \la A\ra$. Then there is a sequence of
  pairs $$(x,x_1),(x_1,x_2),\dots,(x_{n-1},x_n),(x_n,x')$$ that are
  all in~$A\cup A^T$. Since $A\subset E$, also $A^T\subset E$
  and so $A\cup A^T\subset E$, from which we see that all
  these pairs are in~$E$. By transitivity of $E$, $(x,x')\in E$.
\end{proof}

\begin{Def}
  Let $X$ be any set and let $\E(X)$ be the set of all equivalence
  relations on~$X$. Then $(\E(X), \subset)$ is an ordered set.  Given
  a set of equivalence relations $\E \subset \E(X)$, define
  $$\Land \E=\Cap \E=\{(x_1,x_2)\in X^2\mid (\forall E\in \E)((x_1,x_2)\in E)\}$$
  Then $\Land\E$ is an equivalence relation such that $\Land \E\subset E$
  for all $E\in\E$, so it is a lower bound for~$\E$.
  Define
  \begin{align*}
    \Lor\E=\la \Cup \E\ra= \Cap\{E\supset \Cup \E\mid E\text{ is an
  equivalence relation on }X\}.
  \end{align*}
  Then $E\subset \Lor\E$ for all $E\in \E$, so $\Lor \E$ is an upper
  bound for~$\E$.
\end{Def}

\begin{Prop}
  Let $X,\E(X)$ and $\E\subset\E(X)$ be as above. Then $\Lor \E$ and
  $\Land\E$ are respectively the least upper bound and the greatest
  lower bound of~$\E$. Consequently, $(\E(X),\subset)$ is a complete
  lattice.
\end{Prop}
\begin{proof}
  Suppose $E_1$ is a lower bound for~$\E$. Then, by definition
  $E_1\subset E$ for all $E\in\E$. But then $E_1\subset \Cap \E$, so
  $\Land \E$ is the meet of~$\E$.

  Suppose $E_0$ is an upper bound for~$\E$. Then, by definition
  $E\subset E_0$ for all $E\in\E$. But then $\Cup \E\subset E_0$, and
  since $E_0$ is an equivalence relation, also
  $\la\Cup \E\ra \subset E_0$. So $\Lor \E$ is the join of~$\E$.
\end{proof}

\subsection{Lattice of sufficient equivalence relations}

We will prove in this section that if $(X,U,\tau)$ is an automaton,
the sufficient equivalence relations form a complete sublattice
of~$(\E(X),\subset)$. Given an automaton $\X=(X,U,\tau)$, denote
by $\E^{U,\tau}_\suf(X)\subset\E(X)$ the set of sufficient equivalence
relations on~$X$. When $U$ and $\tau$ are clear from the context, we
write just~$\E_{\suf}(X)=\E^{U,\tau}_{\suf}(X)$.

\begin{Thm}\label{thm:SuffSubLattice}
  Suppose $(X,U,\tau)$ is an automaton and suppose that
  $\E\subset \E_\suf(X)$ is a set of sufficient equivalence relations.
  Then $\Land \E$ and $\Lor\E$ are sufficient.  Thus,
  $(\E_\suf(X),\subset)$ is a complete sublattice of
  $(\E(X),\subset)$.
\end{Thm}
\begin{proof}
  For $\land$: Suppose $x,x'\in X$ are such that
  $(x,x')\in \Land \E$ and $u\in U$.  Since $\Land\E=\Cap \E$, we have
  that $(x,x')\in E$ for all $E\in\E$. Since all $E\in\E$ are
  sufficient, it follows that $(\tau(x,u),\tau(x',u))\in E$ for all $E\in\E$, and
  so $(\tau(x,u),\tau(x',u))\in\Cap \E=\Land\E$.

  For $\lor$: Suppose $x,x'\in X$ are such that $(x,x')\in \Lor \E$
  and $u\in U$. By the definition of $\Lor\E$ there exist
  $z_1,\dots,z_k\in X$ such that $x=z_1$, $x'=z_k$ and for all $i<k$
  there is $E_i\in \E$ such that $(z_i,z_{i+1})\in E_i$.  By the
  sufficiency of each $E_i$, we have then that
  $(\tau(z_i,u),\tau(z_{i+1},u))\in E_i$ and so the sequence
  $\tau(z_1,u),\dots,\tau(z_k,u)$ witnesses that
  $(\tau(x,u),\tau(x',u))\in\Lor E$.
\end{proof}

Suppose that a labeling $h$ is very important for an agent.  For
example, $h$ could be ``death or life'', or it could be relevant for a
robot's task. Suppose that $h$ is not sufficient. The robot may
want to find a sufficient refinement of $h$. Clearly a one-to-one
$h'$ would do. However, assume that the agent has to use resources
for distinguishing between states; thus, the fewer distinctions the
better. This motivates the following definition. Recall
Definition~\ref{def:Refinement} of refinements.

\begin{Def}
  Let $(X,U,T)$ be a transition system and $E_0\subset X\times X$ an
  equivalence relation.  \emph{A minimal sufficient refinement} of $E_0$
  is a sufficient equivalence relation $E$ which is a refinement of
  $E$ such that there is no sufficient $E'$ with $E_0\le_r E'<_r E$.

  Given a labeling $h_0$ of a transition system $(X,U,T)$, a
  \emph{minimal sufficient refinement} of $h_0$ is a labeling $h$ such
  that $E^{h}$ is a minimal sufficient refinement of~$E^{h_0}$ (recall
  Definition~\ref{def:Equiv_h}).
\end{Def}

%\begin{Ex}
%  \begin{enumerate}
%  \item Suppose $X=\{0,1\}^{*}$, $M=\{0,1\}$ and $\mu(x,b)=x\cat b$
%    (concatenation of the binary string $x$ with the bit~$b$). Let
%   $\s(x)$ be equal to $1$ if and only if the number of ones and the
%    number of zeros in are both prime numbers, and otherwise~$0$. Then
%    the only sufficient refinements of $\s$ are one-to-one.
%  \item Let $X$ be as above and let $\s\colon X\to \{0,1\}$ be such
%    that if $|x|$ is divisible by~3, then $\s(x)=1$, and otherwise
%    $\s(x)=0$.  Then $\s$ is not sufficient. Let
%    $\s'\colon t\to \{0,1,2\}$ be such that
%    $$\s'(x)\equiv |x| \mod 3.$$
%    Then $\s'$ is a minimal sufficient refinement of~$\s$. %FIXME: %Picture?
%  \end{enumerate}
%\end{Ex}

\begin{Ex}
  Let $\X=(X,U,\tau)$ be an automaton where
  $X=\{0,1\}^{*}$, $U=\{0,1\}$ and $\tau(x,b)=x\cat b$
  (concatenation of the binary string $x$ with the bit~$b$). Let
  $h(x)=1$ if and only if the number of ones and the
  number of zeros in $x$ are both prime; otherwise~$h(x)=0$. Then
  the only sufficient refinements of $h$ are one-to-one.
\end{Ex}

\begin{Ex}
  Let $\X$ be as above and let $h\colon X\to \{0,1\}$ be such that if
  $|x|$ is divisible by~3, then $h(x)=1$; otherwise, $h(x)=0$.  Then
  $h$ is not sufficient. Let $h'\colon x\mapsto \{0,1,2\}$ be
  such that
  $$h'(x)\equiv |x| \mod 3.$$
  Then $h'$ is a minimal sufficient refinement
  of~$\s$. %FIXME: Picture?
\end{Ex}

\begin{Thm}\label{thm:MinSufRef1} 
  Consider an automaton $\X=(X,U,\tau)$ and let $E_0$ be an
  equivalence relation on $X$. Then a minimal sufficient refinement of
  $E_0$ exists and is unique.
\end{Thm}
\begin{proof}
  Let $\E$ be the set of all sufficient equivalence relations
  $E\subset E_0$. Then $\E$ is non-empty, because the
  identity-relation $\{(x_1,x_2)\in X^2\mid x_1=x_2\}$ is a sufficient
  refinement of~$E_0$. Let $E=\Lor\E$. By Lemma~\ref{lemma:EquivSup}
  we have $E\subset E_0$, so $E$ is a refinement of~$E_0$, and by
  Theorem~\ref{thm:SuffSubLattice} $E$ is sufficient. On the other
  hand if $E'$ is a sufficient refinement of $E_0$, then $E'\in \E$
  and so $E'\subset E$, so $E$ is $<_r$-minimal. The same argument
  proves the uniqueness too: if $E'$ is another minimal sufficient
  refinement of $E_0$, then again $E'\in\E$ and so $E'\subset E$.
  But $E'\subsetneq E$ would contradict the minimality of $E'$,
  so we must have $E'=E$.
\end{proof}

Theorem~\ref{thm:MinSufRef1} fails, if ``automaton'' is replaced by
``transition system'', or if ``sufficient'' is replaced by
``$n$-sufficient'' for $n>1$ (recall Definition~\ref{def:n-suff})

\begin{Ex}[Failure of uniqueness for $n$-sufficiency]
  \label{ex:Failure1}
  Let $X=\{0,1,2,3,4,5\}$, $U=\{u_0\}$ and
  $$\tau(0,u_0)=1,\tau(1,u_0)=2,\tau(2,u_0)=2,$$
  and
  $$\tau(3,u_0)=4,\tau(4,u_0)=5,\tau(5,u_0)=5.$$
  Let $E_0$ be an equivalence relation on $X$ such that the
  equivalence classes are $\{0,1,3,4\}$, $\{2\}$ and $\{5\}$.  Then
  this relation is not $2$-sufficient, because $(0,u_0,1,u_0,2)$ and
  $(3,u_0,4,u_0,5)$ are $(T,E_0,2)$-equivalent, but $2$ and $5$ are
  not $E_0$-equivalent. Let $E_1,E_2\subset E_0$ be an equivalence
  relation with equivalence classes as follows:
  $$E_1:\{0,3\},\{1,4\},\{2\},\{5\},$$
  $$E_2:\{0,4\},\{1,3\},\{2\},\{5\}.$$
  Then $E_1$ and $E_2$ are refinements of $E_0$. They are both
  $2$-sufficient, because there doesn't exist any $(T,E_1,1)$ or
  $(T,E_2,1)$ equivalent $2$-chains.  They are also both
  $\le_r$-minimal with this property which can be seen from the fact
  that they are actually $\le_r$-minimal refinements of $E_0$ as
  equivalence relations (not only as sufficient ones).
\end{Ex}

\begin{Ex}[Failure of uniqueness for transition systems]
  \label{ex:Failure2}
  Let $X=\{0,1,2,3,4\}$, $U=\{u_0\}$ and
  $T=\{(0,u_0,3),(2,u_0),4\}$. Let $E_0$ be the equivalence relation
  with the equivalence classes $\{0,1,2\}$, $\{3\}$ and $\{4\}$.
  Then $E_0$ is not sufficient, because $(0,2)\in E_0$,
  but $(3,4)\notin E_0$. Let $E_1$ and $E_2$ be the refinements
  of $E_0$ with the following equivalence classes:
  $$E_1: \{0,1\},\{2\},\{3\},\{4\},$$
  $$E_2: \{0\},\{1,2\},\{3\},\{4\}.$$
  Now it is easy to see that both $E_1$ and $E_2$ are sufficient
  refinements of $E_0$, and by a similar argument as in
  Example~\ref{ex:Failure1} they are both minimal. The reason why this
  is possible is the odd behaviour of the state $2$ which doesn't have
  out-going connections. Such odd states are the reason why the
  decision problem ``Does there exist a sufficient refinement with $k$
  equivalence classes?'' is NP-complete for finite transition systems.
  %see~\cite{WAFR}.  
\end{Ex}

\begin{Remark}
  It is worth noting that Theorems~\ref{thm:SuffSubLattice} and
  \ref{thm:MinSufRef1} do not assume anything about the cardinality of
  $X$ or of $U$, other structure on them (such as metric or topology)
  nor anything about the function~$\tau$ or the
  relation~$E_0$. Keeping in mind potential applications in robotics,
  $X$ and $U$ could be, for instance, topological manifolds, and
  $\tau$ a continuous function, or $X$ could be a closed subset of
  $\R^n$, $U$ discrete and $\tau$ a measurable function, or any other
  combination of those. In each of those cases, the sublattice of
  sufficient equivalence relations is complete, as per
  Theorem~\ref{thm:SuffSubLattice}, and every equivalence relation
  $E_0$ on $X$ admits a unique minimal sufficient refinement as per
  Theorem~\ref{thm:MinSufRef1}.
\end{Remark}

Recall Definition~\ref{def:E-preserving} of an equivalence relation
preserving function.  We say that an equivalence relation $E$ on $X$
is \emph{closed under $f\colon X\to X$} if for all $x\in X$, we have
$(x,f(x))\in E$. If $E$ is closed under $f$, then $f$ is
$E$-preserving: given $(x,x')\in E$, we have
$(x,f(x)),(x',f(x'))\in E$, because $E$ is closed under~$f$.  Now by
transitivity of $E$ we have $(f(x),f(x'))\in E$, so $f$ is
$E$-preserving.

\begin{Def}\label{def:OrbitEquiv}
  Let $f\colon X\to X$ be a bijection.  The induced \emph{orbit
    equivalence relation} is the relation $E_f$ on $X$ defined by
  $(x,x')\in E_f\iff (\exists n\in\Z)(f^n(x)=x')$, in which $f^n(x)$ is defined
  by induction as: $f^0(x)=x$, $f^{n+1}(x)=f(f^n(x))$,
  $f^{n-1}(x)=f^{-1}(f^n(x))$.
\end{Def}

\begin{Thm}\label{thm:Automorphism1}
  If $f$ is an automorphism of the automaton $(X,U,\tau)$, then
  $E_f$ is a sufficient equivalence relation.
\end{Thm}
\begin{proof}
  Suppose $(x,x')\in E_f$. Then there is $n$ such that
  $x=f^{n}(x')$. Now
  $$\tau(x,u)=\tau(f^{n}(x'),u)=f^{n}(\tau(x',u)).$$
  The last equality follows from the fact that $f$ is an automorphism.
  By definition, this means that $(\tau(x,u),\tau(x',u))\in E_f$.
\end{proof}

\begin{Thm}\label{thm:Automorphism2}
  Let $\X=(X,U,\tau)$ be an automaton and $E$ be an equivalence
  relation on~$X$. Suppose $f\colon X\to X$ is an automorphism 
  such that $E$ is closed
  under~$f$. Let $E'$ is the minimal sufficient refinement
  of~$E$. Then $E'$ is closed under~$f$ and $E\le_r E'\le_r E_f$.
\end{Thm}
\begin{proof}
  Since $E$ is closed under $f$, $E_f$ is a refinement of~$E$.  By
  Theorem~\ref{thm:Automorphism1}, $E_f$ is also sufficient, so by
  $\le_r$-minimality of $E'$, we have $E_f\subset E'$ which implies
  that $E'$ is closed under~$f$.
\end{proof}

% FIXME: Discussion of emergence and emergent properties.
% Connections to filters and O'Kane & Shell..

% Problem of finding minimal sufficient refiniment and its connection
% to DFA-minimization, filtering, and finding maximal bisimulation.

% Also discuss connection to enactivism: Sufficiency is a global-level
% property, but describes something that it usually called
% "prediction". Emergence.

%\begin{Ex}
  % FIXME: Give the example of the 3-beam thing. The 4-state solution
  % is a minimal suffiecient refinement.
%\end{Ex}

\begin{Ex}\label{ex:Mini}
  Consider the environment which is a one-dimensional lattice of
  length five, $E=\{-2,-1,0,1,2\}$, in which the corners ``smell
  bad''; thus, we have a sensor mapping $h\colon E\to S$, $S=\{0,1\}$
  defined by $h(n)=0\iff |n|=2$; see
  Figure~\ref{fig:1D_lattice}(a). Consider two agents in this
  environment.  Both are equipped with the same $h$ sensor, but their
  action repertoires differ. Both have two possible actions. One has
  actions $L=$~``move left one space'' and $R=$~``move right one
  space'', and the other one has actions $T=$~``turn 180 degrees'' and
  $F=$~``go forward one space''. Let $M_0=\{L,R\}$ and
  $M_1=\{T,F\}$. Thus, these agents have a slight difference in
  embodiment. Although both of them can move to every square of the
  lattice in a very similar way (almost indistinguishable from the
  outside perspective), we will see that the differences in embodiment
  will be reflected in that the minimal sufficient refinements will
  produce non-equivalent ``categorizations'' of the environment. The
  structures that emerge from these two embodiments will be
  different.
  
  First, we define the SM-systems that model these agents'
  embodiments in~$E$. The first agent does not have orientation. It can
  be in one of the five states, and the state space is
  $X_0=E$. For the second agent, the effect of the
  $F$ action depends on the orientation of the agent (pointing
  left or pointing right). Thus, there are ten different states the
  agent can be in, yielding $X_1=E\times \{-1,1\}$. The effects of motor
  outputs are specified completely ($L$ means moving
  left, and so on), whereas the agent's internal mechanisms are left
  completely open, so our systems will be quasifilters. According to
  Remark~\ref{rem:LTS_LA}, we can work with a labeled automaton
  instead. Hence, let $\tau_0\colon X_0\times M_0\to X_0$ be defined by
  $\tau_0(x,L)=\max(x-1,-2)$ and
  $\tau_0(x,R)=\min(x+1,2)$.  For the other agent, let
  $\tau_1((x,b),T)=(x,-b)$ and
  $\tau_1((x,b),F)=(\min(\max(x+b,-2),2),b)$.  Now we have
  labeled automata $\X_0=(X_0,M_0,\tau_0,h,S)$ and
  $\X_1=(X_1,M_1,\tau_1,h,S)$.

  It is not hard to see that the one-to-one map
  $h_0\colon X_0\to \{-2,-1,0,1,2\}$ with $h_0(x)=x$ is a sufficient
  refinement of $h$ which is minimal (see
  Figure~\ref{fig:non_isomorphic_quotients}(a)). Thus, every state
  needs to be distinguished by the agent for it to be possible to
  determine the following sensation from the current one. The derived
  information space automaton $\X_0/h_0$ isomorphic to~$\X_0$
  (Proposition~\ref{prop:OneToOneQuotient}).

  For the second automaton, consider the labeling
  $h_1\colon X_1\to \{-2,-1,0,1,2\}$ defined by $h_1(x,b)=b\cdot x$
  (see Figure~\ref{fig:non_isomorphic_quotients}(b)).
  
  \begin{figure}[tbh]
    \centering
    \subfigure[]{\includegraphics[width=0.3\linewidth]{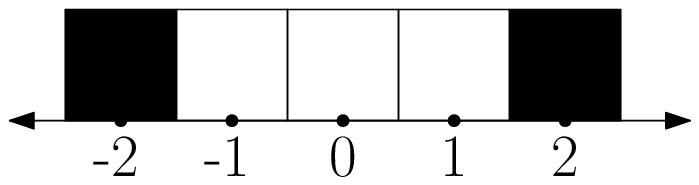}}
    \subfigure[]{\includegraphics[width=0.3\linewidth]{1D_grid_env.eps}}
    \subfigure[]{\includegraphics[width=0.3\linewidth]{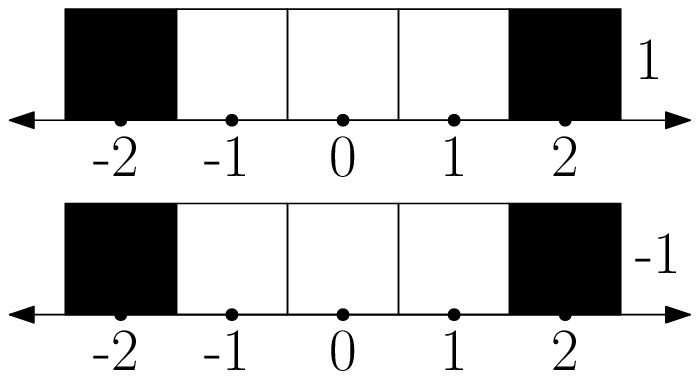}}
    \caption{(a) One-dimensional lattice environment described in
      Example~\ref{ex:Mini}. (b) State space of the agent 0. (c) State
      space of the agent 1.  The states for which the value of the
      sensor mapping is $0$ are shown in black.}
    \label{fig:1D_lattice}
  \end{figure}

  \begin{figure}[tbh]
    \centering
    \subfigure[]{\includegraphics[width=0.33\linewidth]{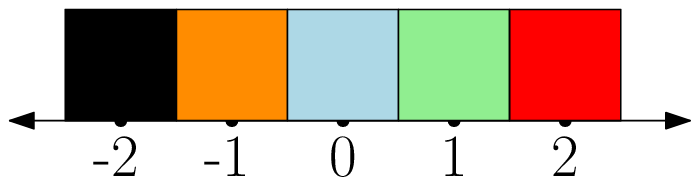}}
    \hspace{2em}
    \subfigure[]{\includegraphics[width=0.4\linewidth]{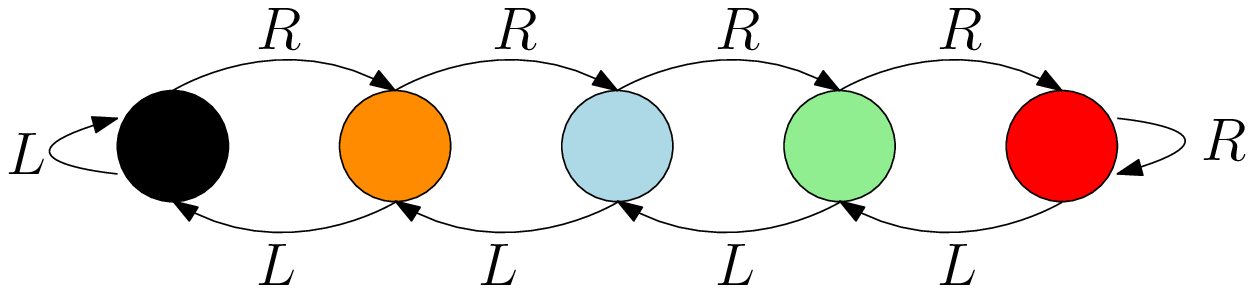}}
    \subfigure[]{\includegraphics[width=0.33\linewidth]{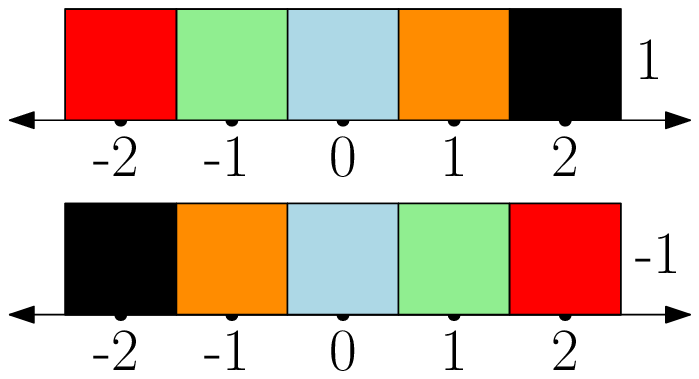}}
    \hspace{2em}
    \subfigure[]{\includegraphics[width=0.4\linewidth]{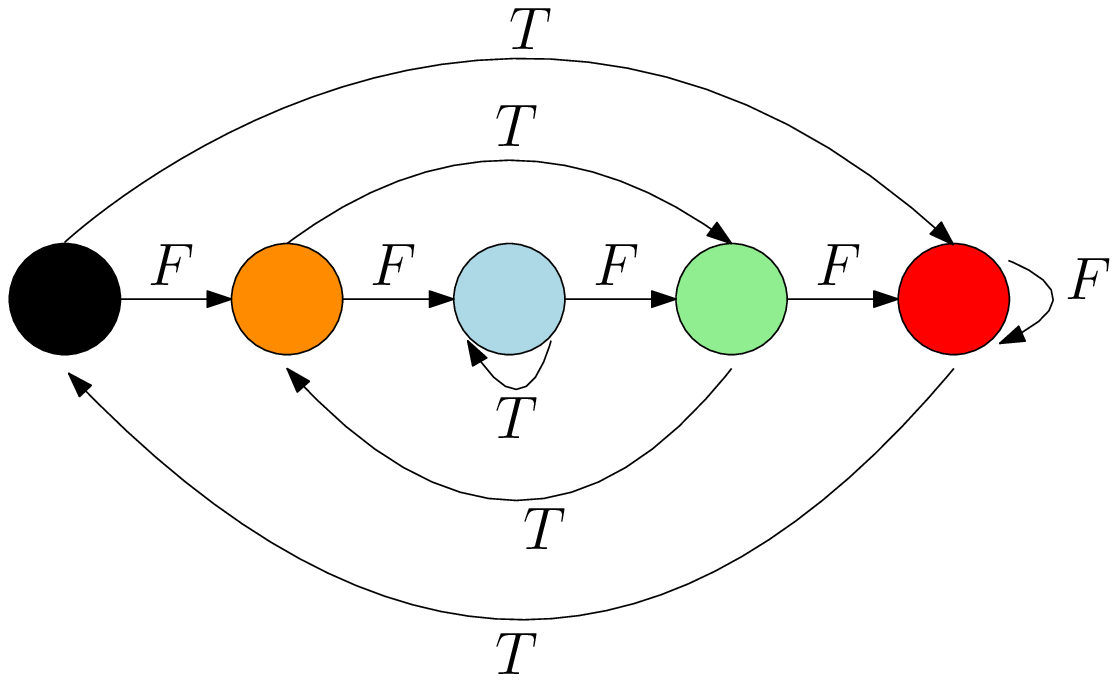}}
    \caption{(a) State space of the agent 0 categorized by $h_0$, states
      that belong to the same class are colored with the same color. (b)
      Resulting quotient $\mathcal{X}_0/h_0$ for the agent 0. (c) State
      space of the agent 1 categorized by $h_1$, states that belong to
      the same class are colored with the same color. (d) Resulting
      quotient $\mathcal{X}_1/h_1$ for the agent 1. }
    \label{fig:non_isomorphic_quotients}
  \end{figure}
  
  \begin{claim}
    $h_1$ is a minimal sufficient refinement of $h$ in $\X_1$.
  \end{claim}
  \begin{proof}
    We will show that the equivalence relation $E^{h_1}$ is the
    mininimal sufficient refinement of $E^h$
    (Definition~\ref{def:Equiv_h}).  Let $f\colon X_1\to X_1$ be
    defined by $f(x,b)=(-x,-b)$. Then $f$ is an automorphism of $\X_1$
    and $E_f=E^{h_1}$, so $E^{h_1}$ is sufficient and any minimal
    sufficient refinement $E$ of $E^h$ must satisfy $E\le_r E^h$.
    
    In the base labeling $h$, moving forward from $(1,1)$ results in a
    different sensation than moving forward from $(1,-1)$, so they
    must be $E$-non-equivalent. But then inductively this also applies
    to their neighbours $(0,1)$ and $(0,-1)$ as well as $(2,1)$ and
    $(2,-1)$ and so on. Thus $E^{h_1}$ is in fact minimal.
  \end{proof}
  Both minimal sufficient labelings, $h_0$ and $h_1$ have five values;
  thus, they categorize the environment into five distinct
  state-types.  However, the resulting derived information spaces are
  different in the sense that the quotients $\X_0/h_0$ and $\X_1/h_1$
  are not isomorphic; compare
  Figure~\ref{fig:non_isomorphic_quotients}(b) with
  Figure~\ref{fig:non_isomorphic_quotients}(d).
\end{Ex}

\vspace*{0.5cm}

\begin{figure}[tbh]
\begin{center}
\begin{tabular}{ccc}
\includegraphics[width=5cm]{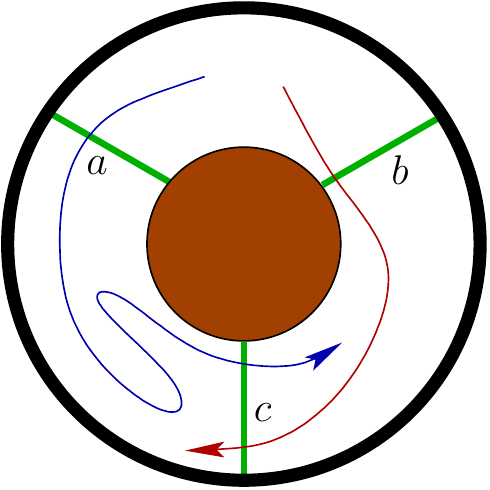} & &
\includegraphics[width=5cm]{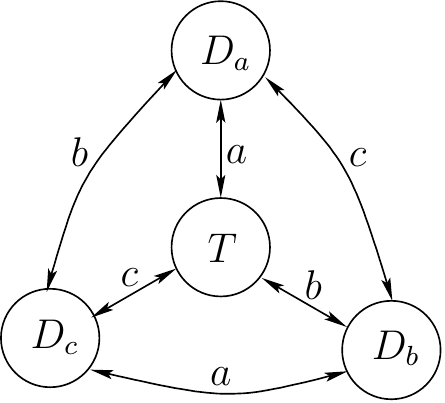} \\
(a) & & (b) 
\end{tabular}
\end{center}
\caption{\label{fig:puzzle} (a) Two point-sized independent bodies move along continuous paths in an annulus-shaped region in the plane.  There are three sensor beams, $a$, $b$, and $c$. When each is crossed by a body, its corresponding symbol is observed.  Based on receiving a string of observations, the task is to determine whether the two bodies are together in the same region, with no beam separating them.  (b) The minimal filter as a transition system has only 4 states: $T$ means that they are together, and $D_x$ means that are in different regions but beam $x$ separates them. Each transition is triggered by the observation when a body crosses a beam.}
\end{figure}

\begin{Ex}
  Figure \ref{fig:puzzle}(a) shows a filtering example from \cite{TovCohBobCzaLav14}.  More complex versions have been studied more recently in \cite{okaneshell}, and are found through automaton minimization algorithms and some extensions.  It can be shown that this example's four-state derived information space depicted
  on Figure~\ref{fig:puzzle}(b) corresponds to the unique minimal sufficient refinement of the the labeling that only
  distinguishes between ``are in the same region'' and 
  ``are not in the same region''. To see this, first note
  that this labeling is sufficient (since it can be represented
  as an automaton, this follows from Theorem~\ref{prop:SufTrAut}).
  It follows from Theorem~\ref{thm:MinSufRef1} that if this
  labeling is not minimal, then there is a minimal one which
  is strictly coarser, and so can be obtained by merging
  the states in the automaton of Figure~\ref{fig:puzzle}(b).
  This is impossible: the state $T$ cannot be merged with 
  anything because it violates the base-labeling; 
  if, say $D_a$ and $D_c$, are merged, then transition $a$
  will lead to inconsistency as it can lead either to $D_b$
  (from~$D_c$) or to $T$ (from~$D_a$). This proves that this
  derived information space is indeed minimal sufficient,
  and by Corollary~\ref{thm:MinSufRef1} there are no others
  up to isomorphism.
\end{Ex}

\subsection{Computing sufficient refinements}
\label{sec:Algos}

% FIXME: Add discussion about O'Kane & Shell, and try to figure out whether their NP-completeness result works for bounded set $U$
% Discuss the "local" approach to finding a minimal sufficient refinement.

This section sketches some computational problems and presents computed examples.  The problem of computing the minimal sufficient refinement in some cases reduces to classical deterministic finite automaton (DFA) minimization, and in other cases it becomes NP-hard \cite{okaneshell}.  

Consider an automaton $(\stateSp,M, \transf)$ and a labeling function
$\labelf_0$, and the corresponding labeled automaton described using
the quintuple $(\stateSp,M, \transf,\labelf_0,L)$.  Suppose that
the automaton $(\stateSp,M,\transf)$ corresponds to that of an
body-environment system.
%such that the agent and the environment are specified.
Hence, $\stateSp$ corresponds to the states of this coupled
system.  Suppose $\labelf_0$ is not sufficient and consider the
problem of computing a (minimal) sufficient refinement of $\labelf_0$,
that is, the coarsest refinement of $\labelf_0$ that is sufficient.

Despite the uniqueness of the minimal sufficient refinement of $\labelf_0$ (by Corollary~\ref{thm:MinSufRef1}), we can argue that the formulation of the problem, in particular, the input, can differ based on the level at which we are addressing the problem (for example, global perspective, agent perspective or something in between). Since the labeled automaton corresponding to an agent-environment coupling is described from a global perspective, the input to an algorithm that addresses the problem from this perspective is the labeled automaton $\mathcal{A}=(\stateSp,\transf,M,\labelf_0,L)$ itself. Then, the problem is defined as given $\mathcal{A}$ compute
$\mathcal{A}'=(\stateSp,M,\transf,\labelf,L)$ such that $\labelf$ is the minimal sufficient refinement of $\labelf_0$.

A special case of this problem from the global perspective occurs if the preimages of $\labelf_0$ partition
$\stateSp$ in two classes which can be interpreted as the ``accept''
and ``reject'' states, for example, goal states at which the agent
accomplishes a task and others. Furthermore, suppose that the initial
state of the agent is known to be some $x_0 \in \stateSp$. Then,
computing a minimal sufficient refinement becomes identical to
minimization of a finite automaton, that is, given a DFA $(\stateSp,M,\transf,x_0,F)$ in which $x_0$ is the initial state and $F$ is the set of accept states find
$(\stateSp',M,\transf',x_0',F')$ such that no DFA with fewer states
recognizes the same language. Existing algorithms, for example
\cite{hopcroft1971n}, can be used to compute a minimal automaton.

Here, we also consider this problem from the agent's perspective for
which the information about the environment states is obtained through
its sensors, more generally, through a labeling function. Note that by agent's
perspective we do not necessarily imply that the agent is the one
making the computation (or any computation) but it means that no
further information can be gathered regarding the environment other than the actions taken and what is sensed by the agent.
At this level we address the following problem; given a set $M$ of actions, a domain
$\stateSp$, and a labeling function $h_0$ defined on $\stateSp$, compute
the minimal sufficient refinement of $h_0$. The crux of the problem is that unlike the global perspective described above, the labeled automaton $\mathcal{A}$ is not given, in particular, the state transitions are not known a priory. Instead, the information regarding the state transitions can only be obtained locally by means of applying actions and observing the outcomes, that is, through sensorimotor interactions. Hence, the current body-environment state is also not observable.
%Furthermore, the sensations correspond to the sensory data such that the actual state is not observable.  
To show that an algorithm exists to compute a sufficient refinement of $h_0$ at this level,
%this is indeed possible, 
we propose an iterative
algorithm (Algorithm~\ref{alg:suff_ref}) that explores $\stateSp$
through agent's actions and sensations by keeping the history
information state, that is, the history of actions and sensations
(labels). We then show, by empirical results, that the sufficient refinement computed by Algorithm~\ref{alg:suff_ref} is minimal for the selected problem.

\begin{algorithm}
\caption{}%Compute sufficient refinement of $\labelf_0$}
\begin{algorithmic}[1]
\State \textbf{Input:} $\labelf_0$, $l_0$, $M$
\State \textbf{Initialize:} $H \gets\emptyset$, $\labelf \gets \labelf_0$, $\obs \gets \obs_0$ %$H$
\For{each step}
\State $\action \gets \text{policy}(\obs)$
\State apply action $m$ and obtain resulting $\obs'$
%\State $\obs' \gets \text{new sensation}$
%\State $\obs' \gets \text{newObservation}(\obs,\action)$
\State add $(\obs,\action,\obs')$ to $H$
%\State $H\text{.append}((\obs,\action,\obs'))$
%\If{$H \neq \emptyset$}
    %\If{$\neg \text{isConsistent}(H, (\obs,\action, \obs')) $}
    \If{$\exists (\obs, m, \obs'') \in H$ such that $s' \neq s''$}\label{algl:consistent_hist} %$\labelf^{-1}(\obs')\neq \labelf^{-1}(\obs'')$}
    \State $\labelf \gets \text{split}(\labelf,s)$
    %\State \red{$H \gets \text{purge}(H,\obs)$}
    \EndIf
%\EndIf
\If{there are labels that can be merged}
%\If{condition to check for mergeable labels \textcolor{red}{e.g., $\log_n(\text{step})=0$}}
\label{algl:merge_cond}
\State $\labelf \gets merge(\labelf,H, \labelf_0)$\label{algl:merge}
\EndIf
\State $\obs \gets \obs'$
\EndFor
\end{algorithmic}
\label{alg:suff_ref}
\end{algorithm}

The functioning of Algorithm~\ref{alg:suff_ref} is as follows.
Starting from an initial sensation $\obs_0=\labelf(x_0)$, the agent
moves by taking an action%
\footnote{This can either be in a real environment or in a realistic simulation.}
given by the mapping $\text{policy}: L \rightarrow M$. Particularly,
we used a fixed policy which samples an action $m$ from a uniform
distribution over $M$ for each $s \in S$. In principle, any policy
that ensures all states that are reachable from $x_0$ will be visited
infinitely often should be enough. The history information state is
implemented as a list, denoted by $H$, of triples
$(\obs,\action,\obs')$ such that $\obs=\labelf(x)$ and
$\obs'=\labelf(x')$ in which $x'=\transf(x,\action)$. At each step, it
is checked whether the current sensation is consistent with the
history (Line~\ref{algl:consistent_hist}). Current sensation is
inconsistent with the history if there exists a triple
$(\obs,m,\obs'')$ in the history such that $s' \neq
s''$. %$\labelf^{-1}(\obs')\neq\labelf^{-1}(\obs'')$.
If it is not consistent then the label is split, which means that
$h^{-1}(\obs)$ is partitioned into two parts $P$ and $Q$. In
particular, we apply a balanced random partitioning, that is, we
select $P$ and $Q$ randomly from a uniform distribution over the
partitions of $\labelf^{-1}(\obs)$ that have two elements with
balanced cardinalities.  The labeling function is updated by a split
operation as
\begin{equation*}
\labelf(x) :=
    \begin{cases}
    \obs_{Q} & \text{if $x\in Q$}\\
    \obs_{P} & \text{if $x\in P$}\\
    \labelf(x), & \text{otherwise}.
    \end{cases}
\end{equation*}
Recall that labels or subscripts do not carry any meaning from the agent's perspective. 

% \red{Proposition: If $\stateSp$ is finite and the policy ensures
% that the reachable states from $x_0$ will be visited ``frequently''
% then there is a finite number of steps such that
% Algorithm~\ref{alg:suff_ref} computes a suff. refinement of the
% restriction of $h_0$ to reachable states from $x_0$.}

Even a trivial strategy that splits the preimage of the label seen at
each step would succeed computing a sufficient refinement. However,
this would result in $\labelf$ being a one-to-one mapping. Hence, the
finest possible refinement. Splitting only at the instances when an
inconsistency is detected might reach a coarser refinement that is
sufficient but there might be more equivalence classes than the ones
induced by the minimal sufficient refinement of $\labelf_0$.
Therefore, a merge operation is introduced (Line \ref{algl:merge}).
Let $\obs$ and $\obs'$ be two distinct labels for which
$\exists \obs'' \in \labelf_0[\stateSp]$ such that
$\labelf^{-1}(\obs) \subset \labelf_0^{-1}(\obs'')$ and
$\labelf^{-1}(\obs') \subset \labelf_0^{-1}(\obs'')$. Let $t$ denote a
triple in $H$ and let $t_k$, $k=1,2,3$, denote the $k^{th}$ element of
that triple. Suppose $\obs' = \obs$, if there are at least $N$ number of triples in $H$ such that for each triple $t$, $(t_1,t_2)=(\obs,m)$ and
$\forall m \in M$ and $\forall t,t' \in H$ such that $(t_1,t_2)=(t'_1, t'_2)=(\obs,m)$ it is
true that $t_3=t'_3$ then labels $\obs$ and $\obs'$ are merged. The
merge procedure goes through all labels and updates $h$ as
\begin{equation*}
\labelf(x) :=
    \begin{cases}
    \obs & \text{if $\labelf(x)\in\{\obs, \obs'\}$}\\
    \labelf(x) & \text{otherwise}.
    \end{cases}
\end{equation*}
for each pair of labels $\obs$ and $\obs'$ that satisfies the aforementioned condition. Note that in principle, one can merge two labels regardless of the number of occurrences in the history. However, we noticed that this can result in oscillatory behaviour between split and merge operations especially for states that are reached less frequently. At present, we considered $N$ as a tunable parameter and we know that it depends on the cardinality of the state space $X$ such that larger the number of states, larger $N$ should be. The problem of defining $N$ as a function of the problem description remains open.

In the following, we present an illustrative example to show the
practical implications of the previously introduced concepts in Section~\ref{ssec:min_suff_ref}. In particular, we show how a simple algorithm
like Algorithm~\ref{alg:suff_ref} can be used by a computing unit
which relies only on the sensorimotor interactions of an agent to
further categorize the environment such that there are no
inconsistencies in terms of the actions taken by the agent and the
resulting sensations with respect to an initial categorization
induced by $h_0$.

\begin{figure}\label{fig:cheese_maze}
    \centering
    \subfigure[]{\includegraphics[width=0.3\linewidth]{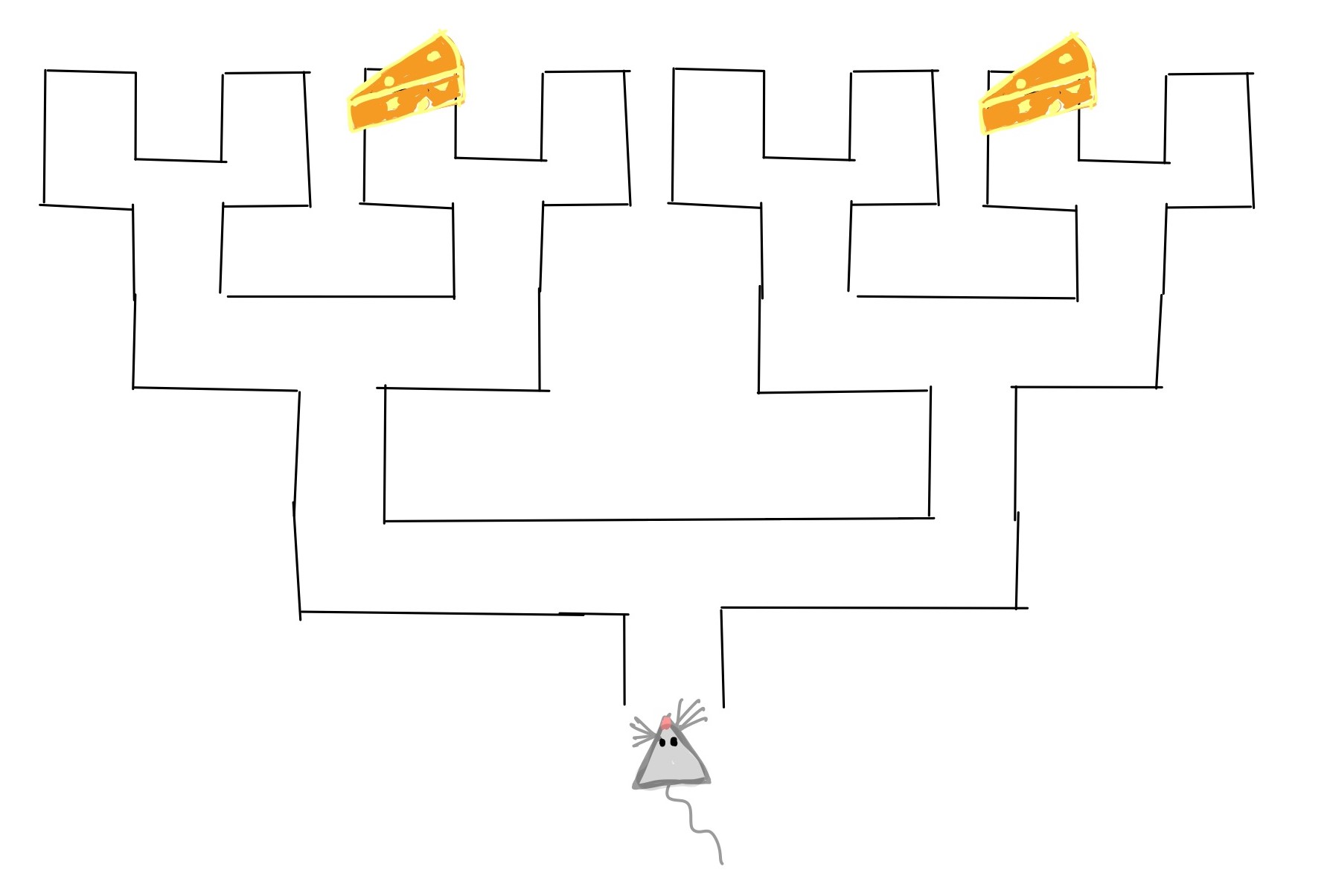}}
    \subfigure[]{\includegraphics[width=0.3\linewidth]{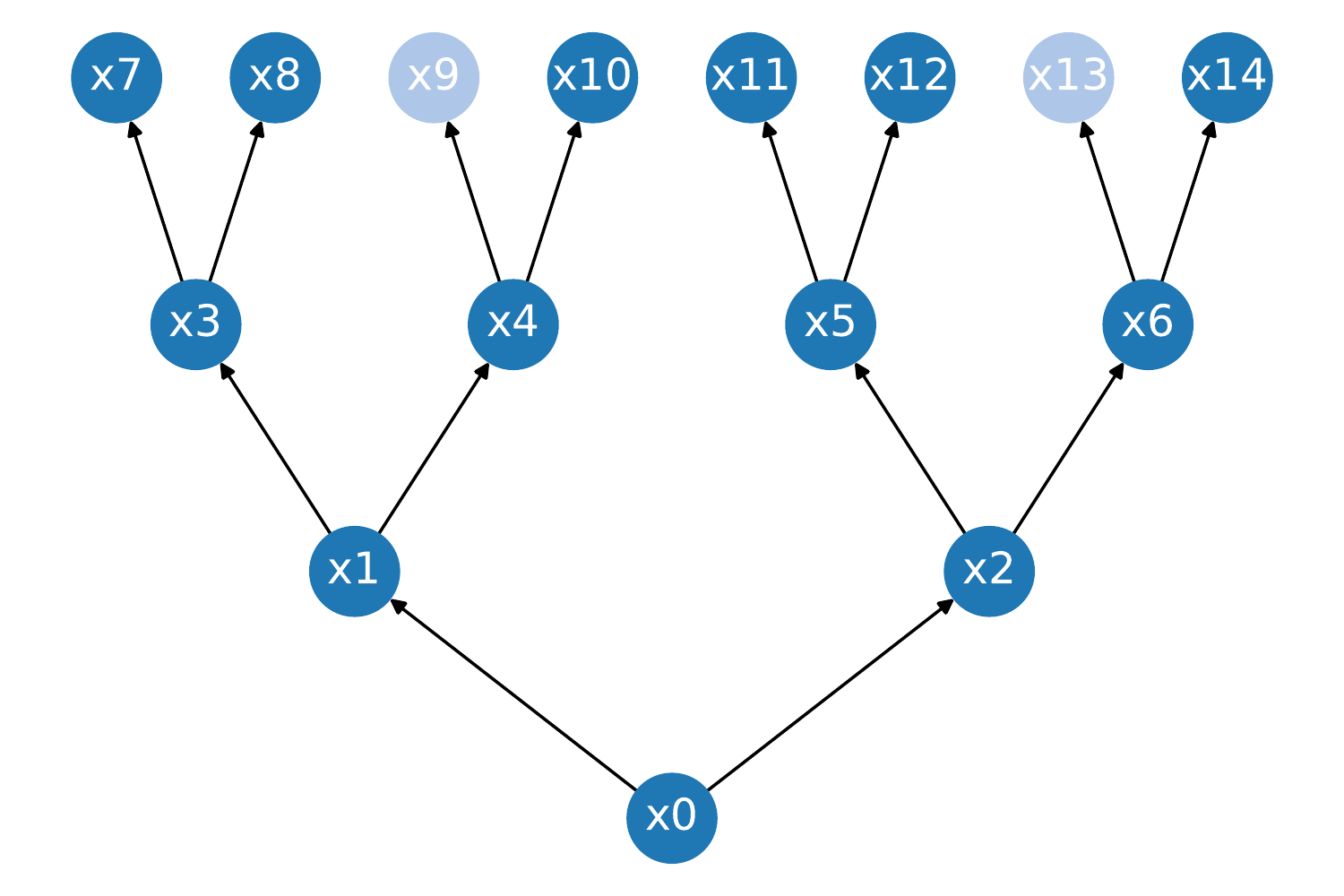}}
    \subfigure[]{\includegraphics[width=0.3\linewidth]{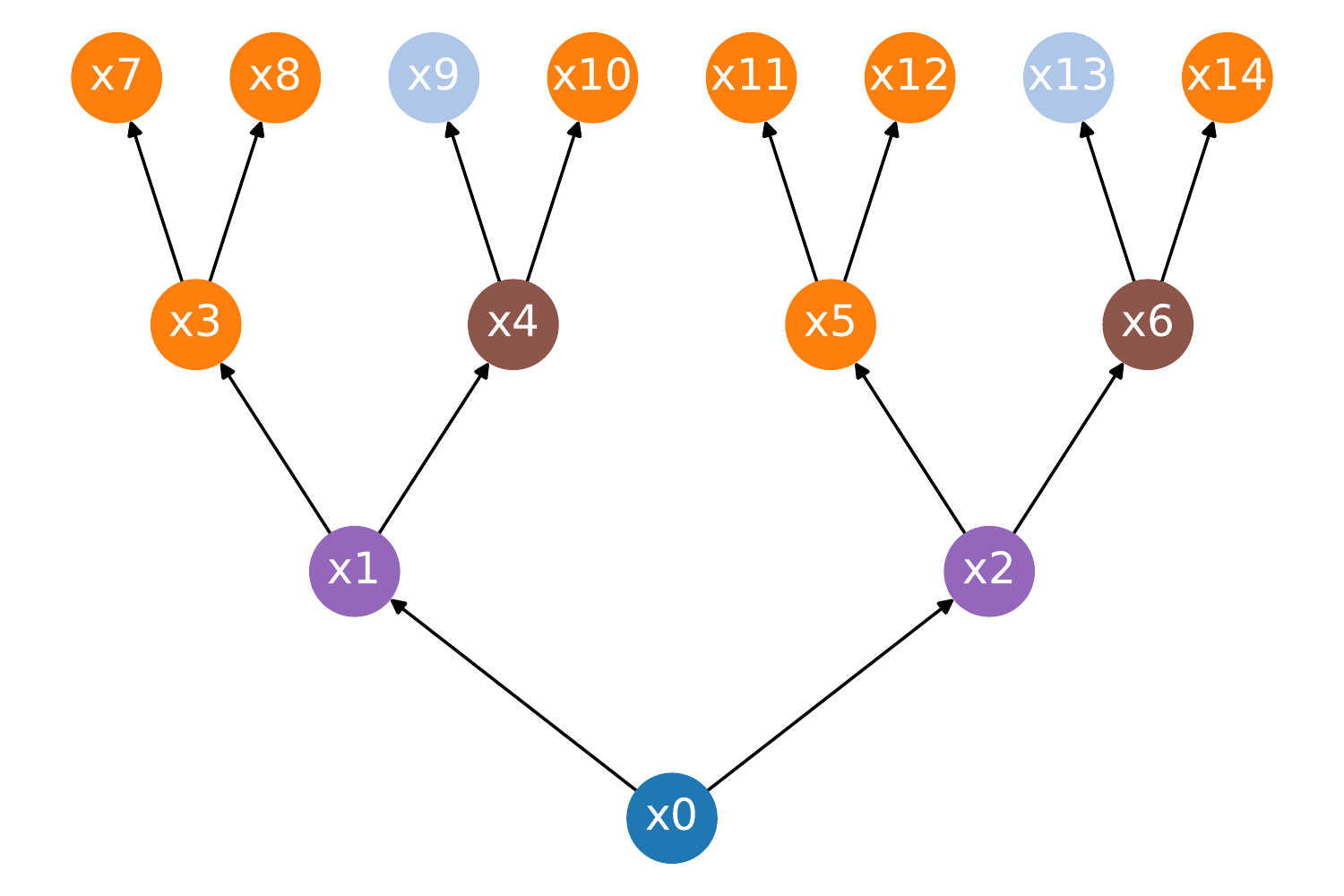}}
    \caption{(a) Cheese maze defined in Example~\ref{ex:cheese_maze}
      (b) Labeled automaton with initial labeling $h_0$ corresponding to
      the cheese-maze example. (c) Minimal sufficient refinement
      of~$h_0$. Self-loops at the leaf nodes are not shown in the
      figure.}
    \label{fig:initial_final_refinement}
\end{figure}

\begin{figure}
    \centering
    \subfigure[]{\includegraphics[width=.24\linewidth]{plot0.pdf}}
    \subfigure[]{\includegraphics[width=.24\linewidth]{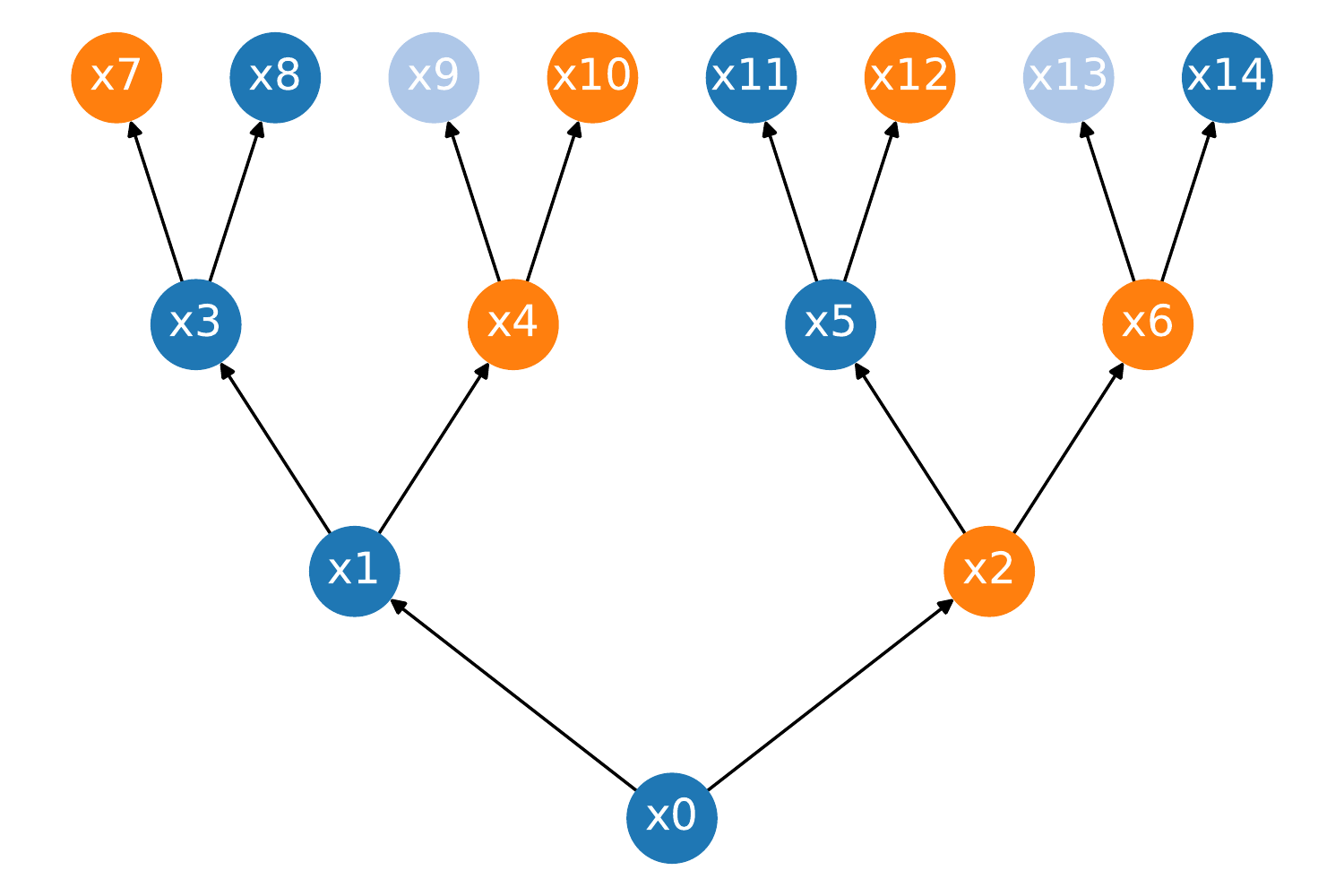}}
    \subfigure[]{\includegraphics[width=.24\linewidth]{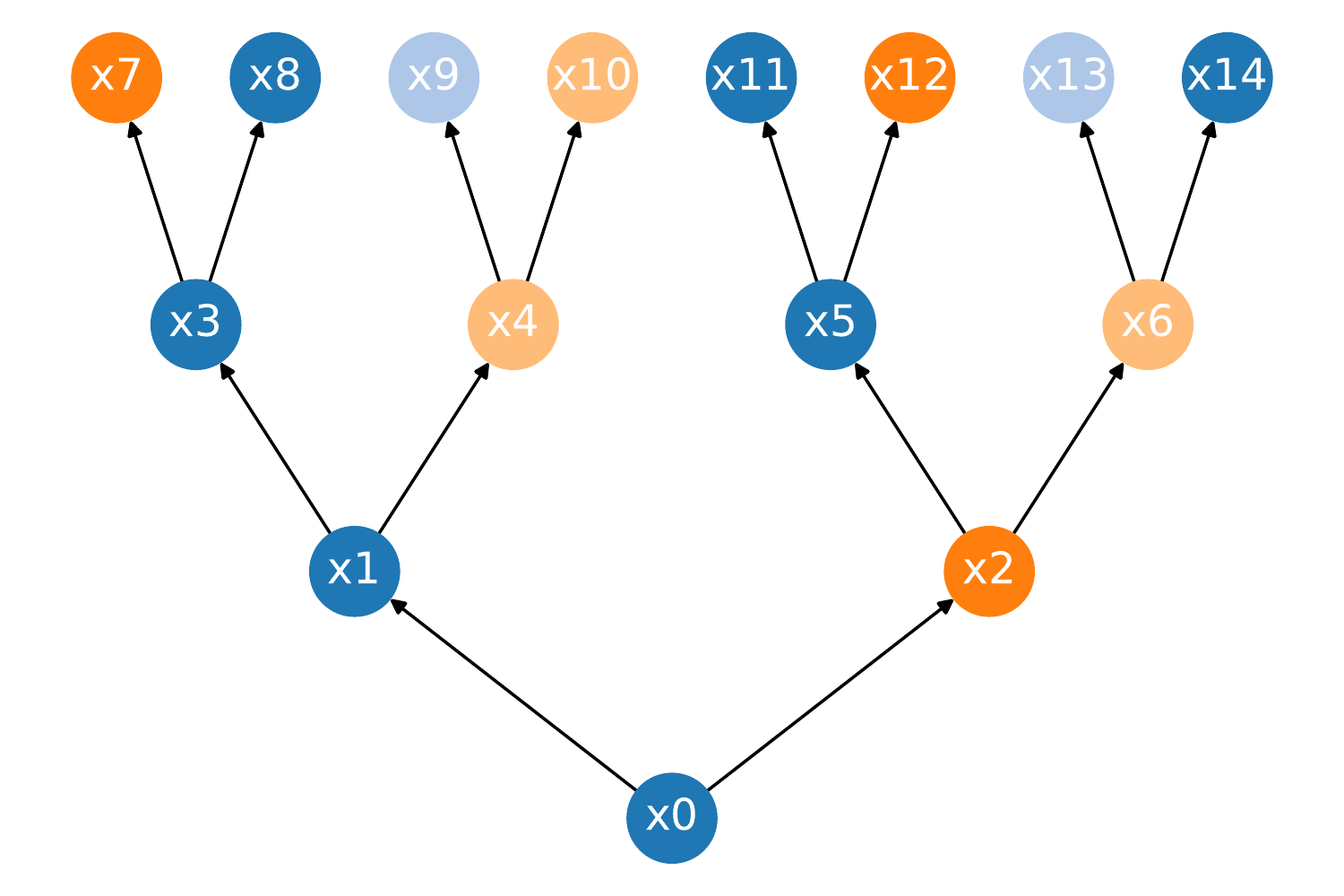}}
    \subfigure[]{\includegraphics[width=.24\linewidth]{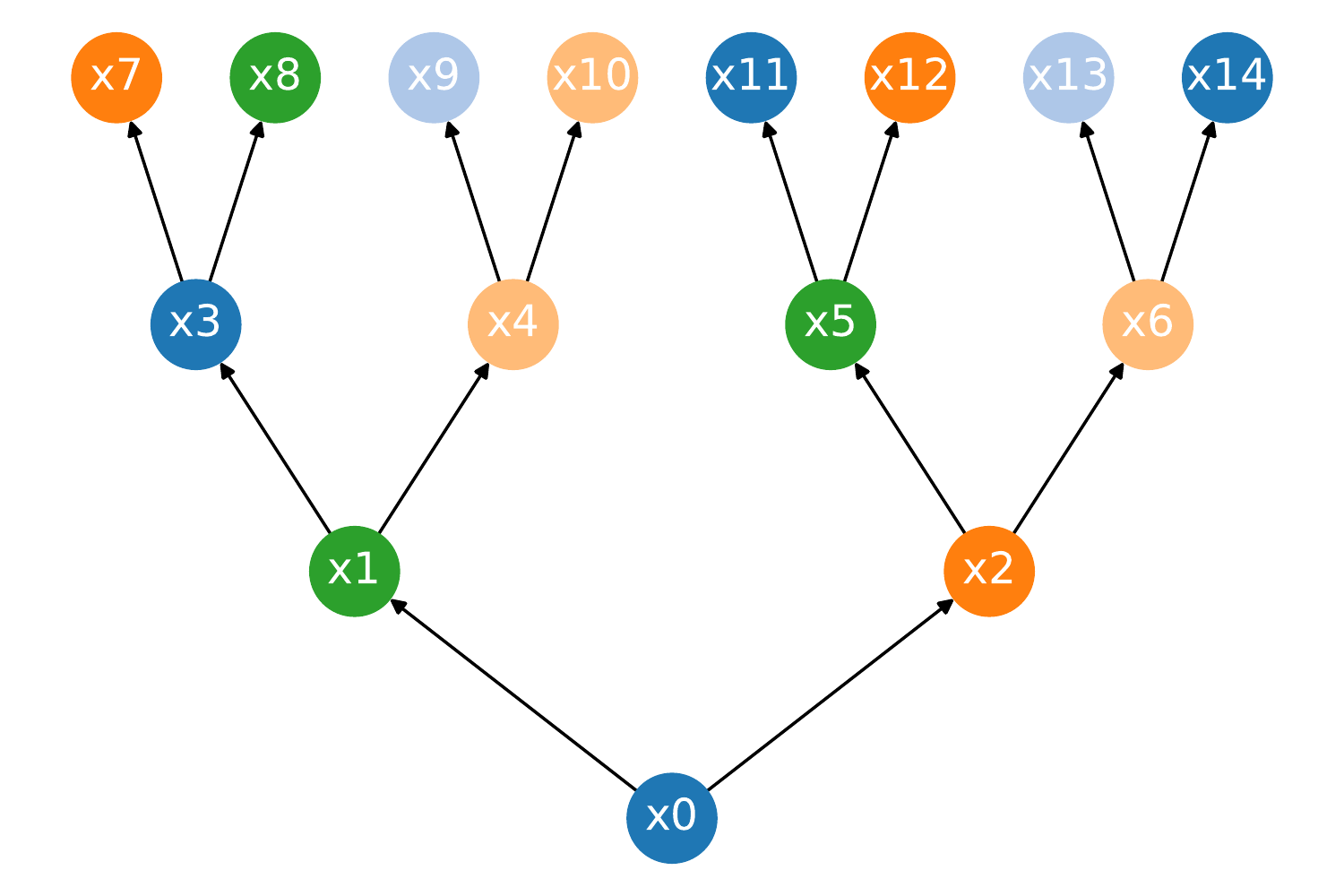}}\\
    \subfigure[]{\includegraphics[width=.24\linewidth]{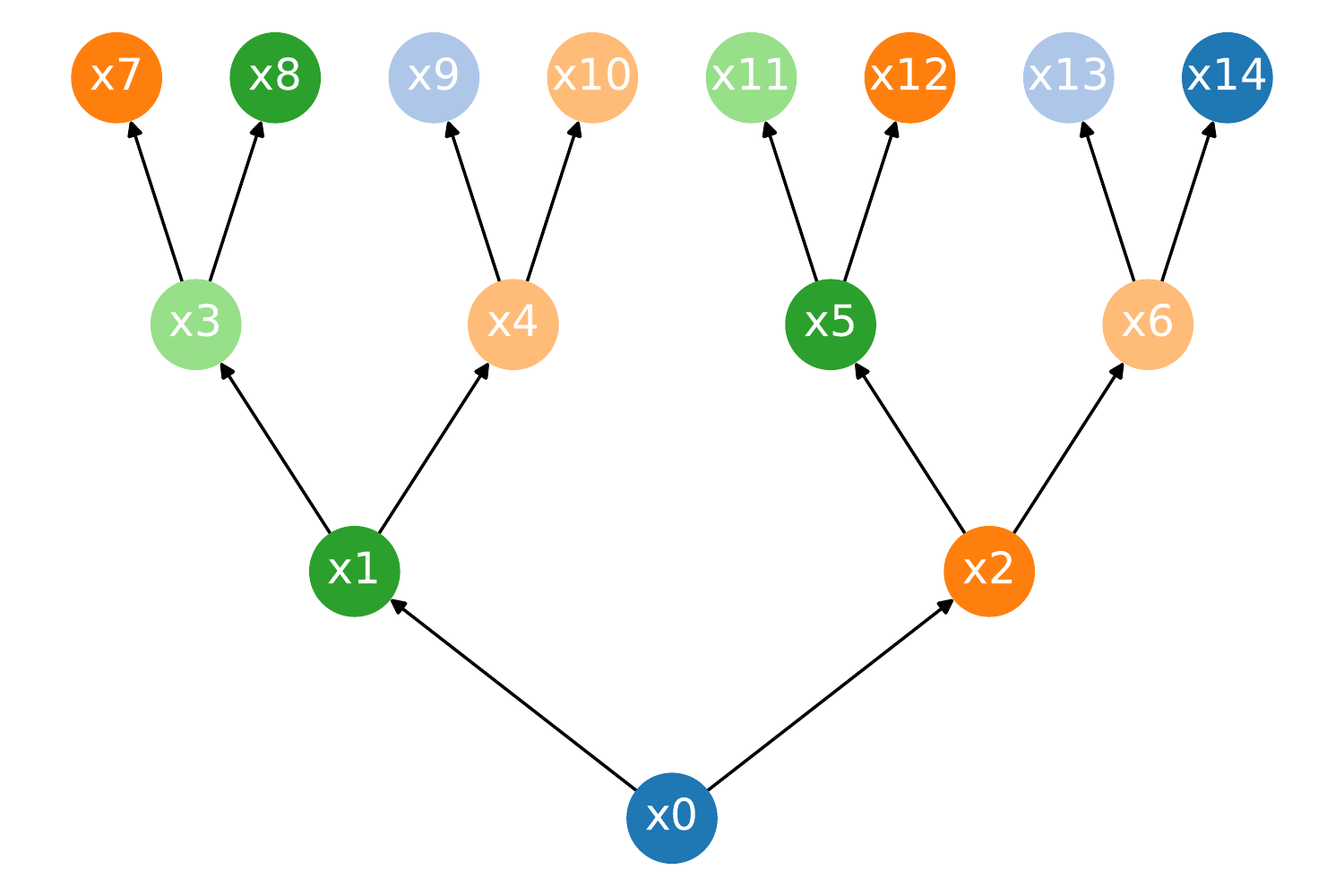}}
    \subfigure[]{\includegraphics[width=.24\linewidth]{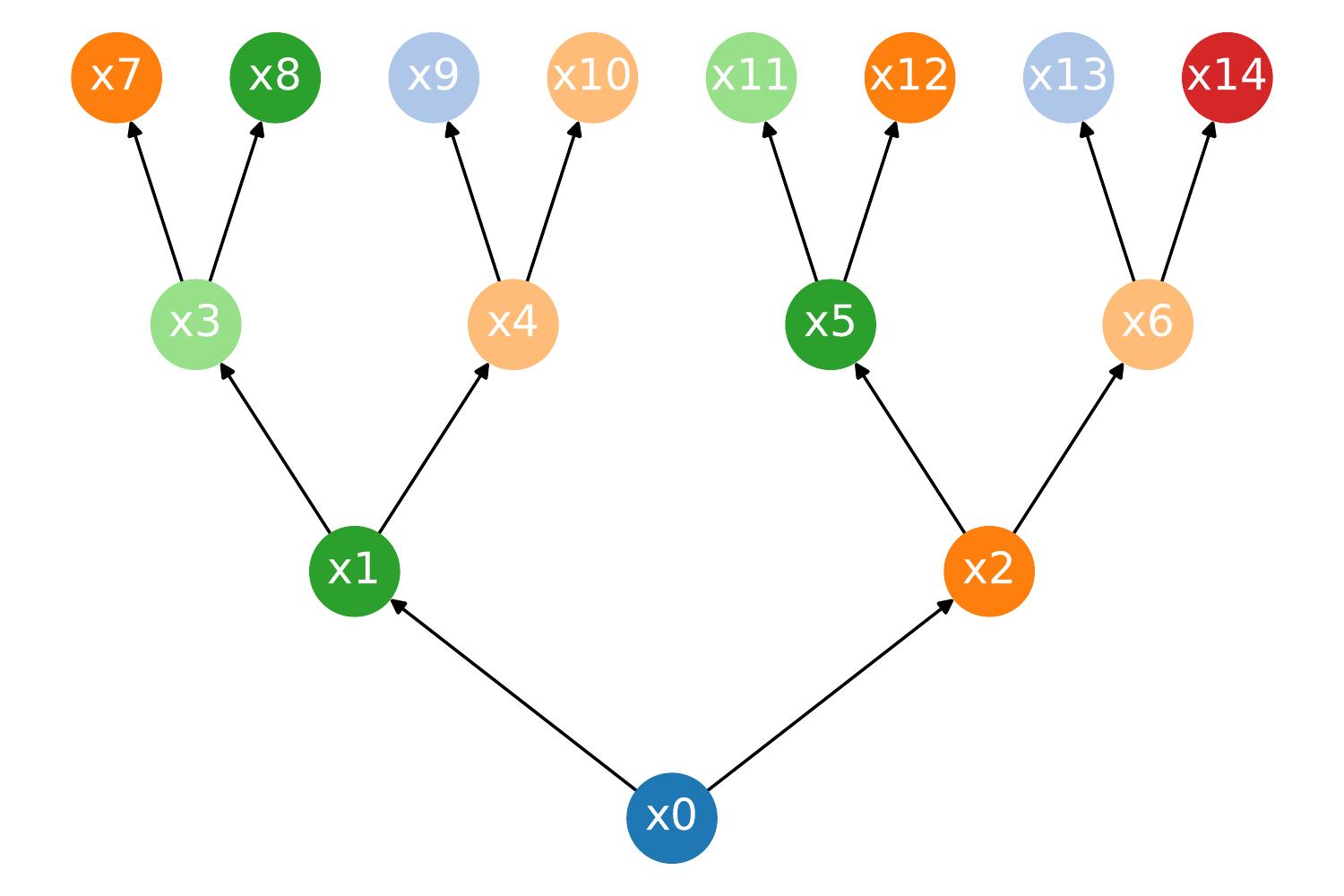}}
    \subfigure[]{\includegraphics[width=.24\linewidth]{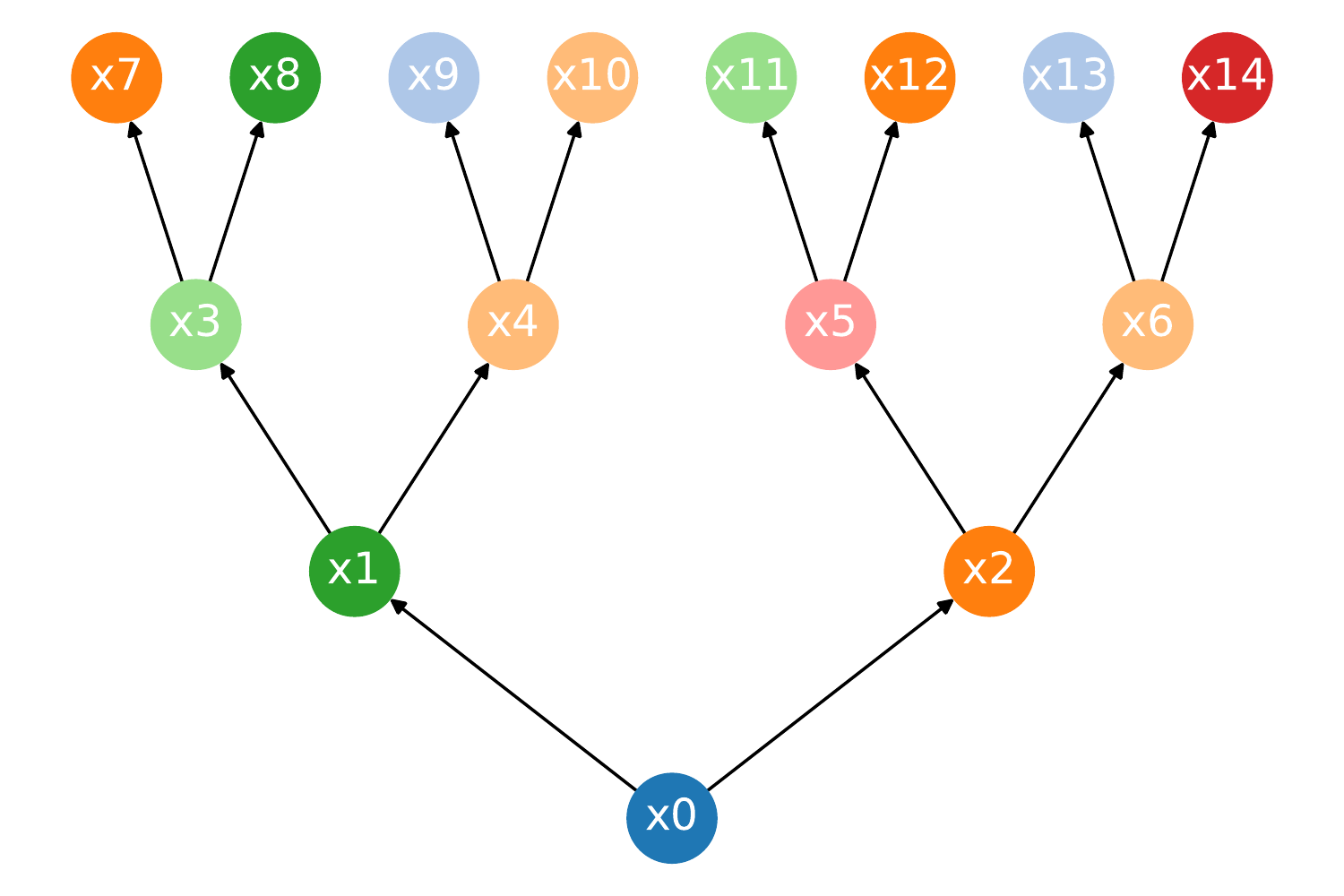}}
    \subfigure[]{\includegraphics[width=.24\linewidth]{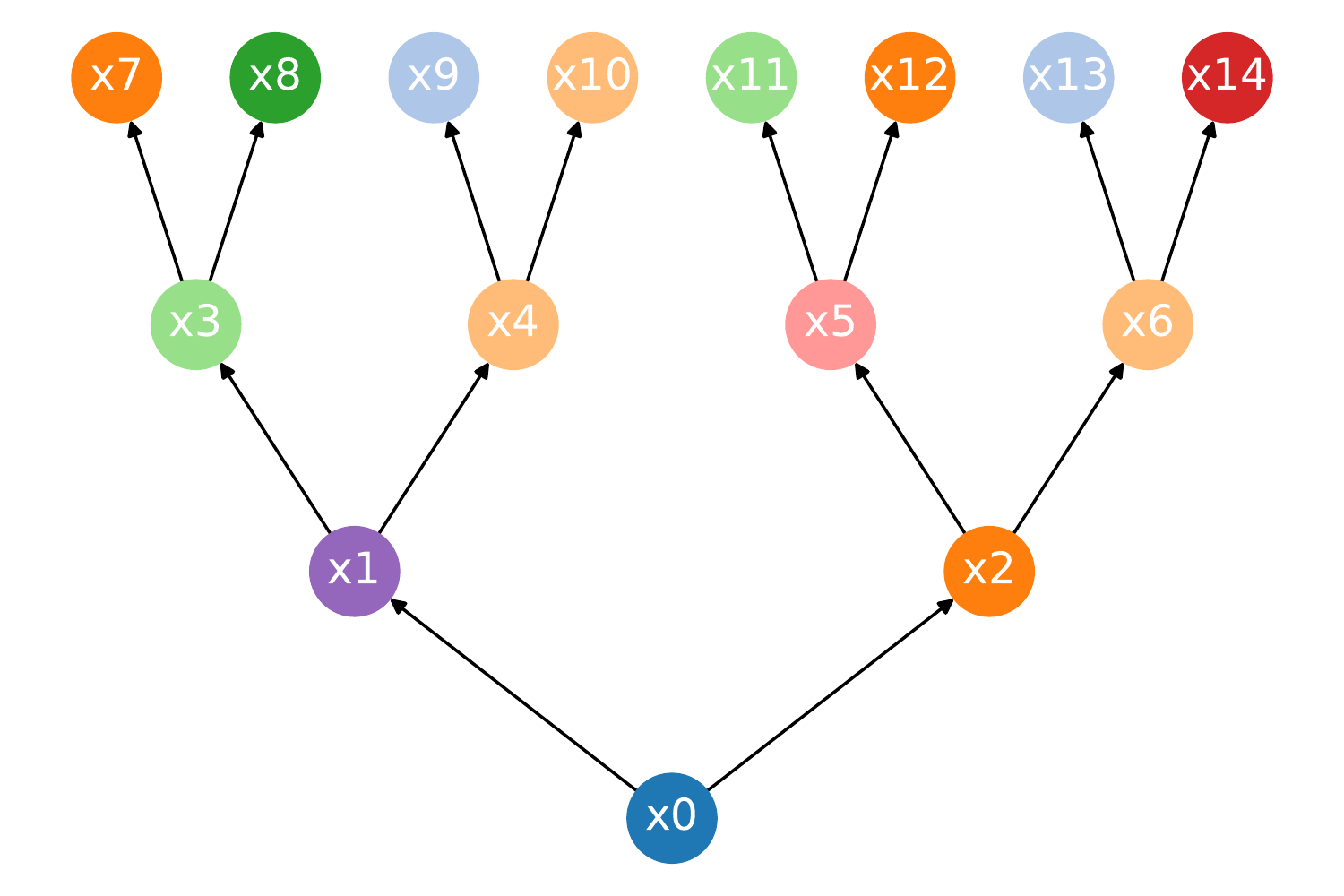}}\\
    \subfigure[]{\includegraphics[width=.24\linewidth]{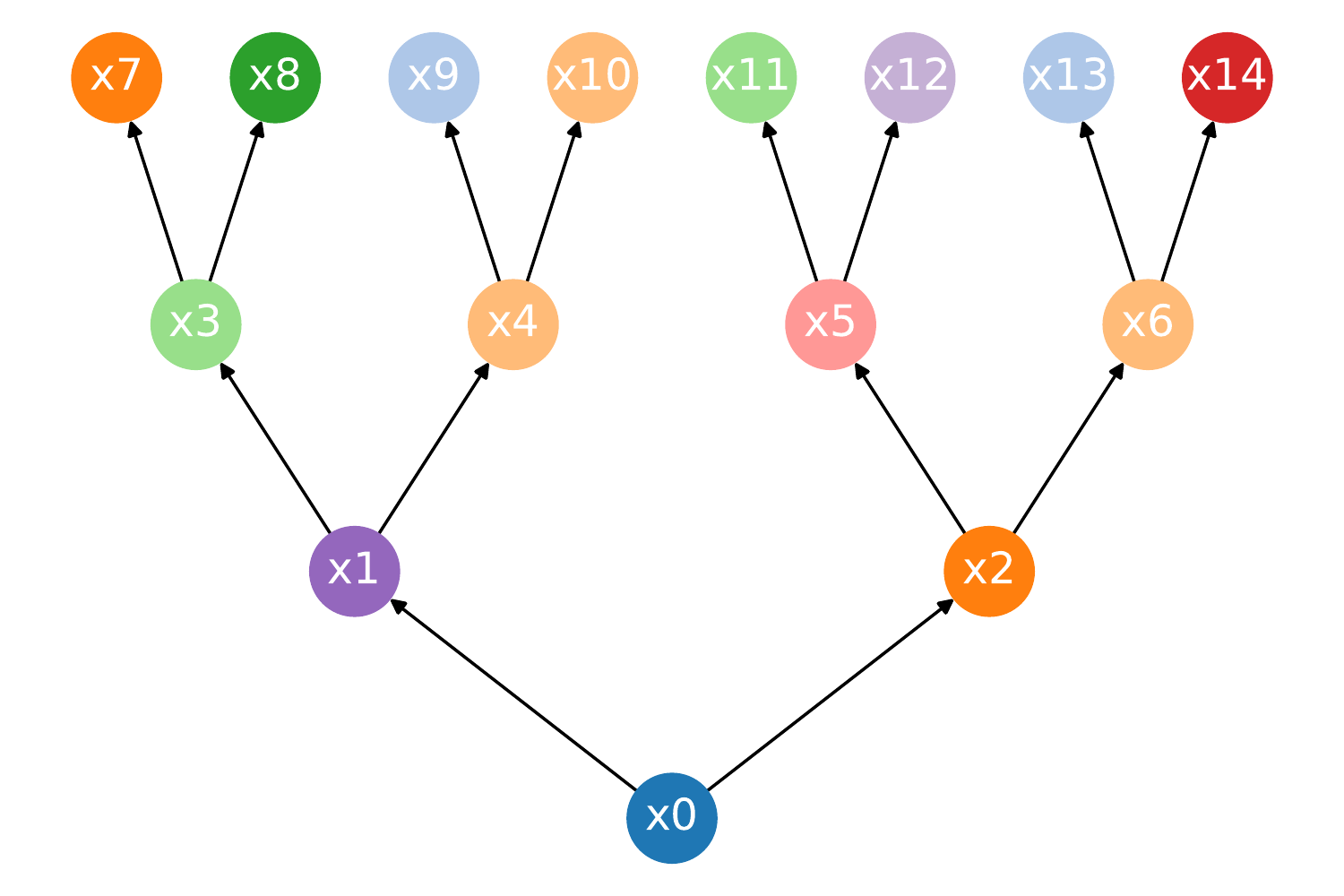}}
    \subfigure[]{\includegraphics[width=.24\linewidth]{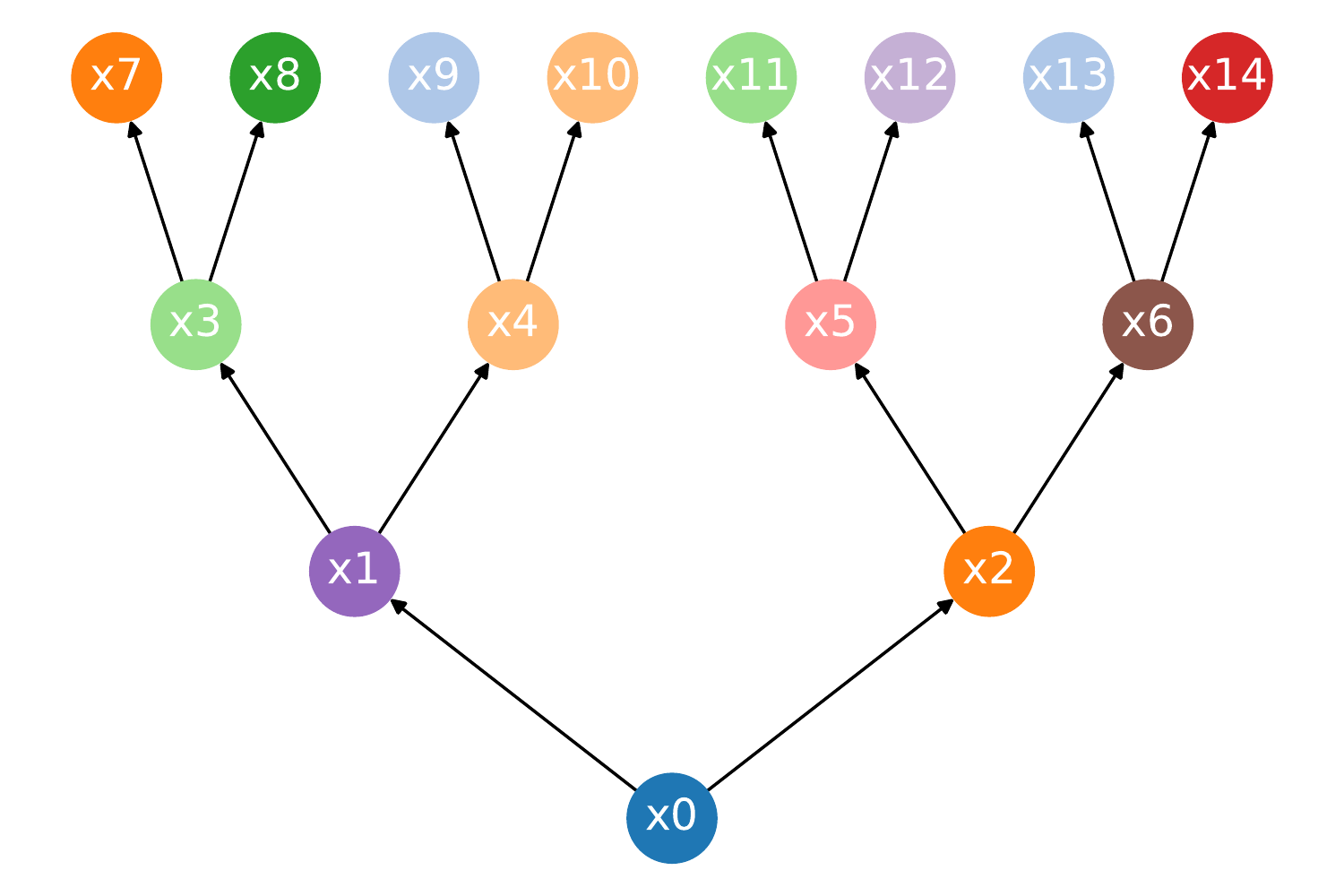}}
    \subfigure[]{\includegraphics[width=.24\linewidth]{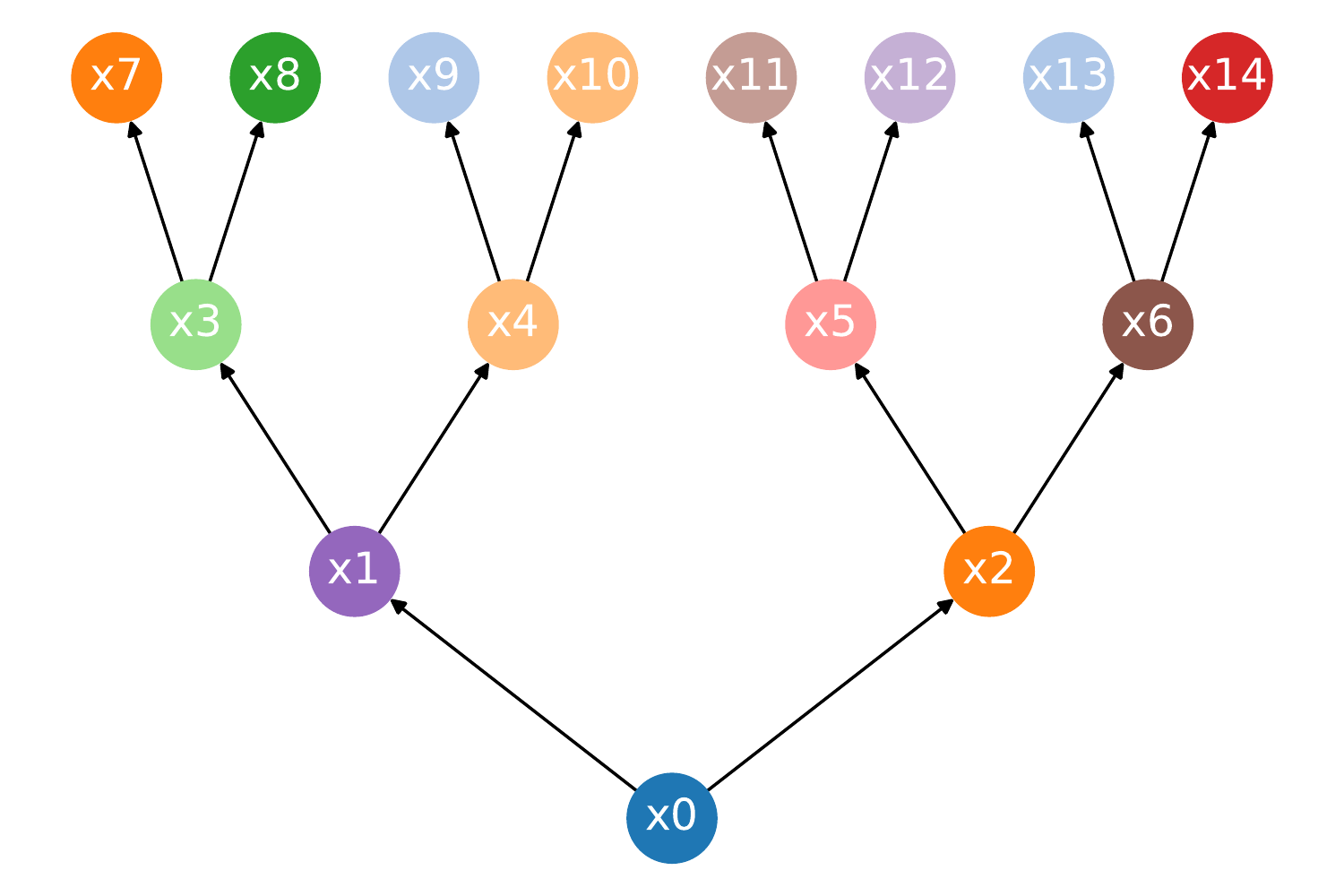}}
    \subfigure[]{\includegraphics[width=.24\linewidth]{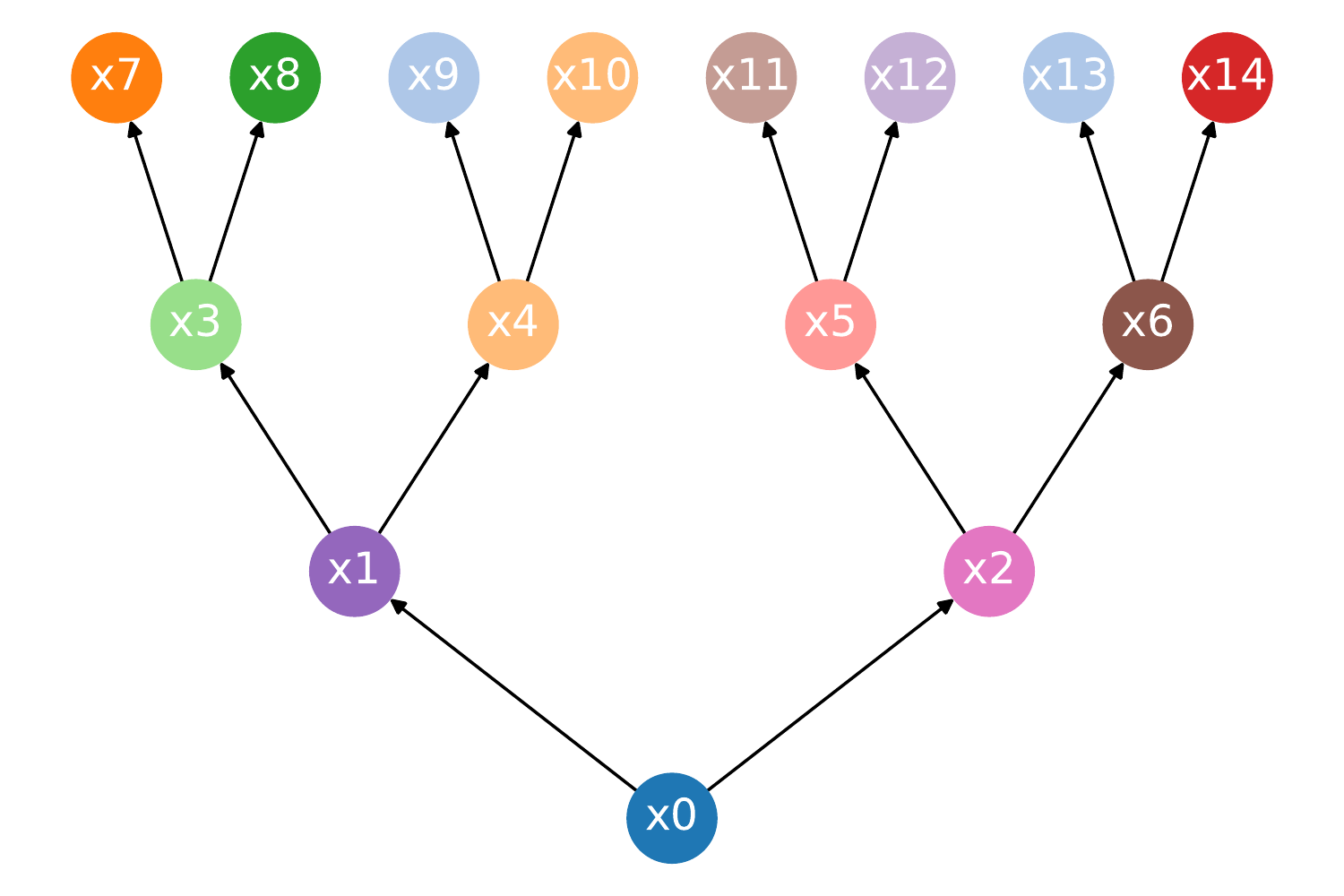}}\\
    \subfigure[]{\includegraphics[width=.24\linewidth]{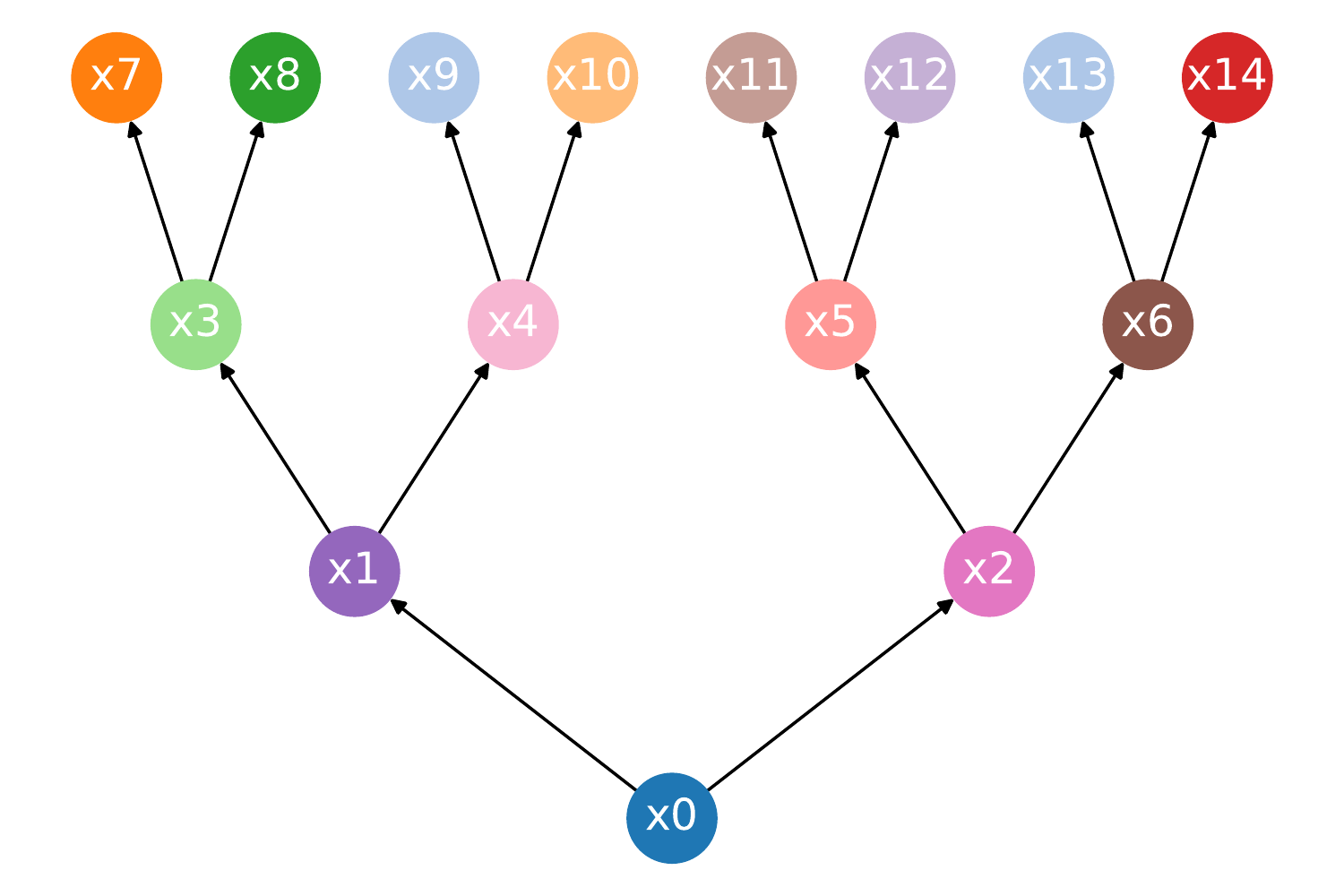}}
    \subfigure[]{\includegraphics[width=.24\linewidth]{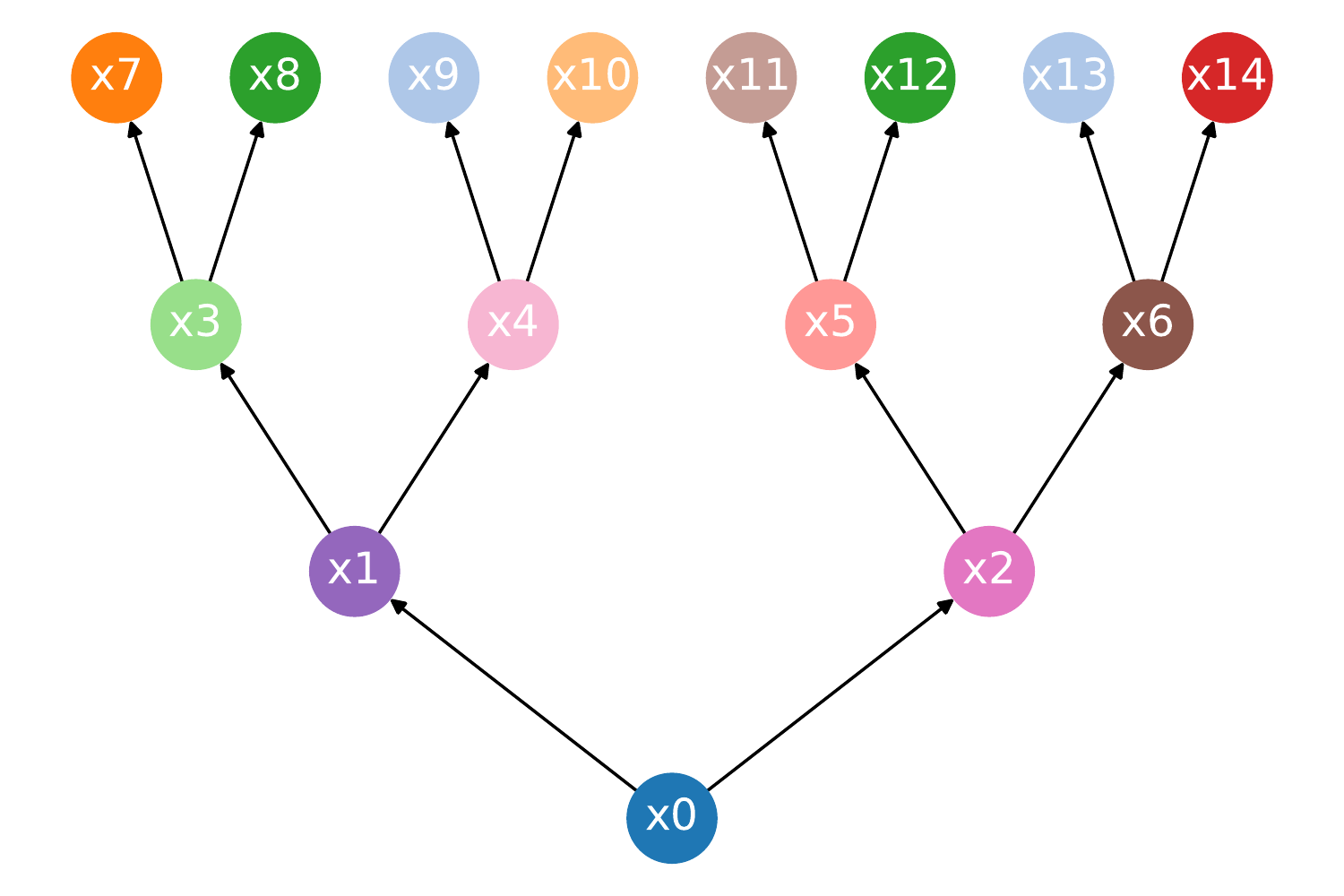}}
    \subfigure[]{\includegraphics[width=.24\linewidth]{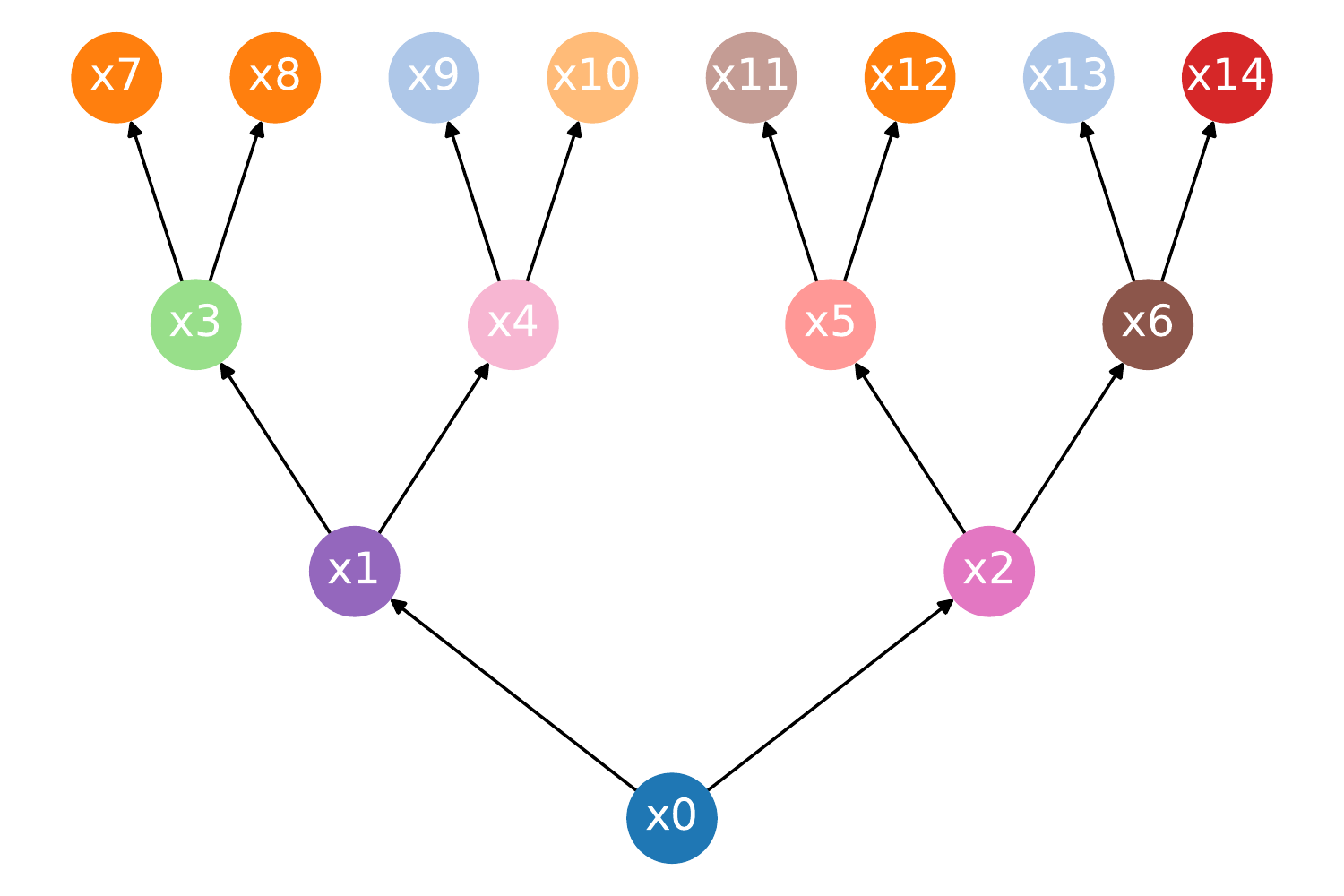}}
    \subfigure[]{\includegraphics[width=.24\linewidth]{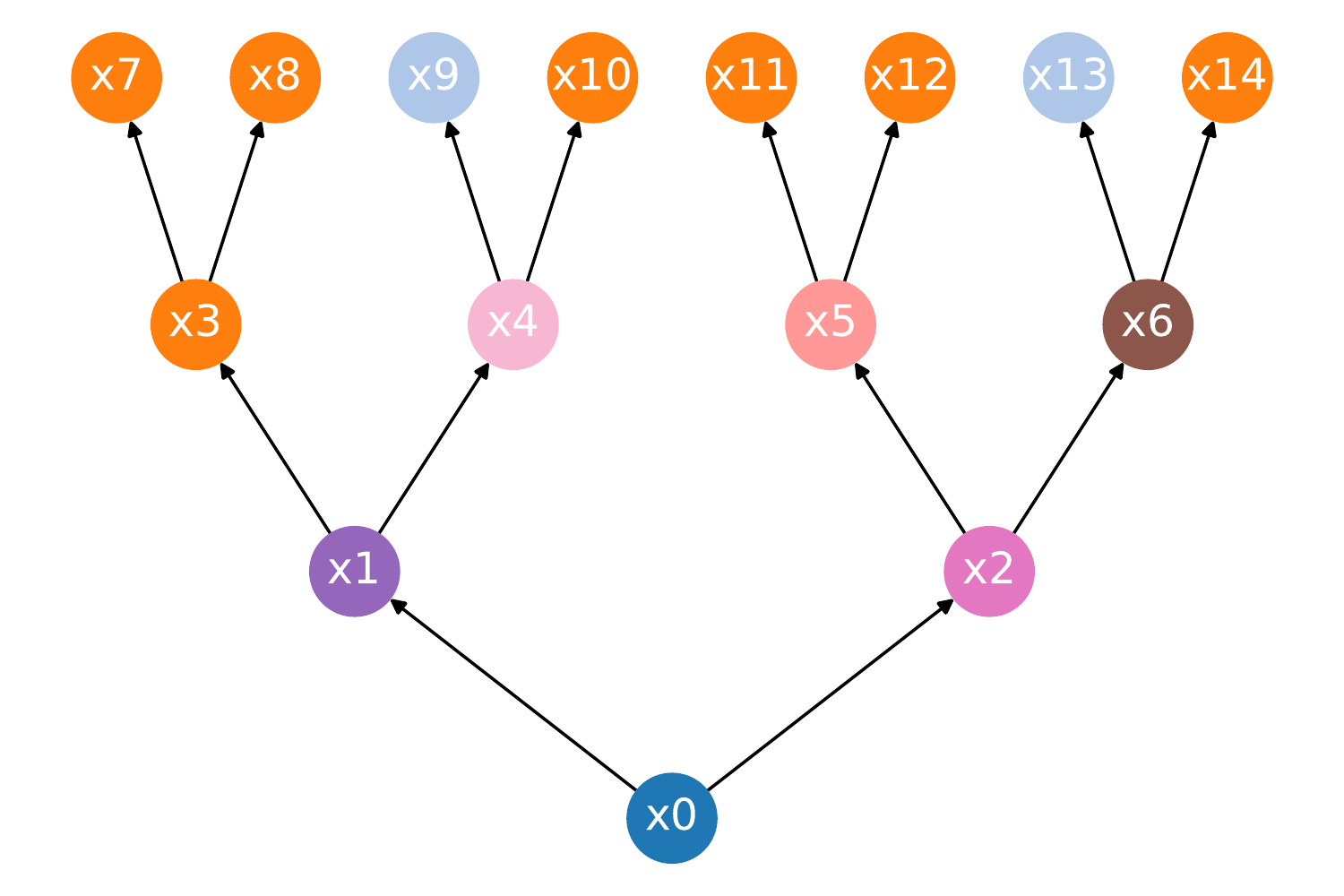}}\\
    \subfigure[]{\includegraphics[width=.24\linewidth]{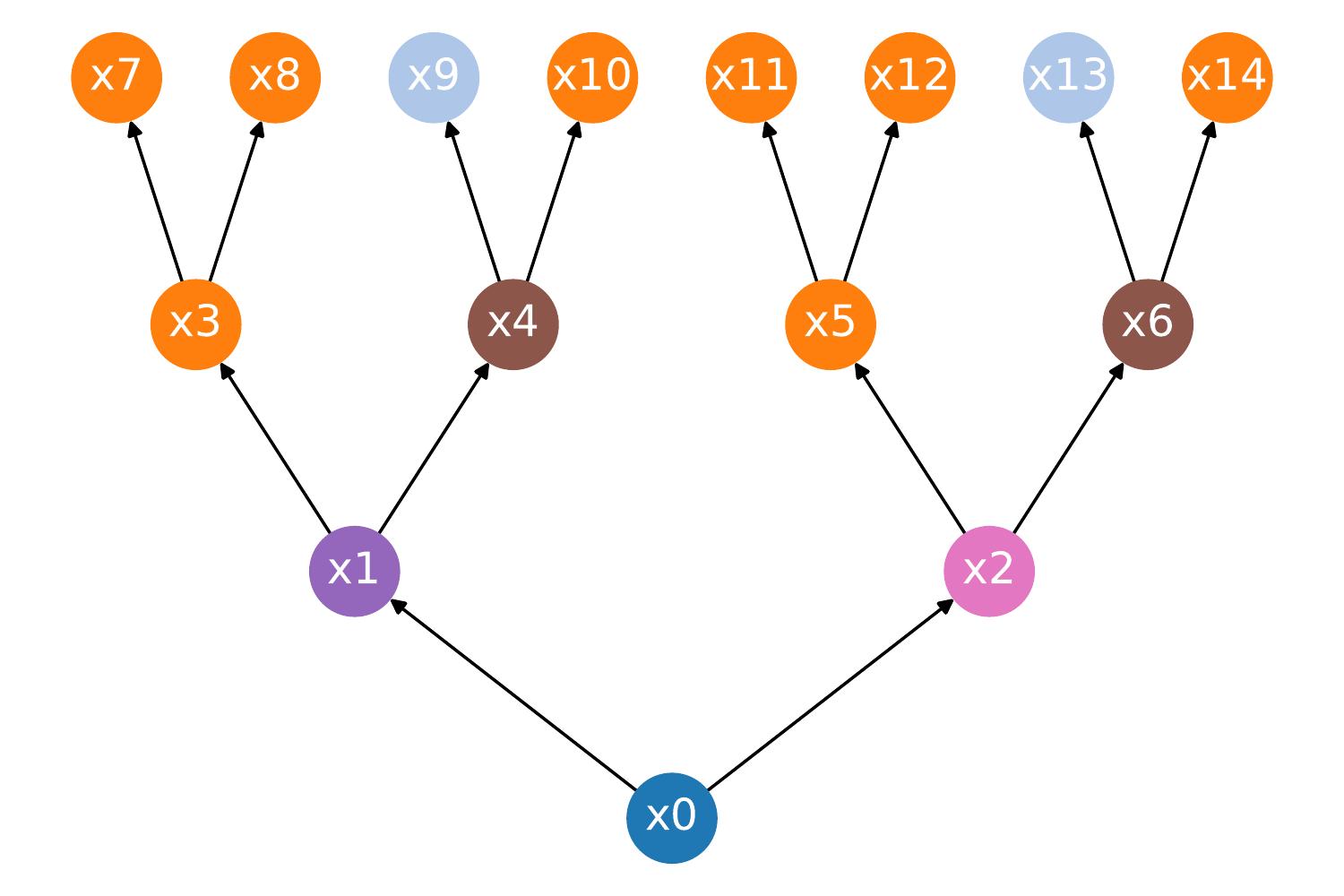}}
    \subfigure[]{\includegraphics[width=.24\linewidth]{plot17.pdf}}
    \caption{(a) Labeled automaton with labeling function $h=h_0$;
      same colored states belong to the same equivalence class. (b-q)
      Updating $h$ by Algorithm~\ref{alg:suff_ref} through splitting
      and merging of the labels. (r) Labeled automaton with the
      labeling function $h$ that is the minimal sufficient refinement
      of~$h_0$.}\label{fig:cheese_maze_updates}
\end{figure}

\begin{Ex}\label{ex:cheese_maze}
  Consider an agent (a mouse) that is placed in a maze where certain
  paths lead to cheese and others do not (see
  Figure~\ref{fig:cheese_maze}(a)). At each intersection the agent can
  go either left or right and it can not go back. Hence, at each step
  the agent takes one of the two actions; go right or go
  left. Figure~\ref{fig:cheese_maze}(b) shows the corresponding
  automaton with $15$ states describing the agent-environment system
  together with the initial labeling $h_0$ that partitions the state
  space into states in which the agent has reached a cheese (light
  blue) and others (dark blue).  The initial state $x_0$ is when the
  agent is at the entrance of the maze. Once the end of the maze is
  reached (a leaf node) the state does not change regardless of which
  action is taken. After a predetermined number of steps the system
  reverts back to the initial state, similar to an episode in the
  reinforcement learning terminology (see, for example,
  \cite{sutton2018reinforcement}). However, despite the system going
  back to the initial state the history information state still
  includes the prior actions and sensations.
  Figure~\ref{fig:cheese_maze_updates} reports the updates of $h$,
  initialized at $h_0$, by Algorithm~\ref{alg:suff_ref} being run for
  1000 steps. It converged to a final labeling
  $h$ (Figure~\ref{fig:cheese_maze_updates}(r)), that is the minimal sufficient
  refinement of $h_0$, in 435 steps. For 20 initializations of Algorithm~1 for the same problem, on average, it took 364 steps to converge to a minimal sufficient refinement of $h_0$.
\end{Ex}

We have also applied the same algorithm to variations of this example with different depths of maze and different number of cheese and cheese placements (varying $h_0$). Empirical evidence shows that the same algorithm was capable of consistently finding the minimal sufficient refinement of the initial labeling. However, it is likely that it might fail for more complicated problems, for example, when the number of actions are significantly larger. It remains an open problem finding a provably correct algorithm for computing the minimal sufficient refinement of $h_0$ from the agent's perspective.   

\subsection{Sufficiency for coupled SM-systems}

Section~\ref{sec:TransSyst} introduced SM-systems, including  the special class of  quasifilters. We showed that
quasifilters can be thought of as labeled transition systems, and we
worked with such systems in Sections~\ref{sec:SRE}
and~\ref{sec:Algos}. Let us see how do the concepts introduced in
those sections work for SM-systems. We also defined
\emph{coupling} of SM-systems
(Definition~\ref{def:Coupling}), but we have not defined what it means
for a coupling to be ``good''. We will use sufficiency to approach
this subject.

Let $\E=(E,(S\times M),T)$ and $\I=(I,S\times M,B)$ be SM-systems.  We
think intuitively of $\E$ as ``the environment'' and $\I$ as the
``agent'', even though they share the set of sensorimotor parameters
$S\times M$. When is the coupling $\E*\I$ ``successful''? Given
another $\I'=(I',S\times M,B')$, how can we compare $\I$ and $\I'$ in
the context of~$\E$? The coupled system $\E*\I$ is not labeled; therefore, we
cannot apply the definition of sufficiency. However, as soon as we apply some
labeling to it, we can. There are many different ways to do it,
intuitively corresponding to the ``agent's perspective'', the
``environment's perspective'' and a ``god's perspective'' (or ``global
perpsective''). 

The first one is the labeling $h\colon E\times I\to I$, which is the
projection to the right coordinate, $h_I(e,i)=i$. The second one is
the projection to the left coordinate $h_E(e,i)=i$, and the third one
is the labeling of states by themselves, $h_G(e,i)=(e,i)$. Clearly,
$h_G$ is a refinement of both $h_E$ and~$h_I$. Yet another option is
to use the sensory data as labelings, which is a coarser labeling than
$h_I$. Or perhaps there was already a labeling $h\colon E\to S$ to
begin with, so then we can ask about the property of
$\hat h\colon E\times I\to S$ defined by $\hat h(e,i)=h(e)$. We
focus on what we called the agent's perspective, $h_I$, for the rest
of this section.

Recall Definition~\ref{def:DegreeOfInsufficiency} of the degree of
insufficiency. Given SM-systems $\E$ (environment) and $\I$
(agent), we can ask what is the degree of insufficiency of $h_I$
in~$\E*\I$?  The smaller the degree, the better the agent is attuned
to the environment. This says something about the way in which the
agent is adapted or attuned to the environment without attributing
contentful states or representations to the agent in alignment with
(EA2) and~(EA4).

Let $\E$, $\I$, and $\I'$ be SM-systems.  When is
$\deginsuf(\E*\I,h_I)<\deginsuf(\E*\I',h_{I'})$?  Of course, if $\I$
is fully constrained (Definition~\ref{def:FullyConstrained}), then
$\deginsuf(\E*\I)=\infty$. This corresponds to the agent never
engaging in any sensorimotor interaction with the environment. No
wonder that it can always ``predict'' the result of such passive
existence. Assume, however, that there some constraints on the
coupling.  For example, we may demand that the agent must regularly
visit states of some particular type to survive.  Subject to
such constrains, what can we say about $\deginsuf(\E*\I)$?  This seems
to be a good preliminary notion\footnote{Further research will
  indicate how much of this will be accepted by the most radical enactivists.} of attunement.

\section{Discussion}
\label{sec:Discussion}

In the introduction we defined our basic enactivist tenets:
\begin{enumerate}
\item[(EA1)] Embodiment and the inseparability of the brain-body-environment system,
\item[(EA2)] Grounding in sensorimotor interaction patterns, not in   
   contentful representations.
\item[(EA3)] Emergence from embodiment, enactment of the world,
\item[(EA4)] Attunement, adaptation, and skill as possibilities
to reliably engage in complicated patterns of activity with the 
environment.
\item[(EA5)] Perception as sensorimotor skills.
\end{enumerate}
We developed a model of sensorimotor systems and coupling
for which the purpose is to account for cognition mathematically, but in
congruence with the principles (EA1)--(EA5). The principle (EA1)
is intrinsic in the ways SM-systems are supposed to
model brain-body and body-environment dynamics. The central
ingredient is the control set $S\times M$ in all of those systems
which include ``motor'' and ``sensory'' part; it is \emph{impossible}
in our framework to model say the environment without acknowledging
the way in which the body is \emph{part of} it. The approach
that the actions of an agent depend solely on the history of its
sensorimotor interactions with the environment, our approach
is well in the scope of (EA2). We do not assume any representational
or symbolic content possessed by the SM-systems. We do not evaluate
them by the ``correctness'' of their internal states, but rather
by the ways in which they are, or can be, coupled to the environment
and whether their sensory apparatus generates a sufficient sensor 
mapping or not. Coupling of SM-systems is defined so that two systems 
constrain each other. Thus, when an agent is coupled to the environment,
they constrain each other, thereby creating new global properties of
the body-environment system. 
\iflongversion
The idea of emergence of global properties
from local ones is made precise in two ways
to account for (EA3). One is a logical framework very briefly outlined in
Section~\ref{sec:Logic}, and the other one is a dynamical systems approach,
which was alluded to in Remark~\ref{rem:DynSyst1} and
in Section~\ref{ssec:StrSuf}. 
\else
The (EA3) tenet of emergence of global properties
from local ones naturally arises in the framework through notions of coupling and sufficiency via sensorimotor activity, as well as
the dynamical systems appriach which was alluded in Remark~\ref{rem:DynSyst1} % This is in Section 2
\fi
The principle (EA4) is mostly
discussed in connection with minimal sufficient refinements.
Given a labeling, or a categorization, or an equivalence relation on 
the state space, one can ask how well does this labeling 
``predict itself''. The interpretation of this labeling can be anything
from a sensor mapping to the labeling of environmental states by
the internal states of the agent which coincide with them
(this is not representation, this is mere co-occurence; see enactivist
interpretation of the place cells in \cite{huttomyin2} for comparison).
A sufficient sensor mapping can be achieved in many different
ways. In Section~\ref{sec:Algos} we present a way in which the
agent ``develops'' new sensors to be better attuned to the environment
and in that way finds a sufficient sensor mapping. Another way for
the agent would be to learn to act in a way that excludes ``unpredictability''. Both are examples of situations where the
agent ``structures'' its own body-environment reality and gains skill.
Finally, perception (EA5) can be understood as sensorimotor patterns on
a microlevel. Can the agent engage in a sensorimotor activity
locally without making big moves, such as moving the eyes without moving
the body? The result of such sensorimotor interaction is another
labeling function on a macro level.
\iflongversion
The way to mathematically capture this idea is briefly outlined in Section~\ref{sec:Logic}.
\fi

In this paper, we not only presented mathematical definitions, but proved
a number of propositions and theorems about them. There would be (and we hope there will be!) 
much more of them, but they did not fit in this expository work
for which the main purpose was to demonstrate the connection of the mathematics
in question with the enactive philosophy of mind.
\iflongversion
\ 
\else
   We have already developed more concepts and theorems on top of this framework, including notions of \emph{degree of insufficiency}, \emph{universal covers}, \emph{hierarchies}, and \emph{strategic sufficiency}, but these are omitted here due to space limitations.
\fi
In other, more 
mathematical work, we plan to concentrate on working out
mathematical and logical details of the proposed theory as
well as applying the ideas to fundamental questions in robotics and autonomous systems, control theory, machine learning, and artificial intelligence.

\bibliography{ref}{}
\bibliographystyle{apacite}

\end{document}